\documentclass[letterpaper]{aa}

\usepackage{graphicx}
\usepackage{times}
\usepackage{psfig}
\usepackage{epsfig}
\usepackage{natbib}
\bibpunct{(}{)}{;}{a}{}{,}
\newcommand{\sqdegr}{\raisebox{0.65ex}{\tiny\fbox{$ $}}\,$^{\circ}$}

\def\hide#1{}

\begin{document}

\title{Galactic structure from the Calar Alto Deep Imaging Survey (CADIS)}
\author{S. Phleps\inst{1,2}, S. Drepper\inst{1}, K. Meisenheimer\inst{1}\and B.Fuchs\inst{3}}
\offprints{S.Phleps (sp@roe.ac.uk)}

\institute{
 Max-Planck-Institut f\"ur Astronomie, K\"onigstuhl 17,
   D-69117 Heidelberg, Germany
\and Institute for Astronomy, University of Edinburgh, Royal
 Observatory, Blackford Hill, Edinburgh EH9 3HJ, U.K.
\and Astronomisches Rechen-Institut, M\"onchhofstr. 12--14,
 D-69115 Heidelberg, Germany}
\titlerunning{Galactic Structure}
\date{Received 2. 6.  2004 / Accepted ?. ?. 2004}

\abstract{We used 1627 faint ($15.5\leq R\leq 23$) stars in five fields
  of the Calar Alto Deep Imaging Survey (CADIS) to estimate the
  structure parameters of the Galaxy. The results were derived by
  applying two complementary methods: first by fitting the density
  distribution function to the measured density of stars perpendicular
  to the Galactic plane, and second by modelling the {\it observed} colors
  and apparent magnitudes of the stars in the field, using Monte Carlo
  simulations. The best-fitting model of the Galaxy is then determined
  by minimising the C-statistic, a modified $\chi^2$. Our
  model includes a double exponential for the 
  stellar disk with scaleheights $h_1$ and $h_2$ and a power law halo with
  exponent $\alpha$. $24\,480$ different parameter
  combinations have been simulated. Both methods yield consistent
  results: the best fitting parameter combination is $\alpha=3.0$ (or
  $\alpha=2.5$, if we allow for a flattening of the halo with an axial
  ratio of $(c/a)=0.6$),  
  $h_1=300$\,pc,  $h_2=900$\,pc, and the contribution of thick disk
  stars to the disk stars in the solar neighbourhood is found to be between 4  and 10\,\%.

}
\maketitle
\section{Introduction}
The stellar structure of our own Galaxy has been studied intensively
for almost 400 years now, and although, or perhaps exactly because of
the fact that the subject to be analysed is our own neighbourhood and
surrounds us, there are still a large number of unknowns and the exact
structure parameters are still a topic of debate.

In the {\it standard model}
 (\citealp{BahcallSoneiraII80,BahcallSoneiraI80})
the Galaxy is of Hubble type $Sbc$,  consisting of an exponential disk,
 a central bulge,  and a spherical halo.
The vertical
structure of the disk follows an exponential law ($\rho\propto
e^{-z/h_z}$) with scaleheight $h_z=325$\,pc. However, as \citet{GilmoreReid83}
showed, the data can be fitted much better by a superposition of two
exponentials with scaleheights $h_1$ of $90$ to $325$\,pc and
$h_2\approx 1300$\,pc \citep{Gilmore84}.
It is not clear whether this deviation from the single
exponential is due to a distinct population of stars, although this is
suggested by the different kinematics and lower metallicities
\citep{Freeman92}, it is referred to as ''thick disk'' in the
literature.

In the last two decades highly efficient, large area surveys have been
carried out, and large quantities of high-quality imaging data have
been produced. The mere existance of a thick disk component is fairly
established today, however, the size of its scaleheights are still
under discussion \citep{Norris99,Siegel02}, as several authors claim to have found a
considerably smaller value for the scaleheight of the thick disk than
the canonical one of $h_2\sim1.3$\,kpc: \citet{Robin00} found 750\,pc,
and \citet{SloanStars01} found the scaleheight to be between 580 and
750\,pc.

In general there are two different approaches to deduce Galactic
Structure: the Baconian, and the Cartesian {\it ansatz}, as
\citet{GilmoreWyse87} call it.

The {\it Baconian ansatz} tries to manage with a minimum of
assumptions: by a pure {\it measurement} of the stellar
distribution function by means of starcounts. 
Distances are estimated for each star and then their number
density in dependence of distance from the Galactic plane is
calculated \citep{GilmoreReid83,Reid96,Reid97,Gould98}. The
parameters can then be determined by fitting  distribution
functions to the data. In a first paper on CADIS deep star counts
(\citealp{Phleps00}, Paper I in the following) we presented
first results based on $\approx 300$ faint stars ($15.5 \leq R
\leq 23$) in two CADIS fields covering an area of $\sim
1/15~\sq\degr$ in total. From these data we deduced the density
distribution of the stars up to a distance of about 20\,kpc above
the Galactic plane, using the Baconian ansatz. We found $h_1= 280$\,pc, and
unambigously the contribution of the thick disk with $h_2$ on the
order of $1300$\,pc.

On the other hand, as the main structural features of the Milky Way have
been identified, it has been possible to design synthetic models
of the Galactic stellar populations
(\citealp{BahcallSoneira81,Gilmore84,Robin86,ReidMajewski93,MendezvanAltena98,Chen99,SloanStars01}),
assuming the forms of the density distribution function for the
different components of the Galaxy. This method is what
\citet{GilmoreWyse87} call the {\it Cartesian ansatz}.

The exact structure parameters can then be deduced by comparing
model and star count data.

In this paper we use both approaches using the stellar
component of the CADIS data. Deep ($R<23$) multi
color data is now available for 1627 faint stars in five fields at high Galactic
latitude and different Galactic longitudes. 

In Section \ref{CADIS} the Calar Alto Deep Imaging Survey is briefly
described. In Section \ref{starcounts} the results we deduced by applying the
''classic'', Baconian method are
shown, whereas in Section \ref{model} we show the results given by the modelling
approach. A summary and discussion is given in Section \ref{discussion}.

\section{The Calar Alto Deep Imaging Survey}\label{CADIS}
The {\bf C}alar {\bf A}lto {\bf D}eep {\bf I}maging
{\bf S}urvey was established in 1996 as the extragalactic key project of
the Max-Planck Institut f\"ur Astronomie. It combines a very deep
emission line survey carried out with an imaging
Fabry-Perot interferometer with a deep multicolour survey using three
broad-band optical to NIR filters and up to thirteen medium-band
filters. The combination of different observing
strategies facilitates not only the detection of emission line objects but
also the derivation of photometric spectra of all objects in the fields
without performing time consuming slit spectroscopy. 

The seven CADIS fields measure $\approx 1/30~\sq\degr$
 each ($11''\times11''$) and are
located at high Galactic latitudes to avoid dust absorption and
reddening. In all fields the total flux on the IRAS 100\,$\mu$m maps
is less than 2\,MJy/sr which corresponds to $E_{B-V} <0.07$. Therefore
we do not have to apply any colour corrections. As a second selection
criterium the fields should not contain any star brighter than
$\approx 16^{mag}$ in the CADIS $R$ band. In fact the brightest star
in the five fields under consideration  has an $R$ magnitude of
$15.42^{mag}$. 

All observations were performed on Calar Alto, Spain. In
the optical wavelength region  the focal reducers CAFOS (Calar
Alto Faint Object Spectrograph) at the 2.2 m telescope and MOSCA
(Multi Object Spectrograph for Calar Alto) at the 3.5 m telescope were
used. The NIR observations have been carried out using the Omega Prime
camera at the 3.5m telescope.

In each filter, a set of 5 to 15 individual exposures was taken. The
images of one set were then bias subtracted, flatfielded and 
corrected for cosmic ray hits, and then coadded to one deep sumframe. This basic data
reduction steps were done with the MIDAS software package in combination with the
data reduction and photometry package MPIAPHOT (developed by  H.-J.~R\"oser and
K.~Meisenheimer).
\subsection{Object detection and classification}
Object search is done on the sumframe of each filter using the source
extractor software {\bf SE}xtractor \citep{Bertin96}. The
filter-specific object lists are then merged into a master list
containing all objects exceeding a minimum $S/N$ ratio in any of the
bands. Photometry is done using the program {\it Evaluate}, which has
been developed by Meisenheimer and R\"oser (see
\citealp{MeisenroeserEval}, and \citealp{Roeser91}). Variations in seeing
between individual exposures are taken into account, in order to get
accurate colours. Because the photometry is performed on individual
frames rather than sumframes, realistic estimates of the
photometric errors can be derived straightforwardly.

The measured counts are translated into physical fluxes outside the
terrestrial atmosphere by using a set of
''tertiary'' spectrophotometric standard stars which were established in the
CADIS fields, and which are calibrated
with secondary standard stars \citep{Oke90,eso} in photometric
nights.

From the physical fluxes, magnitudes and colour indices (an object's
brightness ratio in any two filters, usually given in units of
magnitudes) can be calculated. The CADIS magnitude system is described
in detail in \citet{Wolf01a} and \citet{Fried01}.

The CADIS color is defined by:
\begin{eqnarray}\label{CADIScolor}
(b-r)=2.5 \log \frac{F_R^\gamma}{F_B^\gamma}~.
\end{eqnarray}
Here $F^\gamma_k$ is the flux outside the atmosphere in units of Photons ${\rm
m}^{-2} ~{\rm s}^{-1} ~{\rm nm}^{-1}$ in the CADIS filter $k$.

$b-r$ can be calibrated to the Johnson-Cousins system (the CADIS $R_C$ is
very close to the Cousins $R$) by using Vega as
a zero point:
\begin{eqnarray} 
(B-R)_C &=& (b-r) + 2.5 \log \frac{F^\gamma_{{\rm Vega}}(\lambda=440 {\rm
nm})}{F^\gamma_{{\rm Vega}}(\lambda=648 {\rm nm})}\nonumber\\
&=& (b-r)+0.725~.
\end{eqnarray}

With a typical seeing of $1\farcs5$ a morphological star-galaxy
separation becomes unreliable at $R\ga 21$ where
already many galaxies appear compact. Quasars have point-like
appearance, and thus can not be distinguished from stars by
morphology. Therefore a classification scheme was developed, which is
based solely on template spectral energy distributions ($SED$s)
\citep{Wolf01a,Wolf01b}. The classification algorithm basically
compares the observed colours of each object with a colour library of
known objects. This colour library is assembled from observed spectra
by synthetic photometry performed on an accurate representation of the instrumental
characteristics used by CADIS. As input for the stellar library we
used the catalogue by \citet{Pickles98}.

Using the minimum variance estimator (for details see \citealp{Wolf01a}), each
object is assigned a type (star -- QSO -- galaxy), a redshift (if it is
not classified as star), and an $SED$.  

Note that we do not apply any morphological star/galaxy separation or
use other criteria. The classification is purely spectrophotometric.

Five CADIS fields have been fully analysed so far (for coordinates see
Table \ref{coordtab}). We identified 1627 stars with $R\leq 23$.
The number of stars per field is given
in Table \ref{coordtab}.

\begin{table}
\caption[ ]{The Galactic coordinates of the five fields investigated so far,
and the number of stars per field, $R\leq 23$.\\\label{coordtab}}
\begin{tabular}{r|r r c  }
CADIS field& $l$& $b$& $N_{stars}$\\ \hline
1\,h&$150^\circ$&$-59^\circ$&$200$\\
9\,h&$175^\circ$&$45^\circ$&$300$\\
13\,h&$335^\circ$&$60^\circ$&$517$\\
16\,h&$85^\circ$&$45^\circ$&$390$\\
23\,h&$90^\circ$&$-43^\circ$&$220$\\
\end{tabular}
\end{table}

\section{The stellar density distribution}\label{densitychapter}
Since none of our fields points towards
the Galactic center and all of them are located at high Galactic latitudes,
neither a bulge component nor spiral arms are needed to describe the
observations. Our model of the density distribution of the stars
includes only a disk and a halo. 
 
\subsection{The distribution of stars in the Galactic disk}\label{densityindisk}
The {\it vertical} density distribution of stars in the disk is described by a sum of
two exponentials (the ``thin'' and a ``thick'' disk component):
\begin{eqnarray}
\rho(z)=\left[n_1 \exp\left(-\frac{z}{h_1}\right)+n_2
  \exp\left(-\frac{z}{h_2}\right)\right]~,
\end{eqnarray}
where $z$ is the vertical distance from the Galactic plane, $n_1$
and $n_2$ are the normalisations at $z=0$, and  $h_1$ and $h_2$
are the scaleheights of thin and thick disk, respectively.

The {\it radial change} of the density can be taken into account by
assuming an
  exponential decrease with scalelength $h_r$. The value of $h_r$ is
still poorly determined, measurements range from values as small as
$h_r=2$\,kpc  (\citealp{Robin92,Ruphy96,Ojha96,Bienayme99}) to
 $h_r=3.5$\,kpc 
  (\citealp{BahcallSoneiraI80,BahcallSoneira81,Wainscoat92}).

For each
  field with Galactic longitude $l$ and latitude $b$, respectively, we
  can describe the  distribution of the stars in the disk by
\begin{eqnarray}
\rho_d(r,l,b)=\left[n_1 \exp\left(-\frac{r \sin b}{h_1}\right)+n_2
  \exp\left(-\frac{r \sin b}{h_2}\right)\right]\\\nonumber
\cdot\exp\left[-\frac{\sqrt{r^2-2
  R_\odot\frac{r \sin b}{\tan b}\cos l+r^2 \sin^2
  b}+R_\odot}{h_r}\right]~.\label{diskequation} 
\end{eqnarray}
The distance
of the sun from the Galacic center is assumed to be $R_\odot=8$\,kpc.

\subsection{The distribution of stars in the Galactic halo}\label{densityinhalo}
The space density corresponding to the $r^{1/4}$ law in
projection has no simple analytic form \citep{Young76}. There are
different possibilities to find an approximation, the simplest of
which is a power law \citep{FuchsHalo} of the form
\begin{eqnarray}
\rho=\rho_\odot\cdot\left[\frac{r_c^2+R_\odot^2}{r_c^2+r^2}\right]^{\alpha/2}~,\label{halofkt}
\end{eqnarray}
with an arbitrarily chosen core radius of $r_c=1$\,kpc,
and
\begin{eqnarray}
r=R_\odot^2+\frac{z^2}{\sin^2 b}-2 R_\odot\frac{z}{\tan b}\cos l+\frac{z^2}{(c/a)^2}~.
\end{eqnarray}

Several authors found that the
stellar halo is significantly flattened, with an axial ratio of $(c/a)$
around $0.6$ (\citealp{Robin00,Lemon04}. 

\section{Star Counts -- a direct measurement}\label{starcounts}
Now, as we have more than twice the number of fields available than in
our first analysis, and due to the increased number of filters used
for the classification of 
the objects, the statistics has
increased considerably, from $\sim 300$ stars in Paper I to
$\sim 1630$ now.

\begin{figure}[h]
\centerline{\psfig{figure=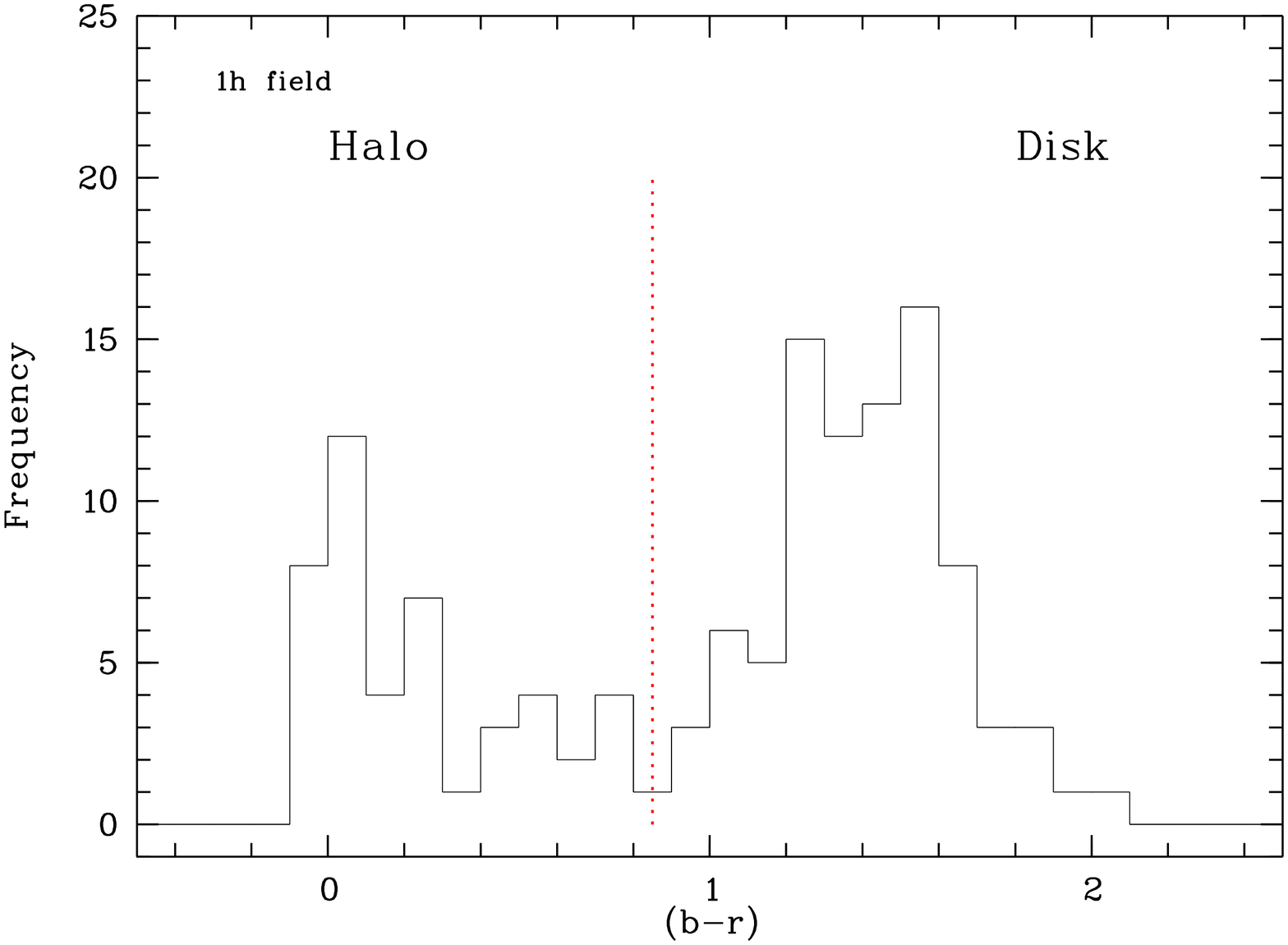,angle=270,clip=t,width=5.6cm}}
\centerline{\psfig{figure=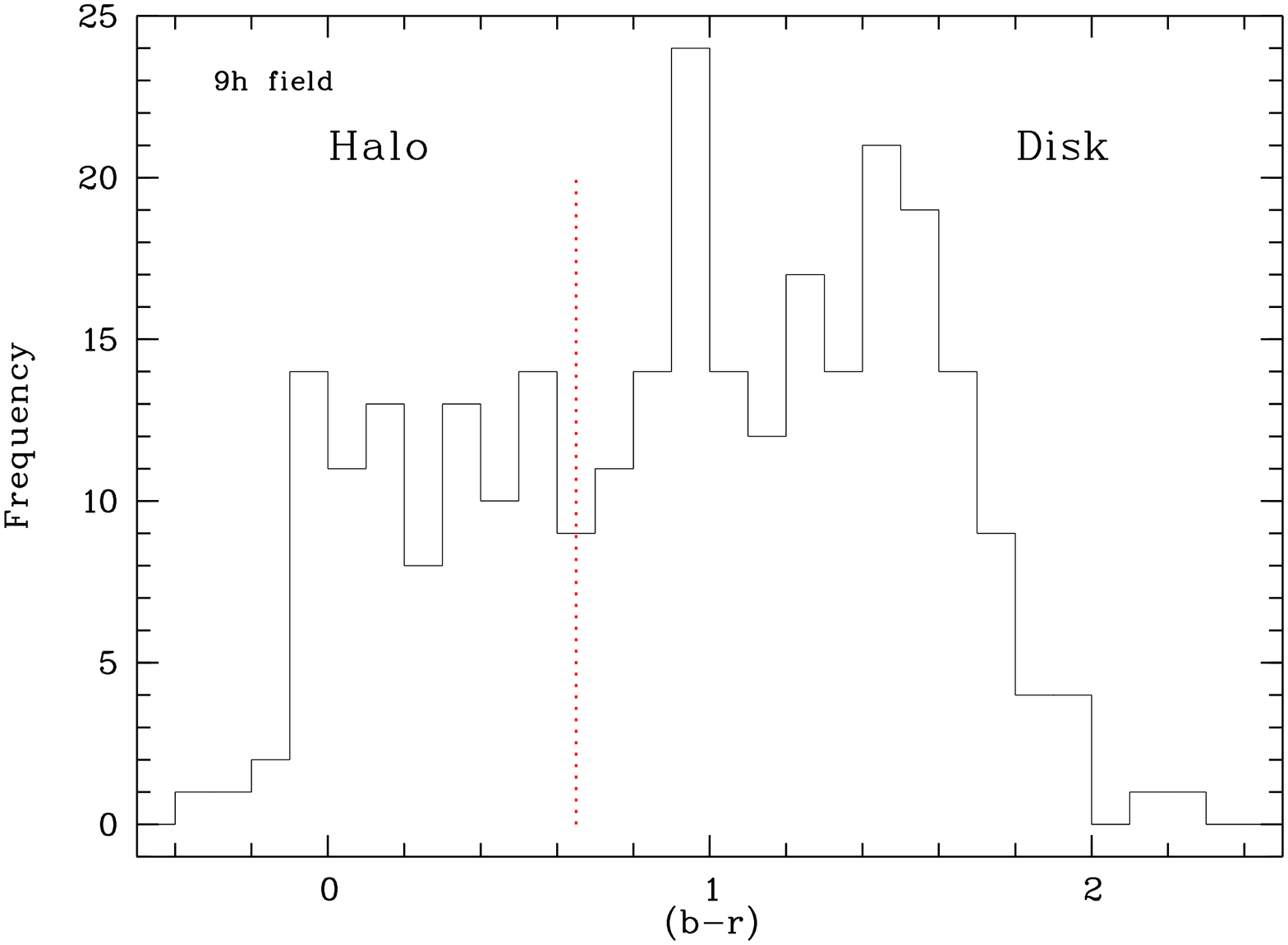,angle=270,clip=t,width=5.6cm}}
\centerline{\psfig{figure=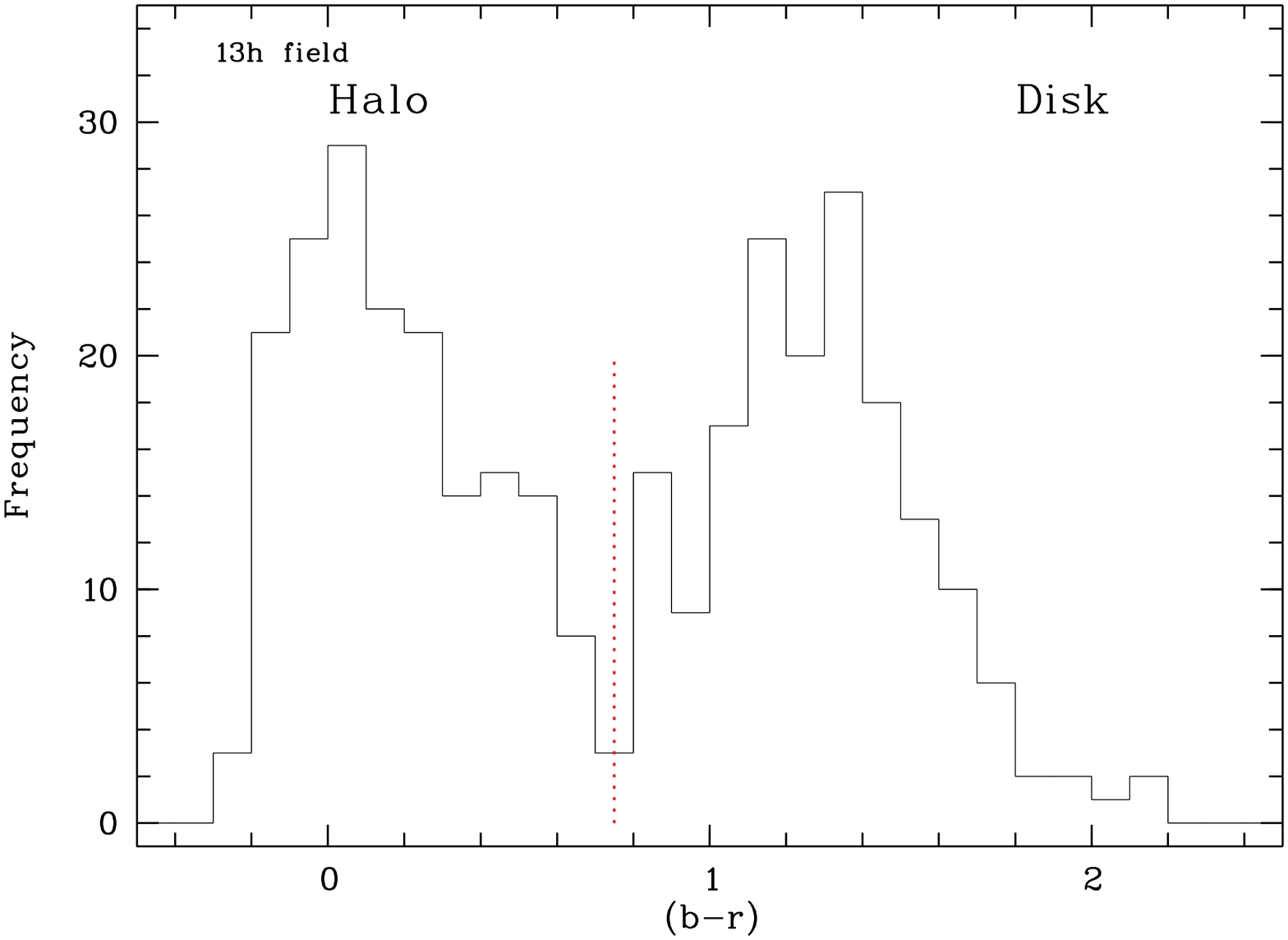,angle=270,clip=t,width=5.6cm}}
\centerline{\psfig{figure=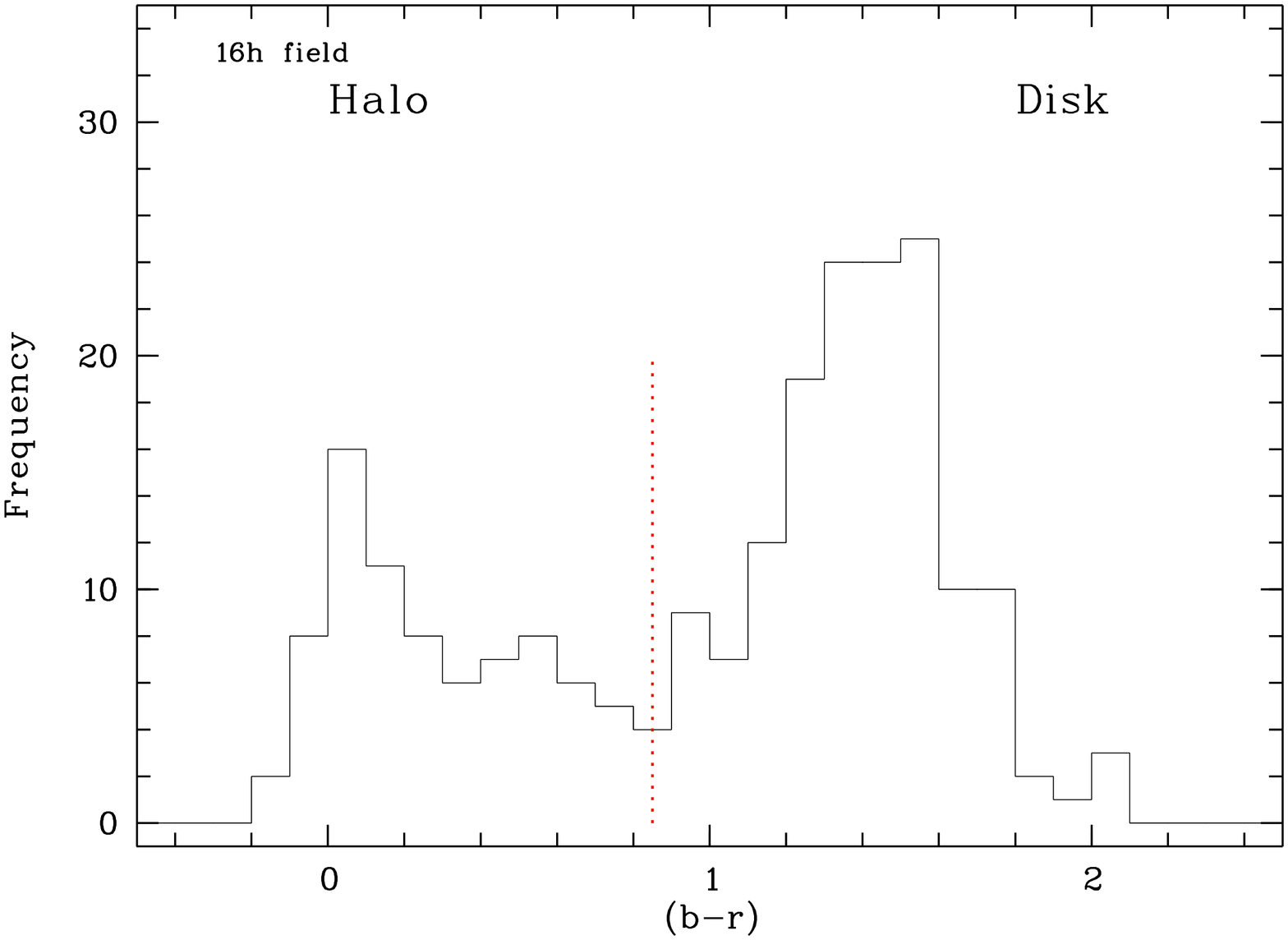,angle=270,clip=t,width=5.6cm}}
\centerline{\psfig{figure=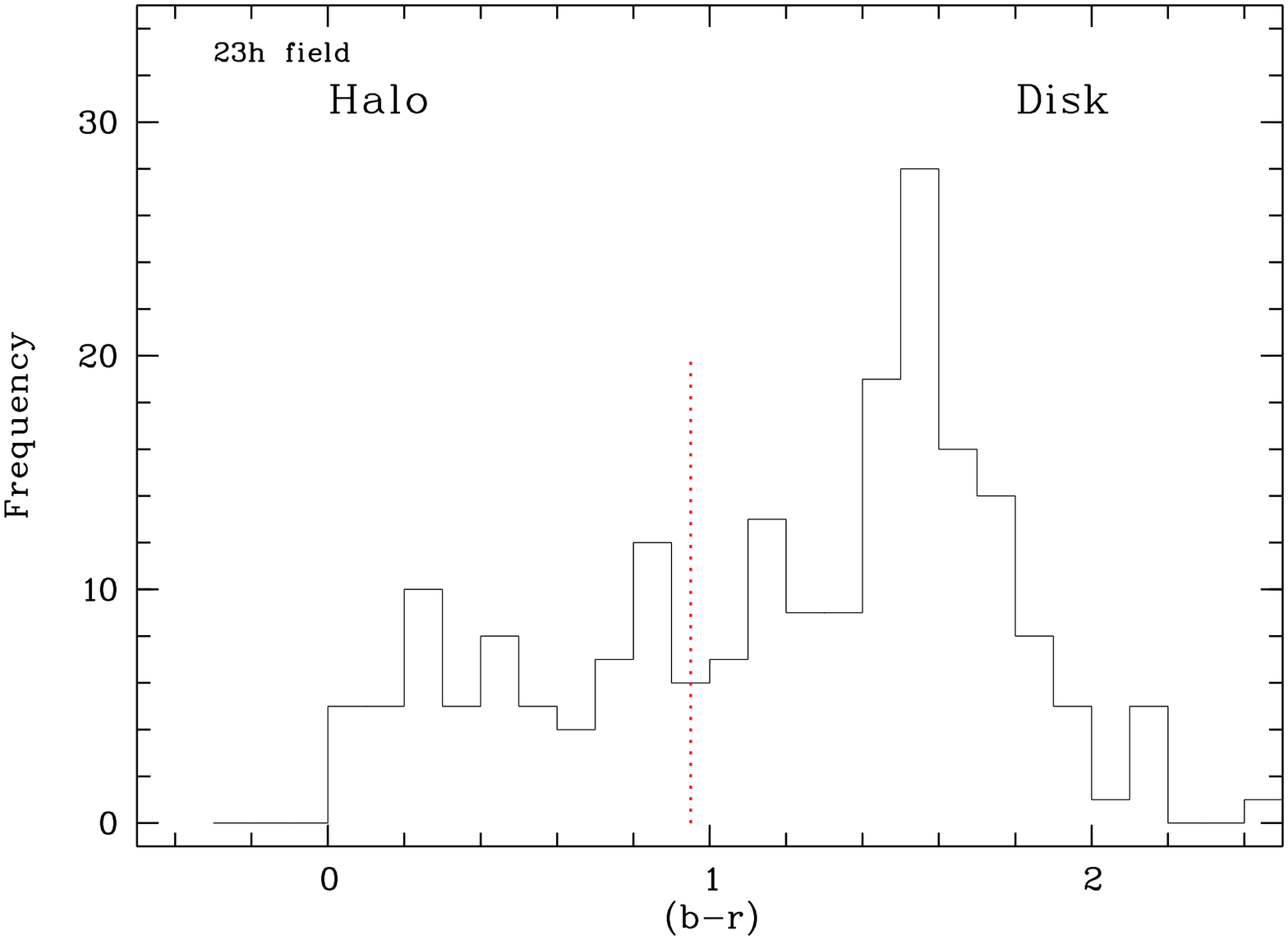,angle=270,clip=t,width=5.6cm}}
\caption[ ]{The distribution of observed $(b-r)_C$ colors in the
fields. The dotted line shows the color cut between halo and
disk. Note that the ordinate has not the same scale in all cases,
whereas the dotted line extends to 20 in each plot. The maxima are not
located at the same colors, but depend on the direction of the field.\label{colorhisto}}
\end{figure}

Figure (\ref{colorhisto}) shows the distribution of measured colors of the selected stars
in the field. As in our first analysis, we use this bimodal
distribution of colors, which comes from the interaction between
luminosity function, density distribution and apparent magnitude
limits ($15.5\leq R \leq 23$), to discriminate between disk and halo
stars. The nearby disk stars are intrinsically faint and therefore
red, whereas the halo stars have to be bright enough to be seen at
large distances, and
are therefore blue. According to the different directions of the
fields the cut is located at slightly different colors, as indicated
in Figure \ref{colorhisto}. Unfortunately it is not possible to
dicriminate between thin and thick disk stars in the same way, so in
the calculation of absolute magnitudes and thus distances we have to 
treat them in the same way.

We repeated the calculation of the photometric parallaxes
of the stars exactly as  described in Paper I. Absolute magnitudes are inferred from a mean
main sequence relation in a color-magnitude diagram
\citep{Lang92}. The main sequence converted into a $M_R$ vs $(b-r)$
relation (see Paper I) can be approximated by a fourth order
polynomial in the range $-2\leq(b-r)\leq1.8$:
\begin{eqnarray}\label{polyfit}
M_R&=&c_0+c_1~(b-r)+c_2~(b-r)^2\nonumber\\
&+&c_3~(b-r)^3+c_4~(b-r)^4,
\end{eqnarray}
the parameters of which are:\\
\begin{center}
$c_0=4.01236$\hspace{0.6cm} $c_1=4.12575$\hspace{0.6cm}
$c_2=-1.89076$\hspace{0.6cm} $c_3=0.762053$\hspace{0.6cm}
$c_4=0.341384$~.\\
\vspace{0.3cm}    
\end{center} 

This main sequence approximation is
valid for all stars in our sample since we can be sure that it is free
from contamination of any non-main sequence
stars, as the faint magnitude intervall we observe ($16 \leq R \leq
23$) does not allow the detection of a giant star.

The mean main sequence
relation is valid strictly only for stars with solar metallicities, whereas our
sample may contain stars spread over a wide range of different
metalicities. The influence of the varying metalicities on the main
sequence color-magnitude relation is taken into account as described
in Paper I: Halo stars are supposed to be metal poor, and are known to
be fainter than  disk stars with the same colors, so the relation is
shifted towards fainter magnitudes by $\Delta M_R=0.75$. This value is the
mean deviation from the mean main sequence defined by the CNS4 stars
\citep{Venice} of a subsample of 10 halo stars for which absolute $R$
magnitudes were available (Jahrei{\ss}, priv. com).\footnote{Note that this artificial
separation may well lead to wrong absolute 
magnitudes for individual stars (e.g. a disk star with $r<1$\,kpc but
$b-r<0.7$), but should be correct on average.}

\begin{figure}[h]
\centerline{\psfig{figure=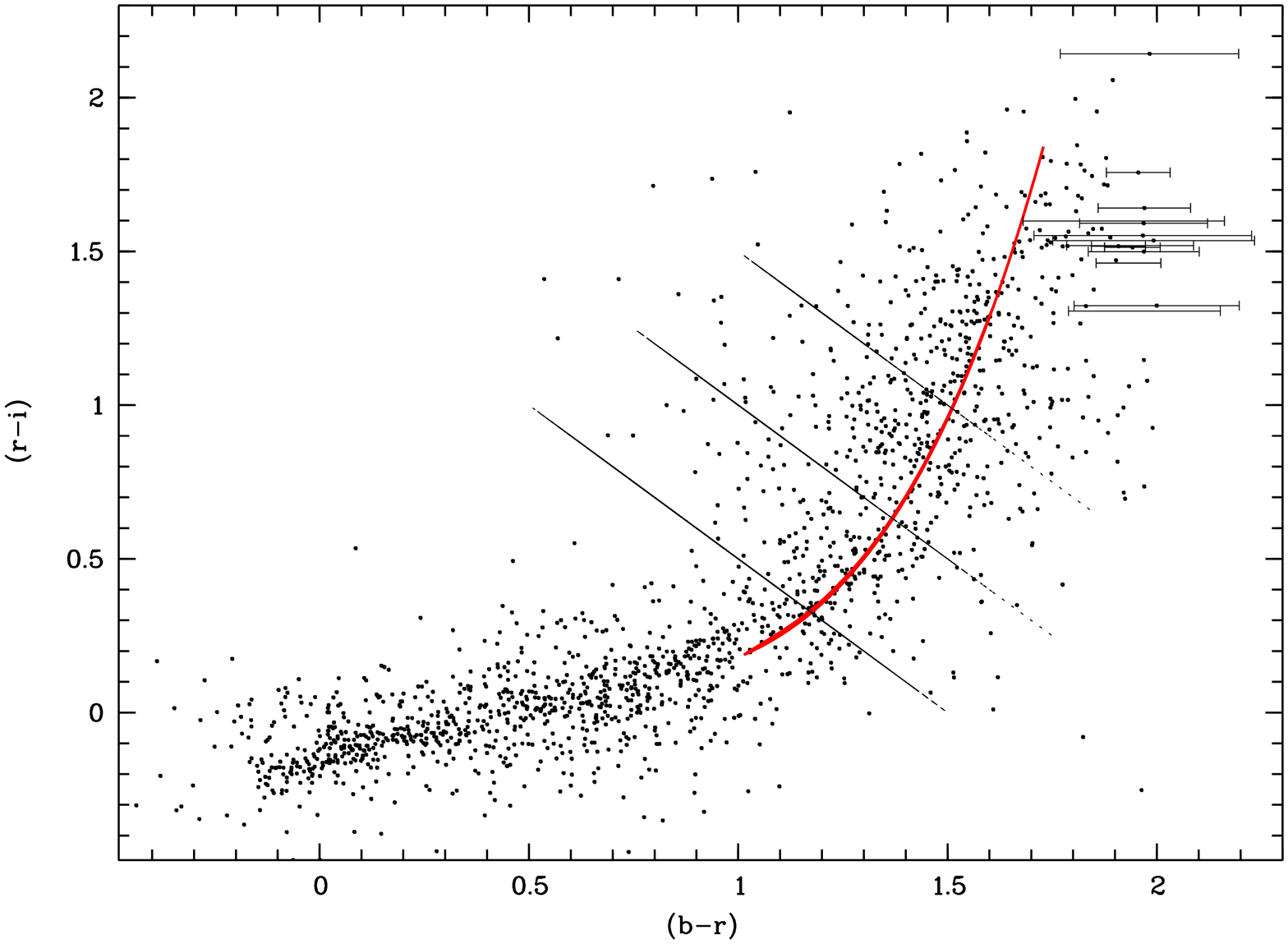,angle=270,clip=t,width=9.cm}}
\caption[ ]{The stars in the five fields in a two color diagram
($(b-r)$ versus $(r-i)$). The spread becomes significant at $(b-r)
\geq 1.0$. The solid line is the mean main sequence track, onto which
the $(b-r)$ colours of the stars are projected along the $R$ isophotes of
varying metallicity (dotted lines). For some extremely red stars, photometric errors are
plotted. \label{metall}}
\end{figure}

The spread in a two
color diagram $(b-r)$ versus $(r-i)$ (that is the CADIS  color between
$R_C$ and $I_{815}$, analog to Eq. \ref{CADIScolor}) becomes significant at 
$(b-r) \approx 1.0$, see Fig. \ref{metall}. The maximal photometric error
for the very faint stars is $0.15^{mag}$.
Here metallicity effects will distort the relation 
between the measured $(b-r)$ colors and the spectral type (temperature) and
thus lead to wrong absolute magnitudes, so we have to correct for
metallicity in order to avoid errors in the photometric parallaxes.

The $R_C$
filter is strongly affected by metallicity effects like absorption
bands of TiO$_2$ and VO molecules in the stars' atmosphere, whereas
the $B_C$ and the medium-band filter $I_{815}$ (the 
wavelength of which was chosen in order to avoid absorption bands in
cool stars) are not. So in a
first approximation 
we can assume the ''isophotes'' of varying metallicity in a $(b-r)$
versus $(r-i)$ two color diagram to be straight
lines with a slope of 
$-1$, along of which we project the measured colors with $(b-r) \geq
1.0$ onto the mean main 
sequence track which in the interval $1.0 \leq (b-r) \leq 1.8$ is defined by
\begin{eqnarray}
(r-i)&=& 0.39~(b-r)_{{\rm corr}}^4  -0.36~(b-r)_{{\rm corr}}^3\nonumber\\
&+& 0.09~(b-r)+0.06~. 
\end{eqnarray}

\begin{figure*}
\centerline{\psfig{figure=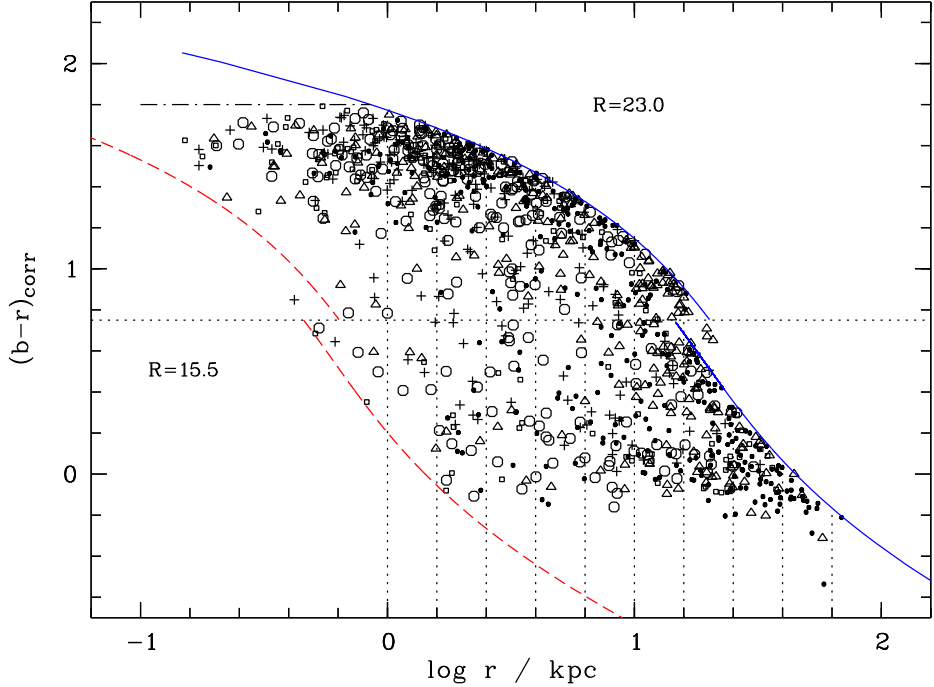,angle=270,clip=t,width=14cm}}
\caption[ ]{Spatial distribution of metallicity corrected
$(b-r)_{{\rm corr}}$.
The solid line represents
the distance dependent upper 
color limit at $R=23^{mag}$, the
dashed line is the lower limit due to the selection criterium of the
fields (no star brighter than $R\approx 15.5$). Since the metal-poor
halo stars are intrinsically fainter, the color limits are shifted
accordingly. The horizontal line at $(b-r)_{\rm corr}=0.75$
denotes the cut between halo and disk stars in the 13h\,field (which
is why stars can be found ''outside the boundaries'': the color cut is
different in each field, see Fig. \ref{colorhisto}), the dashed--dotted line
indicates the cutoff at $(b-r)_{{\rm corr}}=1.8$ due to the metallicity
correction. The different symbols 
refer to the different fields: squares --  1\,h field, triangles --
9\,h field, dots -- 13\,h field, open circles -- 16\,h field, crosses
-- 23\,h field.\label{wichtig}}
\end{figure*}

This projection implies that stars with $(b-r)_{{\rm corr}}\ga
1.8$ cannot exist in
Fig. \ref{wichtig}, which shows the spatial distribution of metallicity
corrected $(b-r)_{{\rm corr}}$ colors (the limit is indicated by the
dashed--dotted line).  Both the upper and lower magnitude limits lead
to selection effects which have to be taken into account.

\subsubsection{Completeness Correction}
We correct for incompleteness in the following
way (for a detailed derivation of the completeness
correction and the corresponding errors we refer the reader to
Appendix 1  of Paper I):
First, we divide
the distance in logarithmic bins of 0.2 and count the stars up to distance dependent
color (that is, luminosity) limit, up to which stars can be detected,
see Fig. \ref{wichtig}.
The nearest bins ($-0.8\leq \log r \leq -0.2$) are
assumed to be complete. For the incomplete bins we multiply
iteratively with a factor given by the ratio of complete to incomplete
number counts in the previous bin, where the limit for the
uncorrected counts is defined by the bin currently under examination ($j$):
\begin{eqnarray}\label{Ncorrect}
N_j^{{\rm corr}}=N_j \prod_{i=1}^{j-1} (1+\frac{N_i^{''}}{N_i^{'}})~,
\end{eqnarray}
where $N_j$ is the number of stars in bin $j$, $N^{'}$ is the number
of stars in the previous bin ($j-1$), up to the
limit given by the bin $j$, and $N^{''}$ is the number of stars from
that limit up to the limit given by bin $j-1$.

With the poissonian errors $\sigma_{N}=\sqrt{N}$,
$\sigma_{N^{'}}=\sqrt{N^{'}}$, and $\sigma_{N^{''}}=\sqrt{N^{''}}$  the
error of the corrected number counts becomes:
\begin{eqnarray}
\sigma^2_{N_j^{{\rm corr}}}&=&\sigma^2_{N_j}\prod_{i=1}^{j-1}
\left(1+\frac{N_i^{''}}{N_i^{'}}\right)^2\nonumber\\
&+&\sum_{i=1}^{j-1}\left[\left(\frac{1}{N_i^{'}}\right)^2\sigma^2_{N_i^{''}}+\left(\frac{N_i^{''}}{N_i^{'}}\right)^2\sigma^2_{N_i^{'}}\right]\nonumber\\
&\cdot&\prod_{i=1 \atop i\not= j}^{j-1}\left(1+\frac{N_i^{''}}{N_i^{'}}\right)
\end{eqnarray}
Note that the completeness correction is done for each field
separately, thus the different color cuts by which we separate disk
from halo stars can be taken into account. Also since the
color-magnitude relation is different for disk and halo stars, we
calculate the correction factors separately for the two
populations. The correction factors are usually in the range $1\leq
C\leq 10$, but can be as large as 20 for the last bin of the halo.

With the corrected number counts the density in the logarithmic spaced
volume bins ($V_j=\frac{1}{3}\omega (r_{j+1}^3 - r_j^3)$) can then be
calculated according to
\begin{eqnarray}
\rho_j=\frac{N^{{\rm corr}}_j}{V_j} =\frac{C \cdot N_j}{V_j}~,
\end{eqnarray}
For every logarithmic distance bin we use the mean height $z$ above
the Galactic plane $<z_j>=\sin b \cdot <r>$, where $<\log r> = \log
r_j+(\log r_{j+1}-\log r_i)/2=\log r_i+0.1 \Rightarrow <r>=1.259
r_i$.

 From the estimated density distribution in the five fields we now deduce the
structure parameters by fitting the distribution function to the
data.

\subsection{The distribution of stars in the Galactic disk}
We first calculate the density of the stars which we assume to be disk stars, that
is, with colors redder than the color cut (see Figure
\ref{colorhisto} in the field under consideration.

As the nearest stars in our fields have still distances of about
200\,pc the normalization at $z=0$ has to be established by other
means: We take stars from the CNS4 (Fourth Catalogue of Nearby Stars,
\citet{Venice}), which
are located in a sphere with radius 20\,pc around the sun.
The stars in our normalization sample
are selected from the CNS4 by their absolute visual magnitudes,
according to the distribution of absolute magnitudes of the CADIS disk
stars ($6.5\leq M_v \leq 14.5$).

Since the thin disk completely dominates the first kpc and there is no indication of
a change of slope within this distance \citep{FuchsWielen93}, it is
justified to fit the density distribution in this range with a single
exponential component (taking of course the radial decrease into
account). The sum of thick and thin disk stars in the solar
neighbourhood, $(n_1+n_2)=6.566\cdot 10^7$\,kpc$^{-3}$ with an
uncertainty of 10\%, is given by
the CNS4 stars. We estimate the scaleheight $h_1$ from this fit, assuming
two extreme values of the 
scalelength: $h_r=2.0$\,kpc, and $h_r=3.5$\,kpc, respectively (see
Table \ref{scaleheights}). For $h_r=2.0$\,kpc we find that the
weighted mean of the scaleheight of the thin disk is 
$\overline{h_1}=281\pm13$, for $h_r=3.5$\,kpc we find
$\overline{h_1}=283\pm13$. These values are indistinguishable from
each other, and also the values of $\chi^2$
indicate that the value adopted for $h_r$ does not influence the
measurement of the scaleheight $h_1$.

\begin{table}
\caption[h]{The scaleheights of the thin disk, assuming a scalelength
of $h_r=2.0 {\rm kpc})$ and $h_1(h_r=3.5 {\rm kpc})$,
respectively. The numbers in brackets are the corresponding values of $\chi^2$. \\\label{scaleheights}}
\begin{tabular}{r|l l}
CADIS field&$h_1(h_r=2.0 {\rm kpc})$&$h_1(h_r=3.5 {\rm kpc})$\\ \hline
1\,h&$249\pm25$ $(0.92)$&$255\pm26$ $(0.92)$\\
9\,h&$336\pm40$ $(1.21)$&$312\pm35$ $(1.21)$\\
13\,h&$249\pm24$ $(0.62)$&$255\pm25$ $(0.63)$\\
16\,h&$321\pm33$ $(0.41)$&$319\pm32$ $(0.41)$\\
23\,h&$319\pm33$ $(1.23)$&$317\pm33$ $(1.21)$\\
\end{tabular}
\end{table}

We then fitted equation (\ref{diskequation})
to the data over the range $0< z < 4$\,kpc, again for $h_r=2.0$\,kpc,
and $h_r=3.5$\,kpc, respectively, keeping the corresponding values of $h_1$ fixed.
With $h_1$ and $(n_1+n_2)$ 
known, the free parameters are
now the scaleheight $h_2$ of the thick disk, and the ratio
$n_2/(n_1+n_2)$. Again the measurement is insensible to the
scalelength assumed in equation (\ref{diskequation}). This is due to
the high Galactic latitude of the CADIS fields ($45\leq b \leq 60$) --
the halo becomes dominant already at a projected radial distance of
only about $7$\,kpc.

\begin{figure*}[hp]
\centerline{\psfig{figure=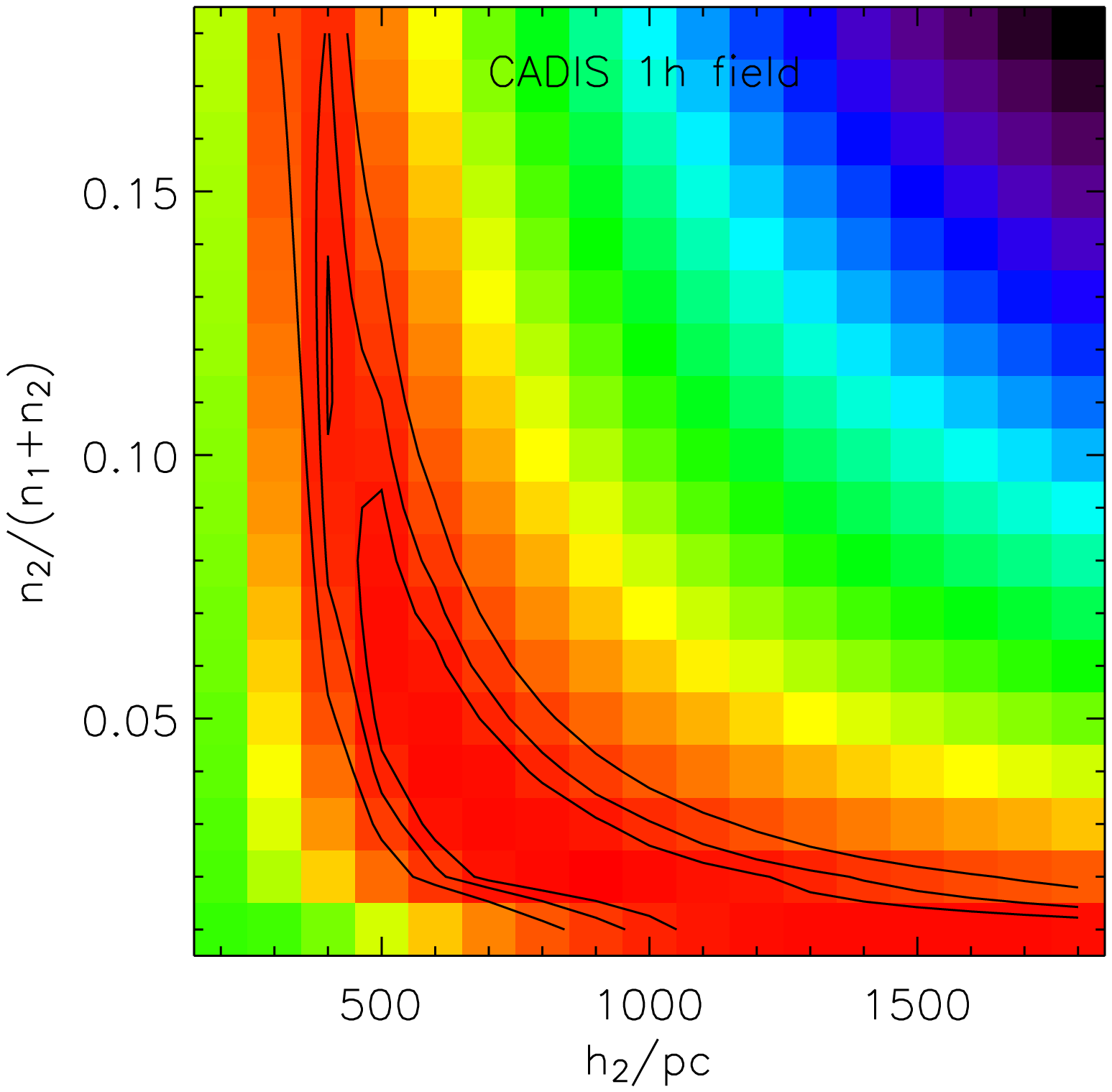,clip=t,width=8.0cm}
\psfig{figure=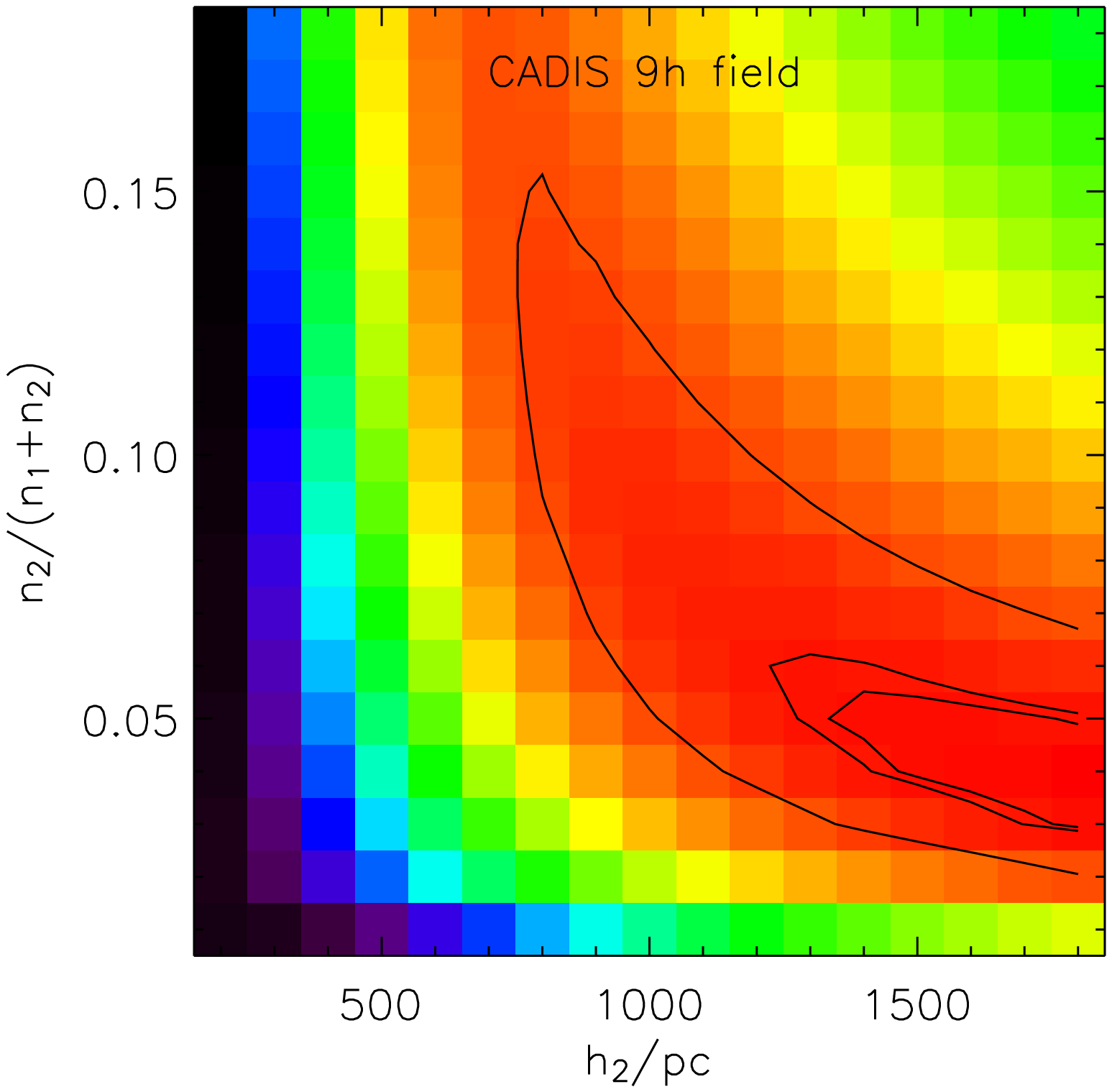,clip=t,width=8.0cm}}
\centerline{\psfig{figure=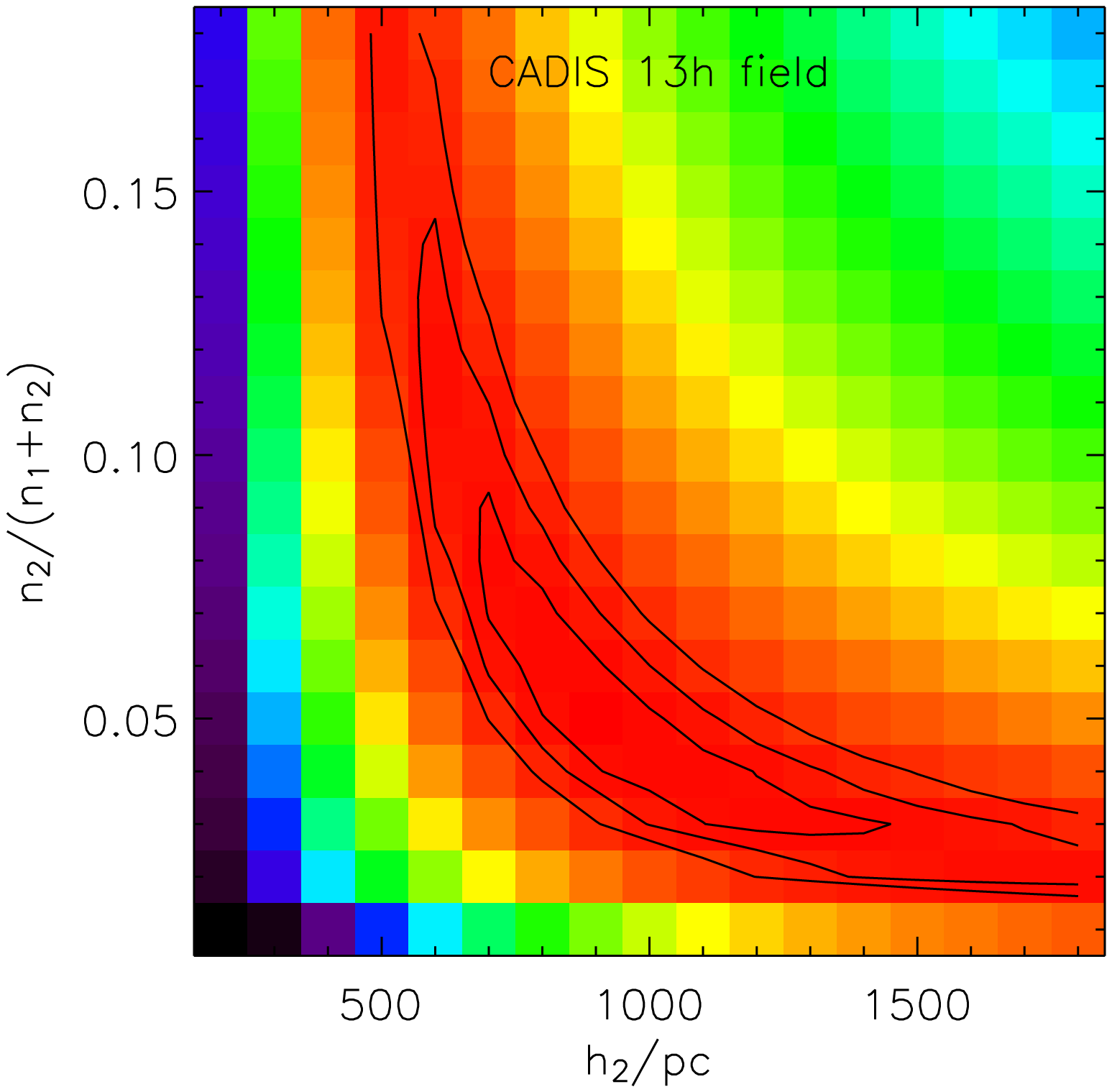,clip=t,width=8.0cm}
\psfig{figure=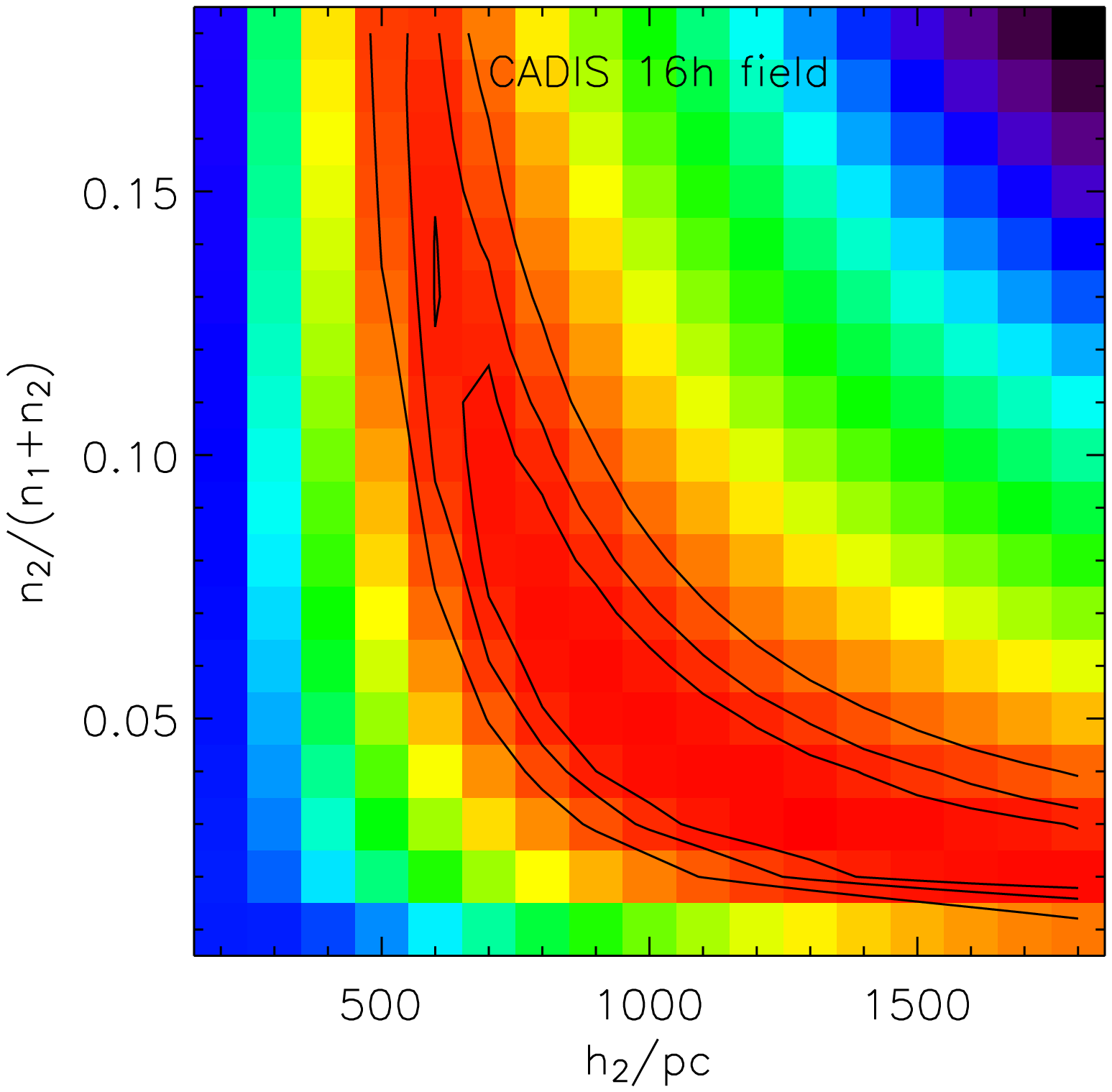,clip=t,width=8.0cm}}
\centerline{\psfig{figure=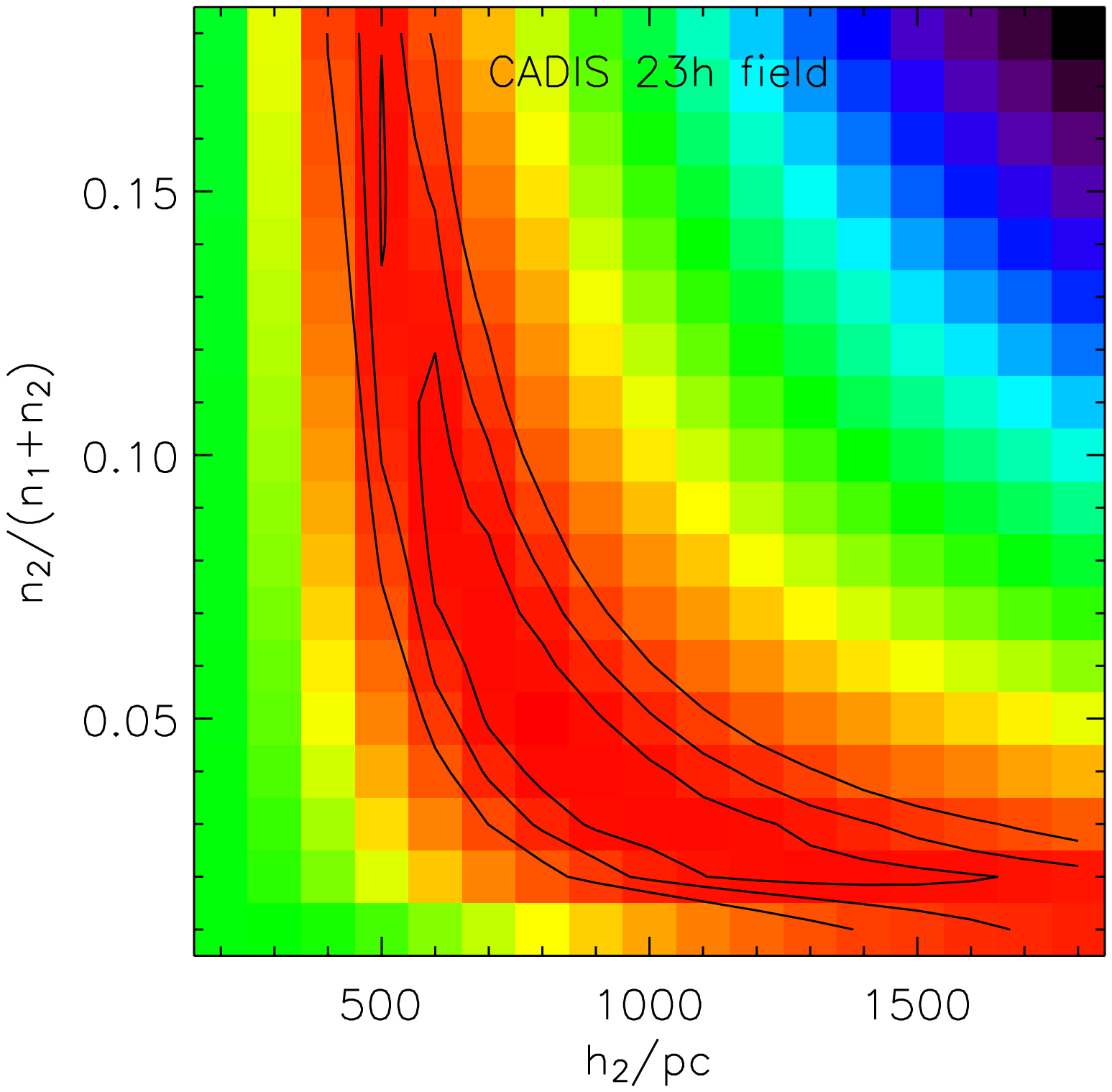,clip=t,width=8.0cm}
\psfig{figure=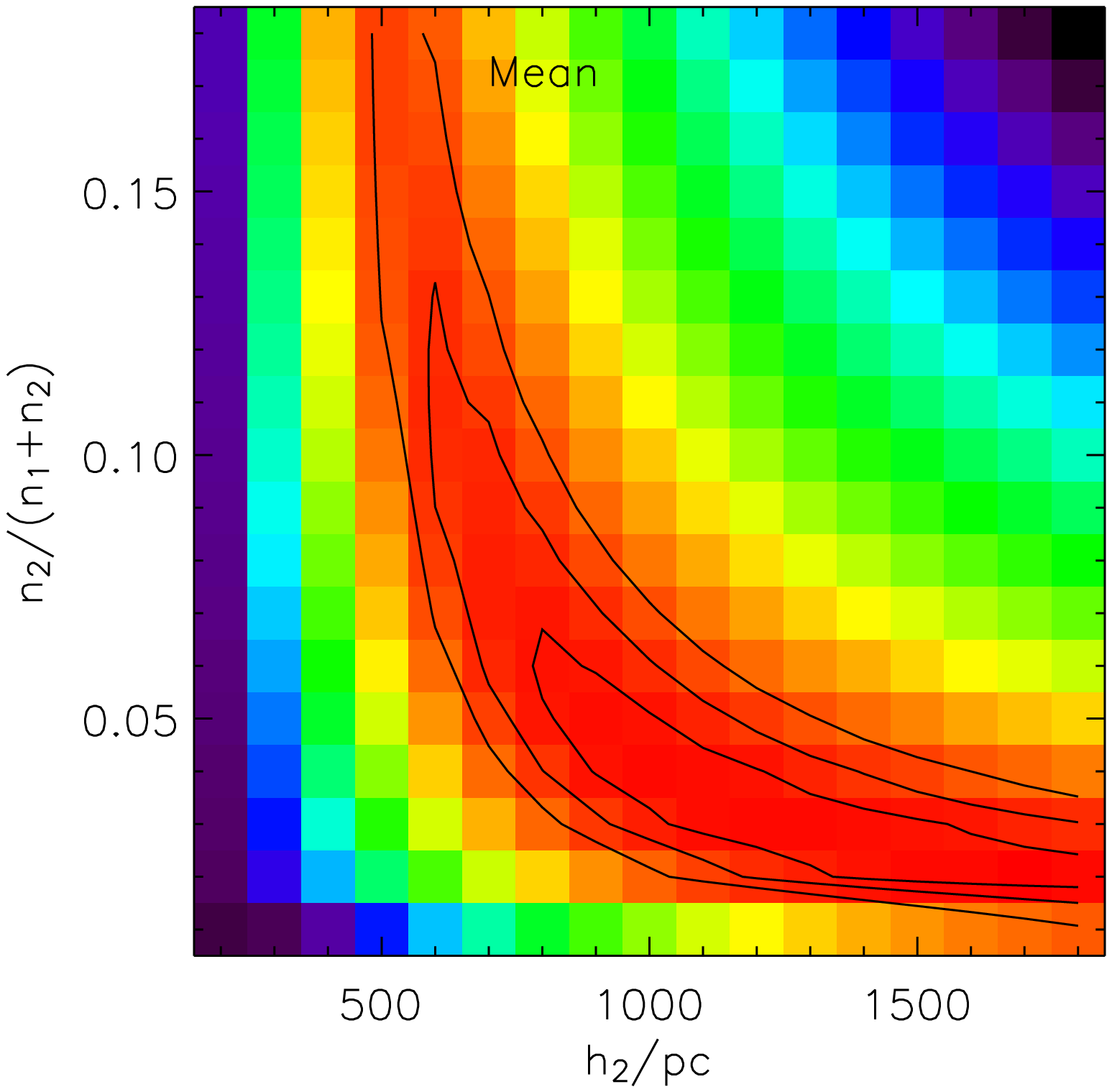,clip=t,width=8.0cm}}
\caption[ ]{$\chi^2$ plane for the parameters $h_2$ and
$n_2/(n_1+n_2)$ for the fit of the measured density distribution in
the five fields, and the mean $\chi^2$ of all five ones (bottom
right). The scalelength is assumed to be $h_r=3.5$\,kpc, and the
 values of $h_1$ are given in Table \ref{scaleheights}. The contour
lines show the $1\sigma$,$2\sigma$ and $3\sigma$   
confidence levels.\label{thickdiskfit}} 
\end{figure*}

Although the number of degrees of freedom is already kept to a minimum, this
fit is highly degenerated, as can be seen from the $\chi^2$
distribution (see Fig. \ref{thickdiskfit}), where we asumed
$h_r=3.5$\,kpc. The contour lines show the $1$,$2$ and $3\sigma$
confidence levels. Neither the ratio 
$n_2/(n_1+n_2)$ nor the value of the scaleheight of the thick disk can
be estimated with high accuracy. Except for the 9\,h field, where the
  $\chi^2$ distribution suggests a very large value of $h_2$, the
minimum can be found at around $900 \la h_2 \la 1100$\,kpc, and 
$n_2/(n_1+n_2)\la 0.05$. The fit of the scaleheight of the thick disk
in the 9\,h field is mainly 
influenced by the one data point at $z\approx 3.6$\,kpc, where the density seems
to be extraordinarily high. Because this value appears to be an
exception, we disregard it in the calculation of the mean. The mean
$\chi^2$, excluding the 9\,h field, is 
$h_2\approx 1000$, $n_2/(n_1+n_2)\approx 0.04$. Including the 9\,h
field, the mean $\chi^2$ (lower right corner) is approximately at $h_2\approx 1200$,
$n_2/(n_1+n_2)\approx 0.04$.

\begin{figure}[h]
\centerline{\psfig{figure=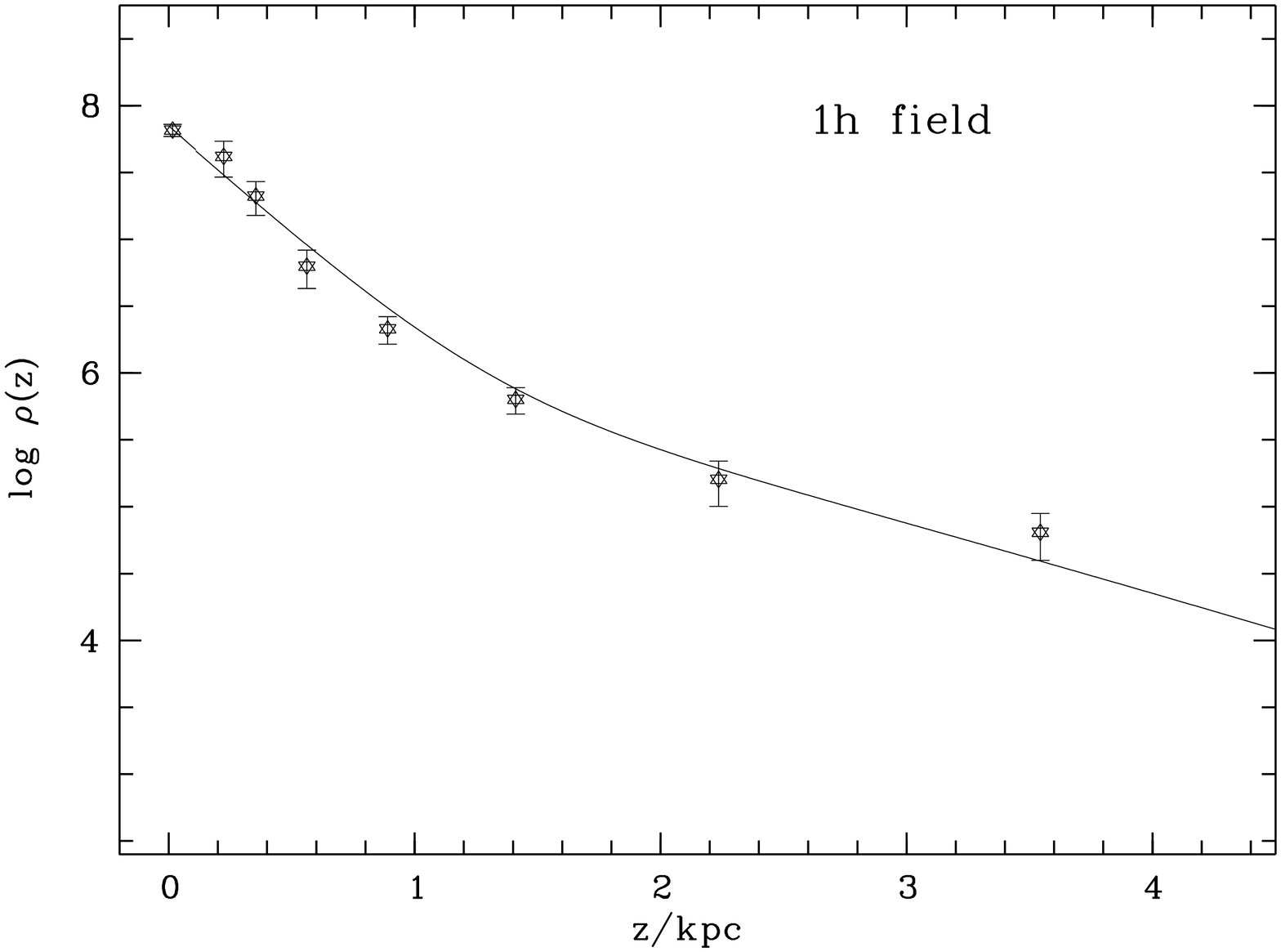,angle=270,clip=t,width=5.6cm}}
\centerline{\psfig{figure=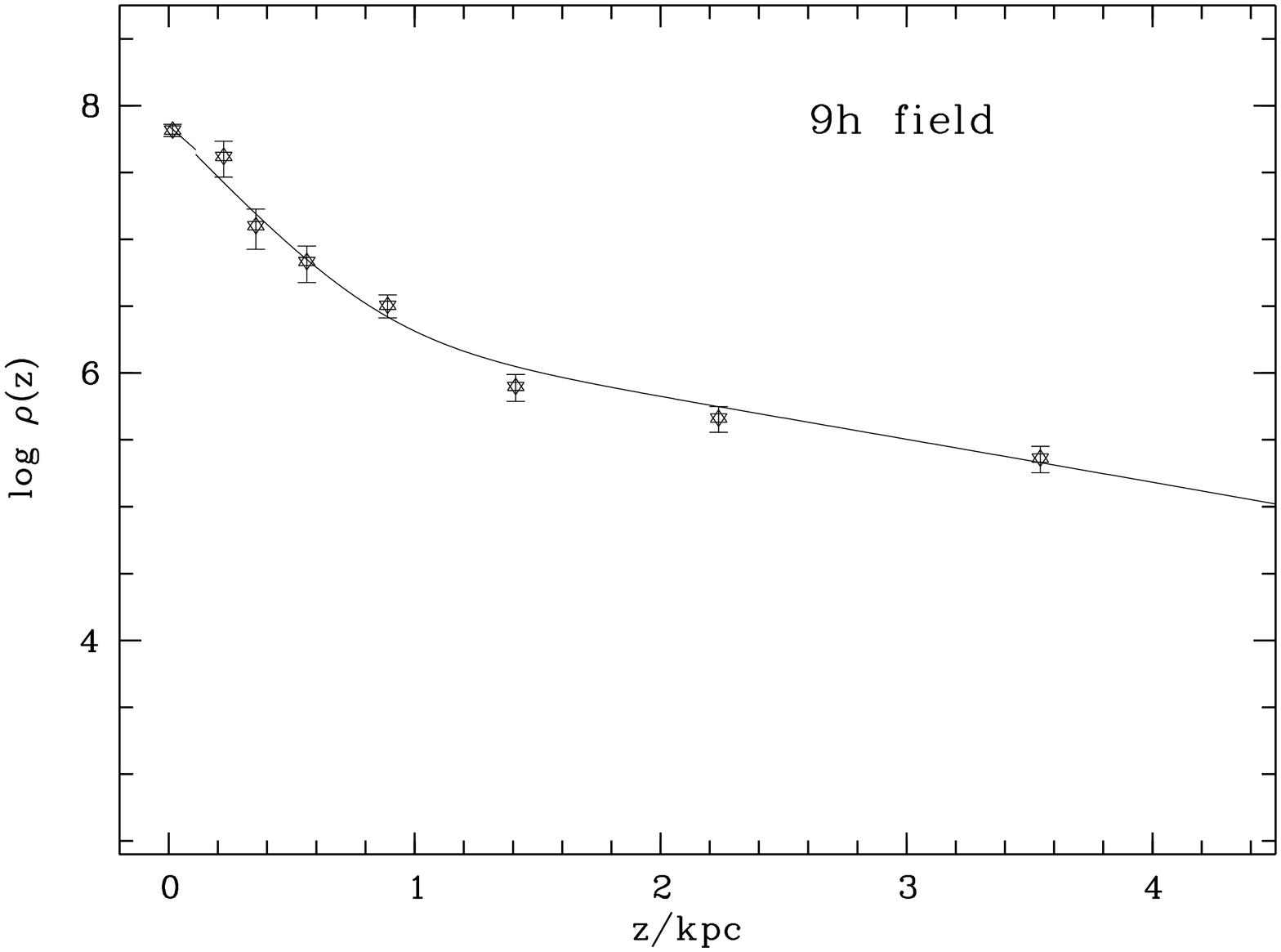,angle=270,clip=t,width=5.6cm}}
\centerline{\psfig{figure=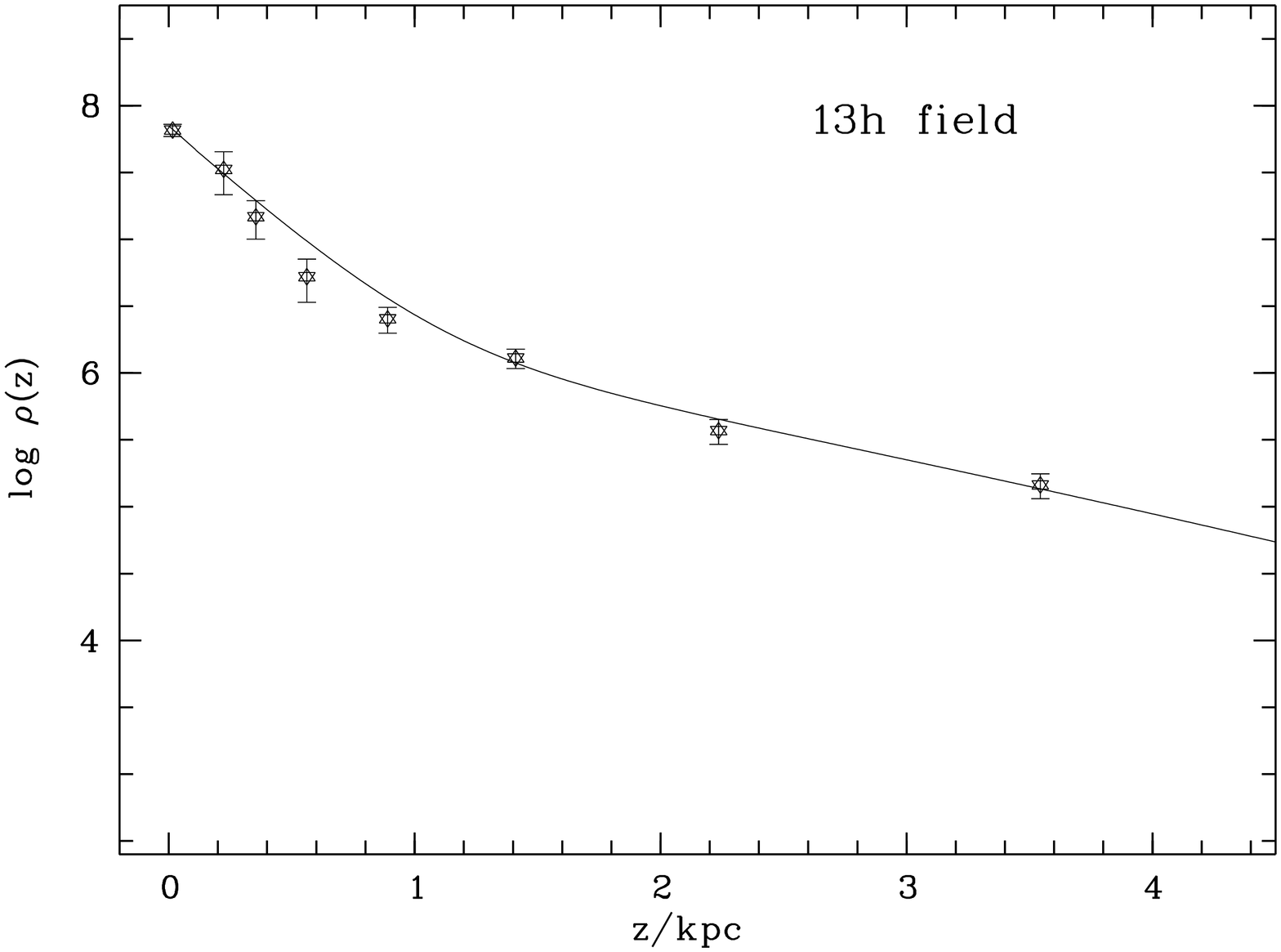,angle=270,clip=t,width=5.6cm}}
\centerline{\psfig{figure=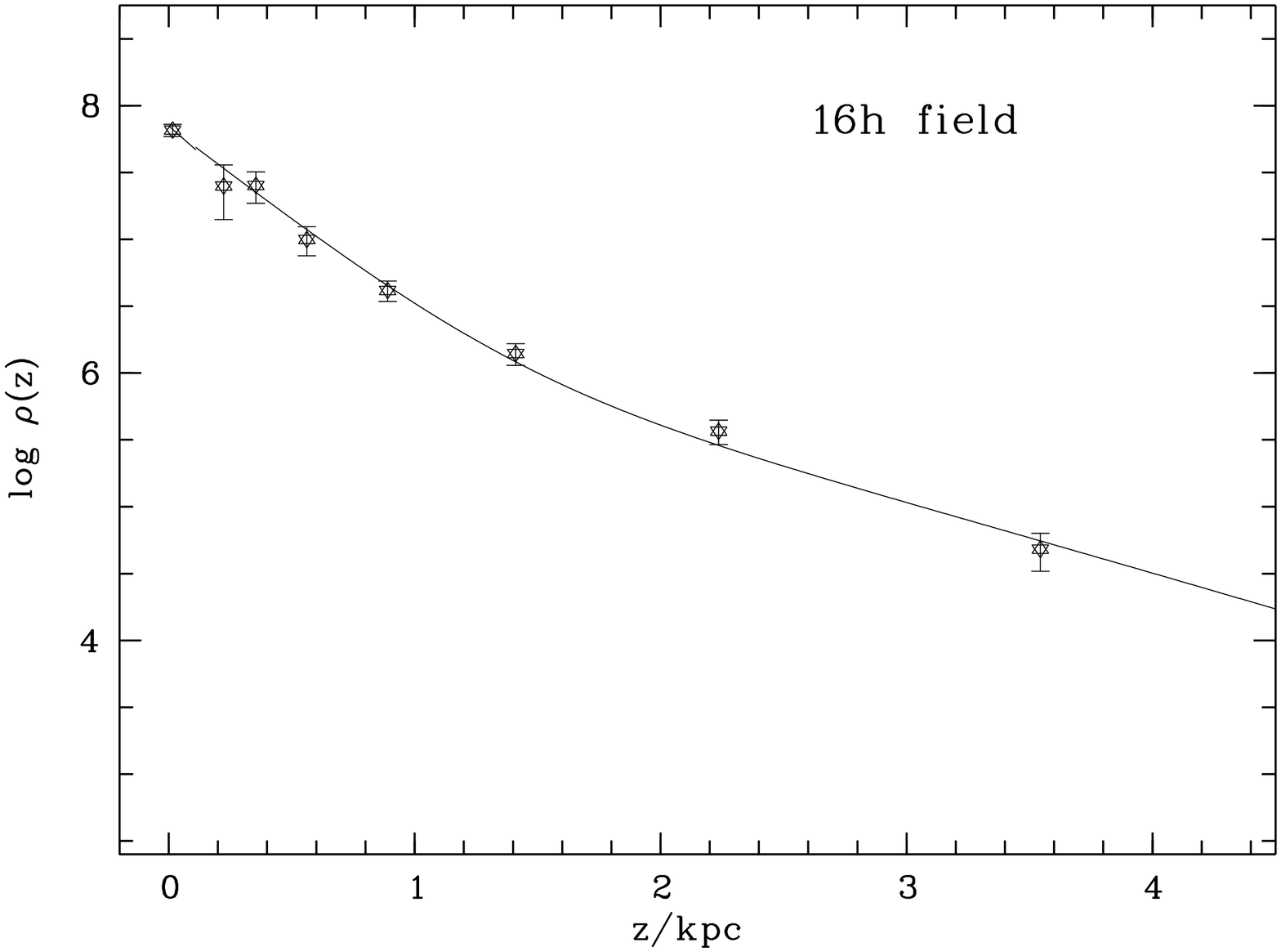,angle=270,clip=t,width=5.6cm}}
\centerline{\psfig{figure=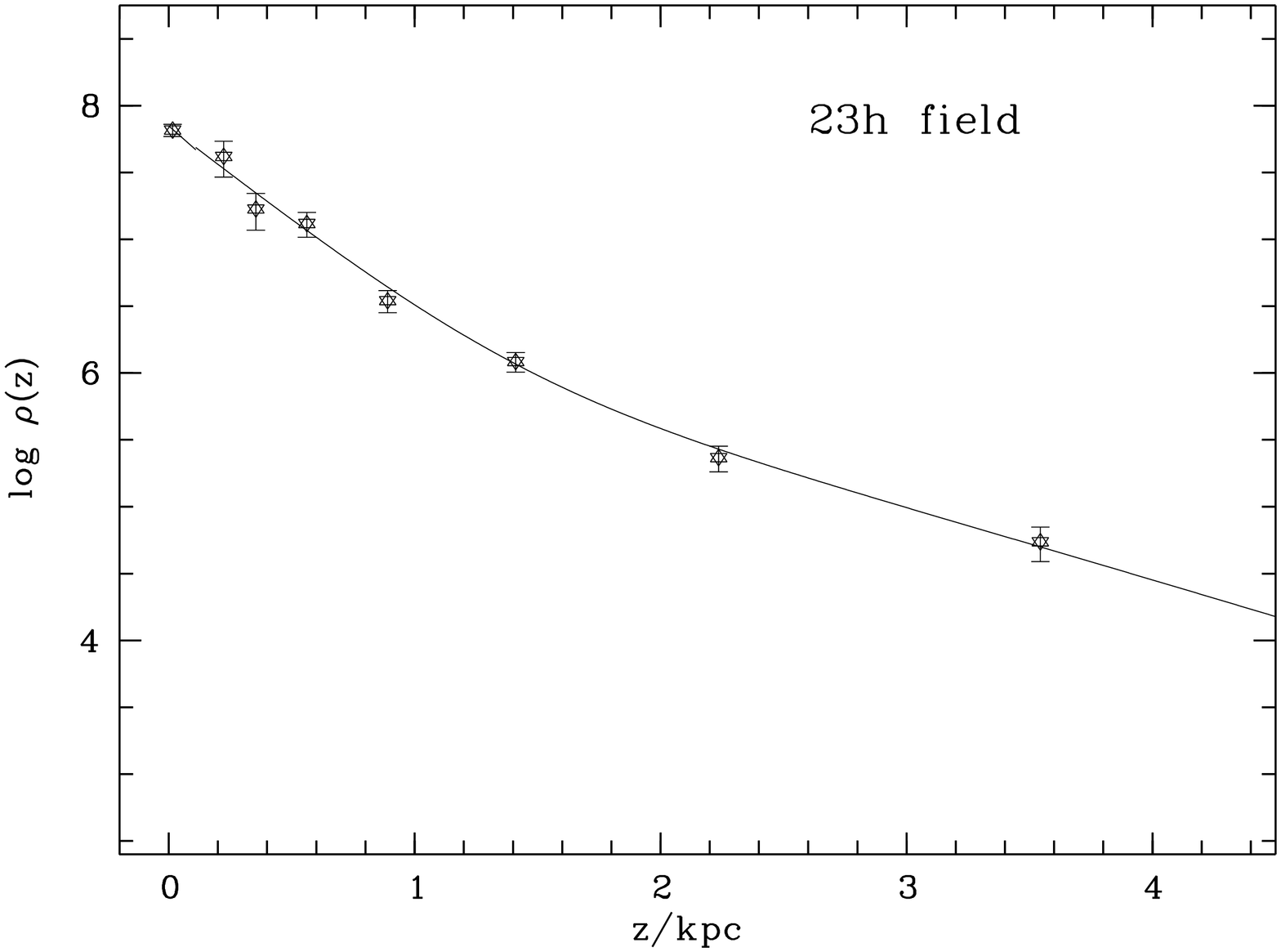,angle=270,clip=t,width=5.6cm}}
\caption[ ]{Density distribution of the stars in the disk. The solid
line is the best fit for a sum of two exponentials.\label{diskplots}}
\end{figure}

Figure \ref{diskplots} shows the density distribution of the disk
stars in the five fields, and the corresponding best fits. The
value for the local normalisation is given by the CNS4, and the values
for $h1$ can be found in Table \ref{scaleheights} (we assumed
$h_r=3.5$\,kpc). The fraction of 
thick disk stars in the local normalisation is assumed to be
$n_2/(n_1+n_2)=0.04$, and the mean value of $h_2=1.0$\,kpc was adopted
for all fields except for the
9\,h field, where a formal value of $h_2=2.6$\,kpc is required to achieve a
satisfying fit.

\subsection{The stellar density distribution of the Galactic halo}

For the local normalisation we take stars from the CNS4, which have
been discriminated against disk stars by their metallicities and
kinematics \citep{FuchsHalo}. From these we selected stars with
$-0.2\leq(b-r)\leq 0.7$ (which corresponds to $3.5\leq M_V \leq
7.5$). All these stars have very low
metallicities ($\left[Fe/H\right]<-1$\,dex) and large space velocities (see Table 1 in
\citet{FuchsHalo}), which clearly 
identifies them as halo stars. The following six stars in a radius of $r\leq 23$\,pc around
the sun satisfy the criteria: Gl~53A, Gl~451A, 
Gl~158A, GJ~1064A, GJ1064B, LHS~2815.  Thus we find
$\rho_0=89125\pm35650$\, stars/kpc$^3$.

\begin{figure}[h]
\centerline{\psfig{figure=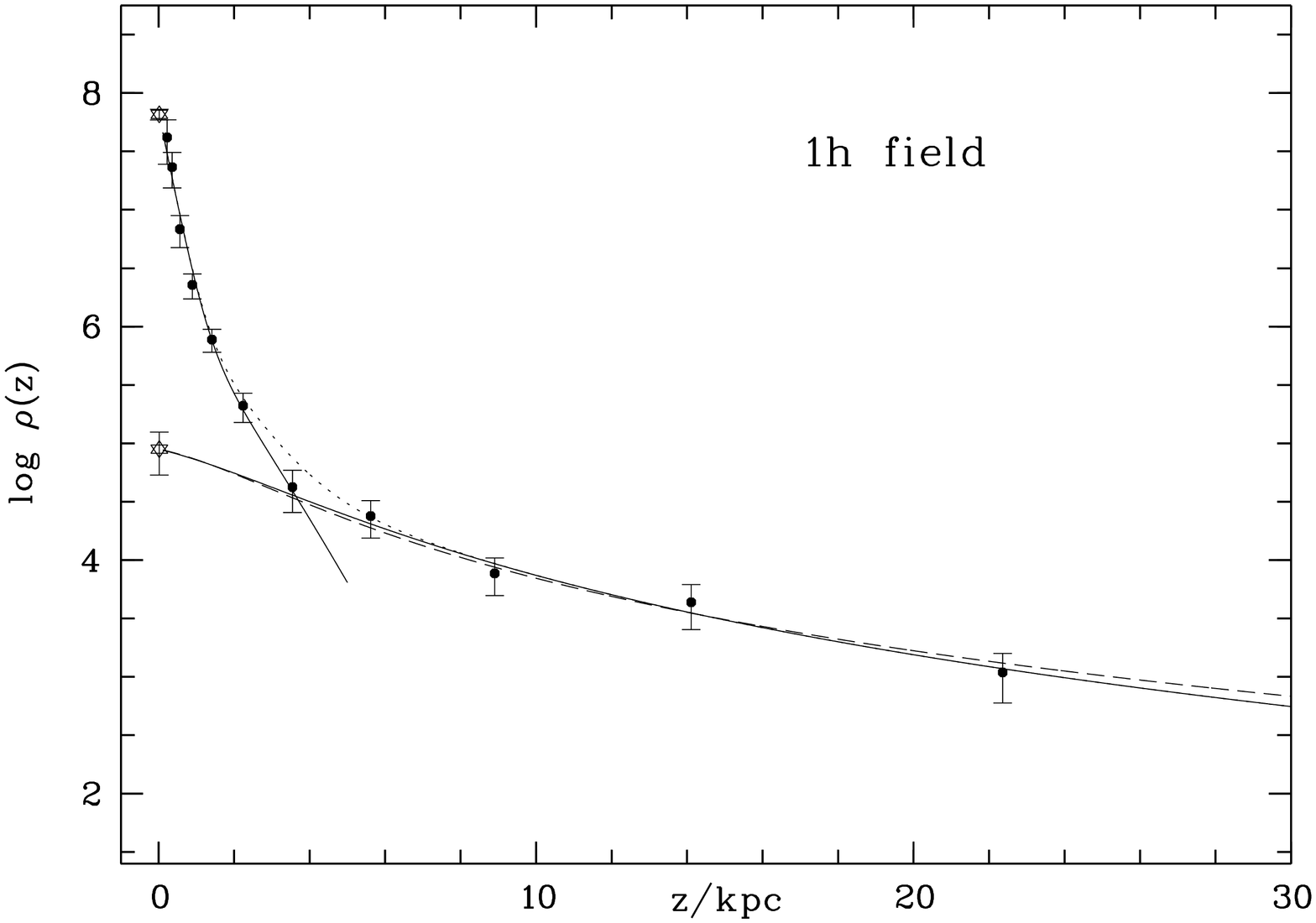,angle=270,clip=t,width=5.6cm}}
\centerline{\psfig{figure=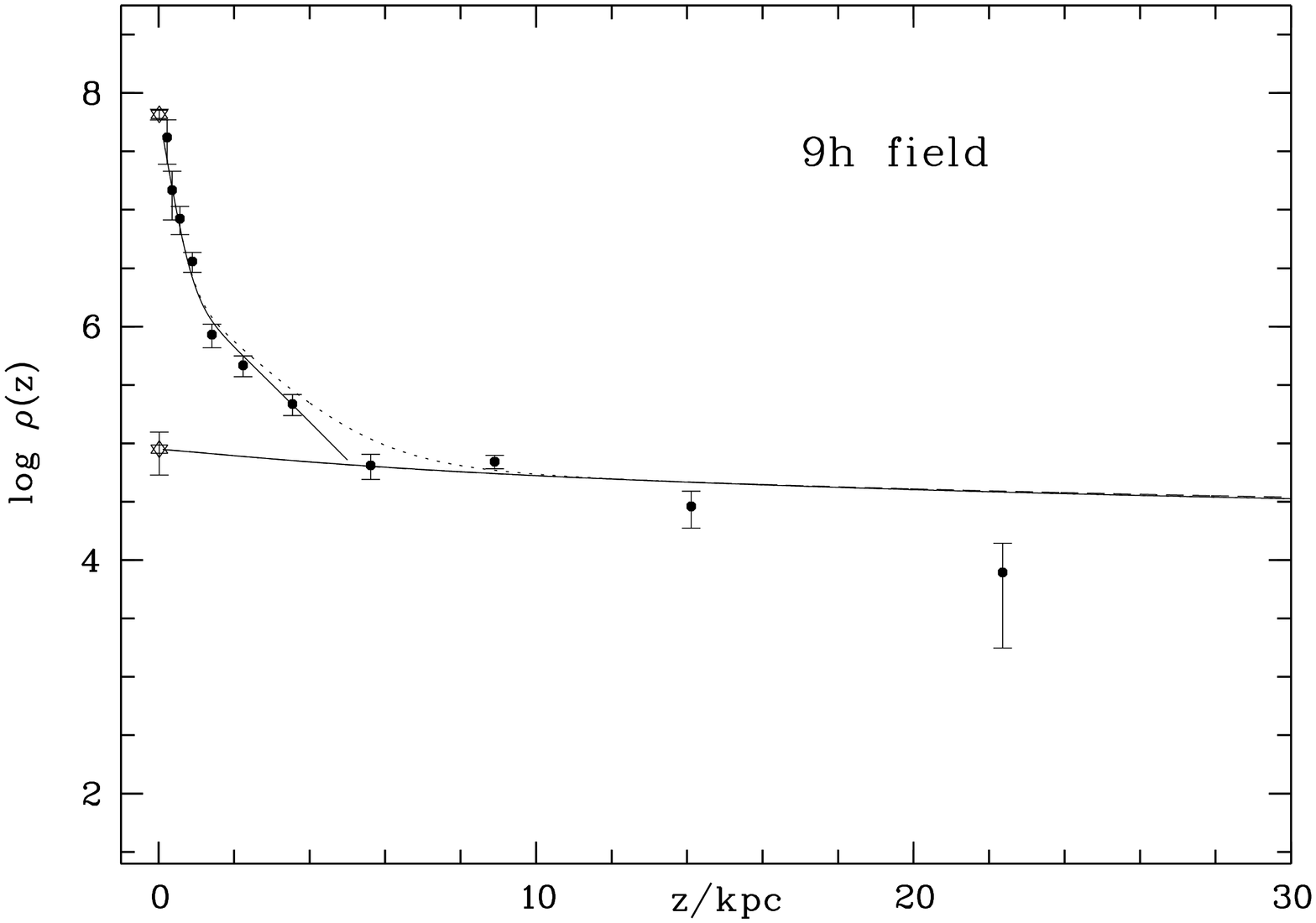,angle=270,clip=t,width=5.6cm}}
\centerline{\psfig{figure=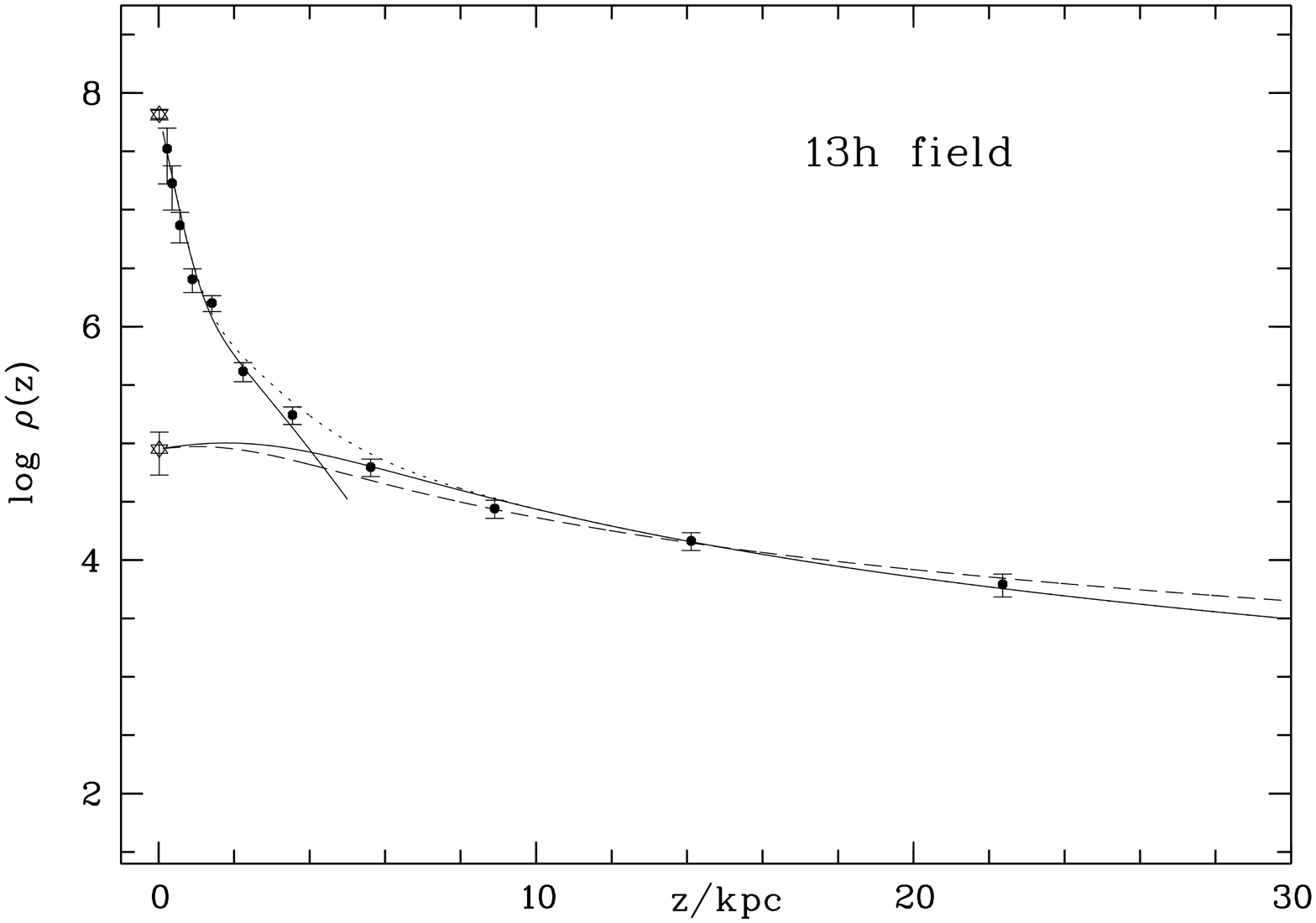,angle=270,clip=t,width=5.6cm}}
\centerline{\psfig{figure=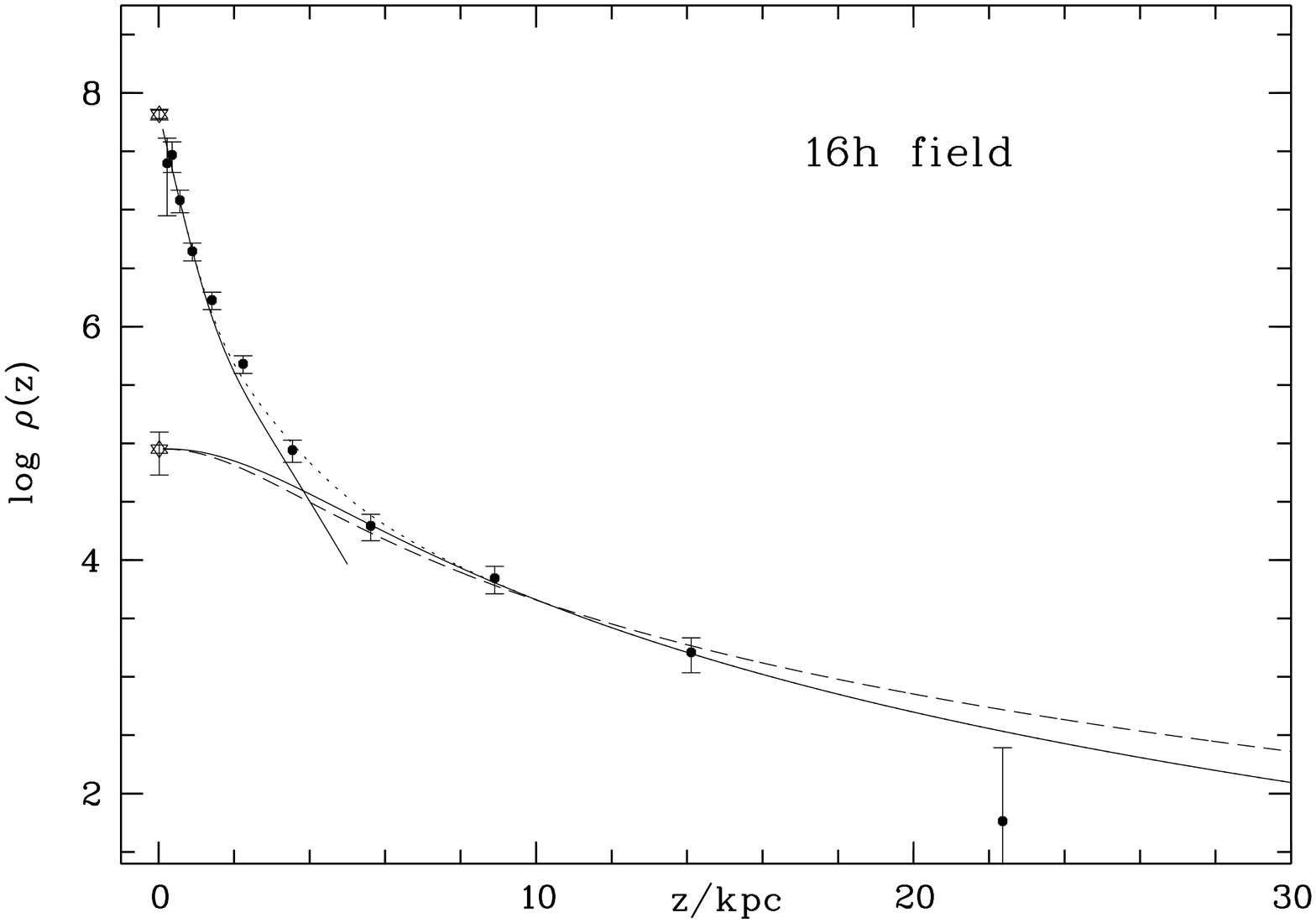,angle=270,clip=t,width=5.6cm}}
\centerline{\psfig{figure=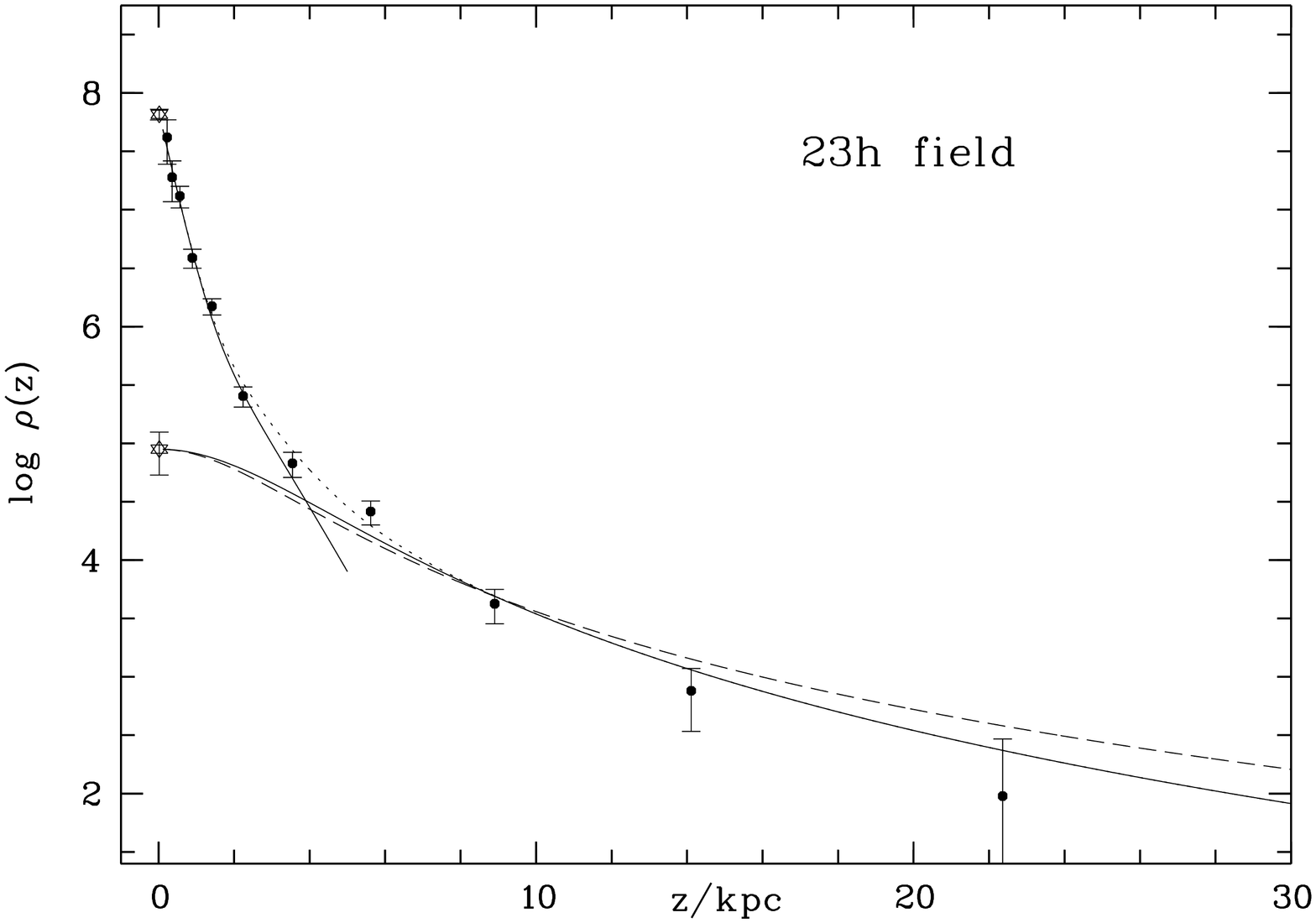,angle=270,clip=t,width=5.6cm}}
\caption[ ]{Density distribution of all stars. The solid
lines are the best fits for the disk (as in Figure (\ref{diskplots})) and
a spherical halo, the dashed line is the fit for a flattened halo with
axial ratio $(c/a)=0.6$. The corresponding exponents $\alpha$ are
given in Table (\ref{alpha})). The dotted 
line is the sum of disk and spherical halo (which is slightly favoured
by the fit). \label{haloplots}}
\end{figure}

Fig. (\ref{haloplots}) shows the completeness corrected density distribution of
{\it all} stars. We fitted equation (\ref{halofkt}) to the last four
data points, while keeping the normalisation at $z=0$ fixed to the value
given by the stars from the CNS4. We estimated the power law index
$\alpha$ for a
flattened halo with an axis ratio of $c/a=0.6$ (dashed line in
Fig. (\ref{haloplots})), and no 
flattening ($c/a=1.0$, solid line). The values of $\alpha$ we deduced
from the fit to the data are listed in Table \ref{alpha}.

\begin{table}
\caption[h]{The exponent $\alpha$ from equation (\ref{halofkt}), for a
flattened halo with
an axis ratio of $c/a=0.6$, and for $c/a=1.0$ (no flattening). The
numbers in brackets are the corresponding $\chi^2$ values.\\\label{alpha}}
\begin{tabular}{r|l l}
CADIS field&$\alpha(c/a=0.6)$&$\alpha(c/a=1.0)$\\ \hline
1\,h&$2.36\pm0.14(1.37)$&$2.8\pm0.16 (1.05)$\\
9\,h&$0.44\pm0.08(13.99)$&$0.5\pm0.1 (13.68)$\\
13\,h&$1.50\pm0.09(3.13)$&$1.98\pm0.09 (1.54)$\\
16\,h&$2.84\pm0.14(3.24)$&$3.52\pm0.18 (1.65)$\\
23\,h&$2.98\pm0.18(11.81)$&$3.68\pm0.22 (7.01)$\\
\end{tabular}
\end{table}

Clearly the fit in the 9\,h field is dominated by one exceptionally high data
point at $z\approx 10$\,kpc, and does not give a reliable result. The
weighted mean, disregarding the 9\,h field, is $\alpha=2.12\pm0.06$ for
$(c/a)=0.6$, and $\alpha=2.51\pm0.07$ for
$(c/a)=1.0$, where the $\chi^2$ values favour a halo without strong
flattening: The sum of the $\chi^2$ values of the fits in the  four
single fields which enter in the determination are
$\sum{\chi^2}=19.55$ for the flattened, and  $\sum{\chi^2}=11.25$ for the
spherical halo.

The axis ratio assumed for the estimation of the power law index
$\alpha$ influences the fitted value: the flatter the halo, the
smaller $\alpha$. But not only the axial ratio influences the
measurement of $\alpha$, also the local normalisation adopted for the
fit. In order to estimate the effect we fitted the data for
$\rho_0^{\rm uplim}=\rho_0+\sigma_{\rho_0}$ and
$\rho_0^{\rm lowlim}=\rho_0-\sigma_{\rho_0}$, respectively, and found
weighted means of 
$\alpha(\rho_0^{\rm uplim})=2.26\pm0.05$ (with $\sum{\chi^2}=10.82$) and
$\alpha(\rho_0^{\rm lowlim})=1.68\pm0.07$
($\sum{\chi^2}=51.29$) for 
$(c/a)=0.6$, and $\alpha(\rho_0^{\rm uplim})=2.95\pm0.07$
($\sum{\chi^2}=16.45$) and
$\alpha(\rho_0^{\rm lowlim})=1.97\pm0.07$
($\sum{\chi^2}=30.12$) for 
$(c/a)=1.0$, respectively (again leaving out the 9\,h field).

The values of $\alpha$ are getting  larger for a larger value
of $\rho_0$, and smaller for smaller $\rho_0$. 
However, the the summed
$\chi^2$ indicate that the normalisation given by the CNS4 stars is
not too far away from the ''true'' normalisation. The
true value is definitely not smaller. If it is slightly larger, then a
flattened halo is favoured. However,  it is not
possible to measure the flattening of the halo accurately with the
current data set. 

\subsection{The stellar luminosity function of the disk}\label{SLFsection}
Knowing the density distribution of the thin disk stars, we can calculate the
stellar luminosity function $\Phi(M_V)$. We restrict the analysis to
stars within a radius of $r\leq1.5$\,kpc, for beyond this range the
contribution by thick disk and halo stars becomes dominant. 
As in Paper I, we calculate absolute visual magnitudes from the $R$
magnitudes to make comparison with literature easier. In the relevant
magnitude intervall ($5 \leq M_V \leq 14$, i.e $-1 < (b-r) <
1.8$) there holds an empirical linear relation:
\begin{eqnarray}
M_{V_J}= 1.058 \cdot M_{R_C}~.
\end{eqnarray}
For every luminosity bin the  effective volume is calculated 
 by integrating the distribution
function along the line of sight, $r$, where the integration limits
are given by the minimum between 1.5\,kpc and the distance modulus
derived for upper and lower limiting apparent magnitude:
\begin{eqnarray}\label{vol}
V_{{\rm max}}^{{\rm eff}}= \omega \int\limits_{R_{{\rm min}}}^{R_{{\rm
max}}}
\nu(r,b) ~r^2 ~dr~,
\end{eqnarray}
where 
\begin{eqnarray*}
R_{{\rm min}}&=&10^{0.2(16^{{\rm mag}}-M_R)-2.0}~,\\
R_{{\rm max}}&=&{\rm min}(1.5~{\rm kpc}, 10^{0.2(23^{{\rm mag}}-M_{\rm
R})-2.0})~.
\end{eqnarray*}
The distribution function 
\begin{eqnarray}\label{distribnorm}
\nu(r,b)= \exp(-r \sin b/ h_1)
\end{eqnarray}
is normalised to unity at $z=0$.

\begin{figure}[h]
\centerline{\psfig{figure=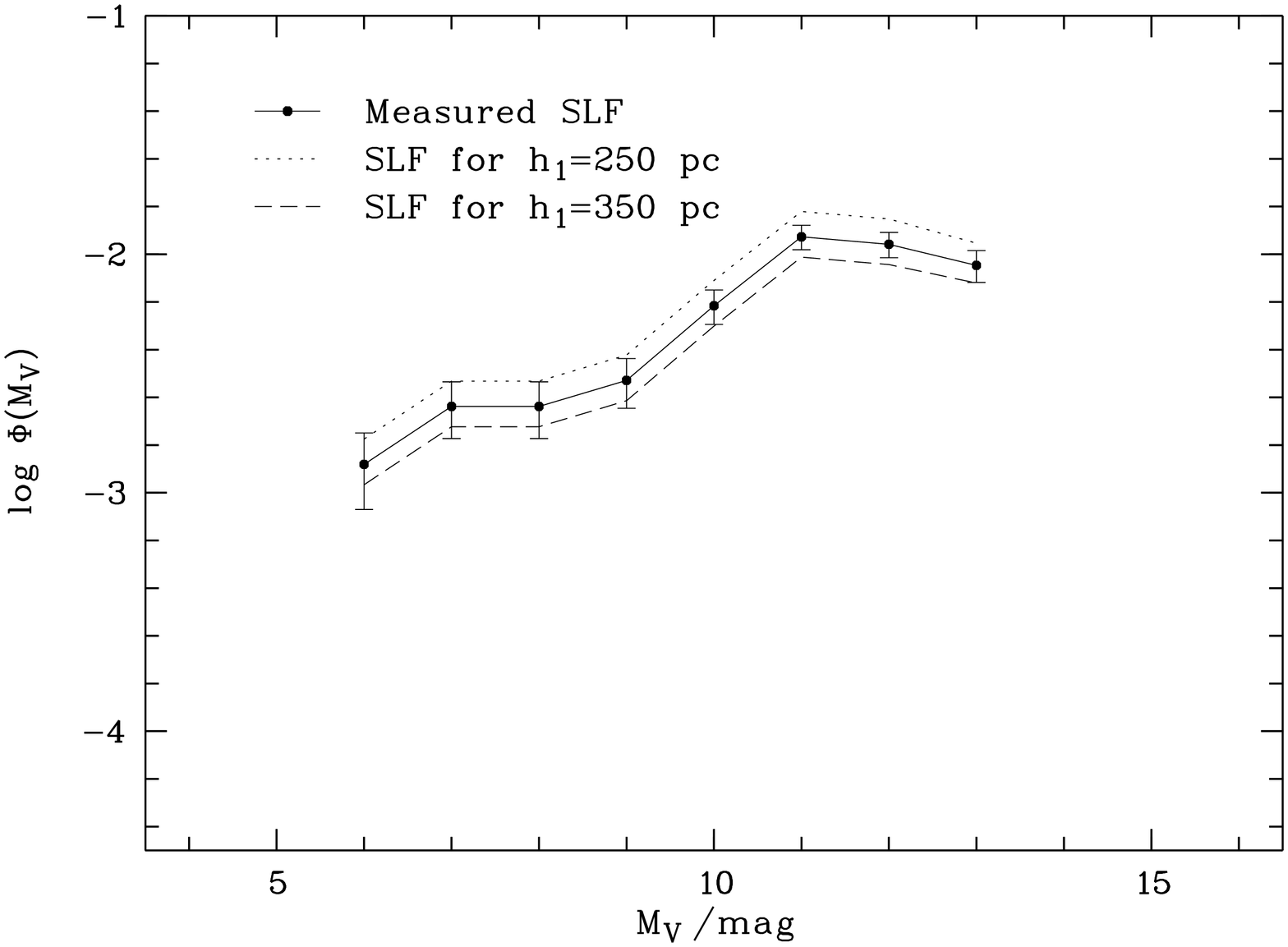,angle=270,clip=t,width=7.0cm}}
\centerline{\psfig{figure=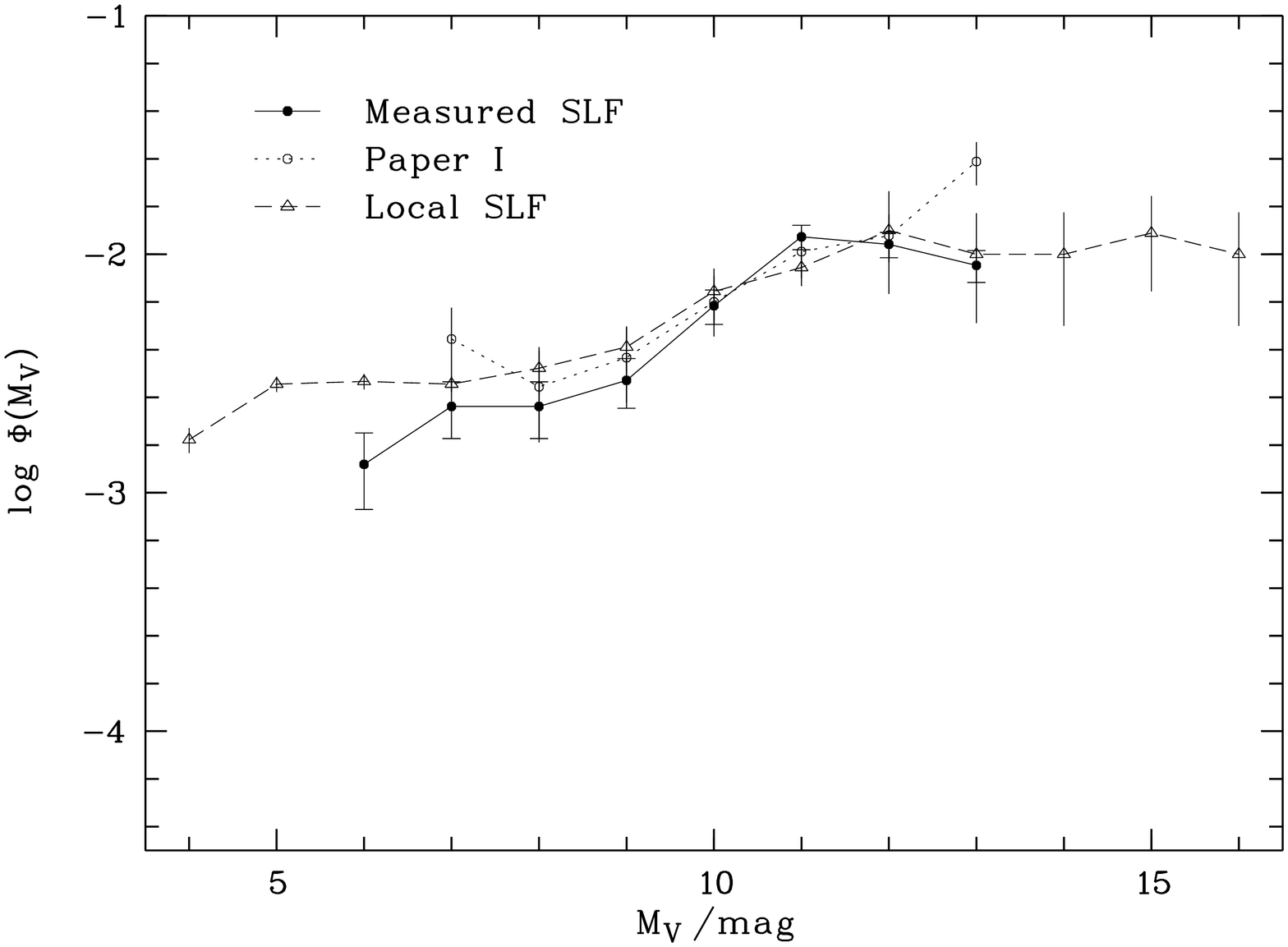,angle=270,clip=t,width=7.0cm}}
\centerline{\psfig{figure=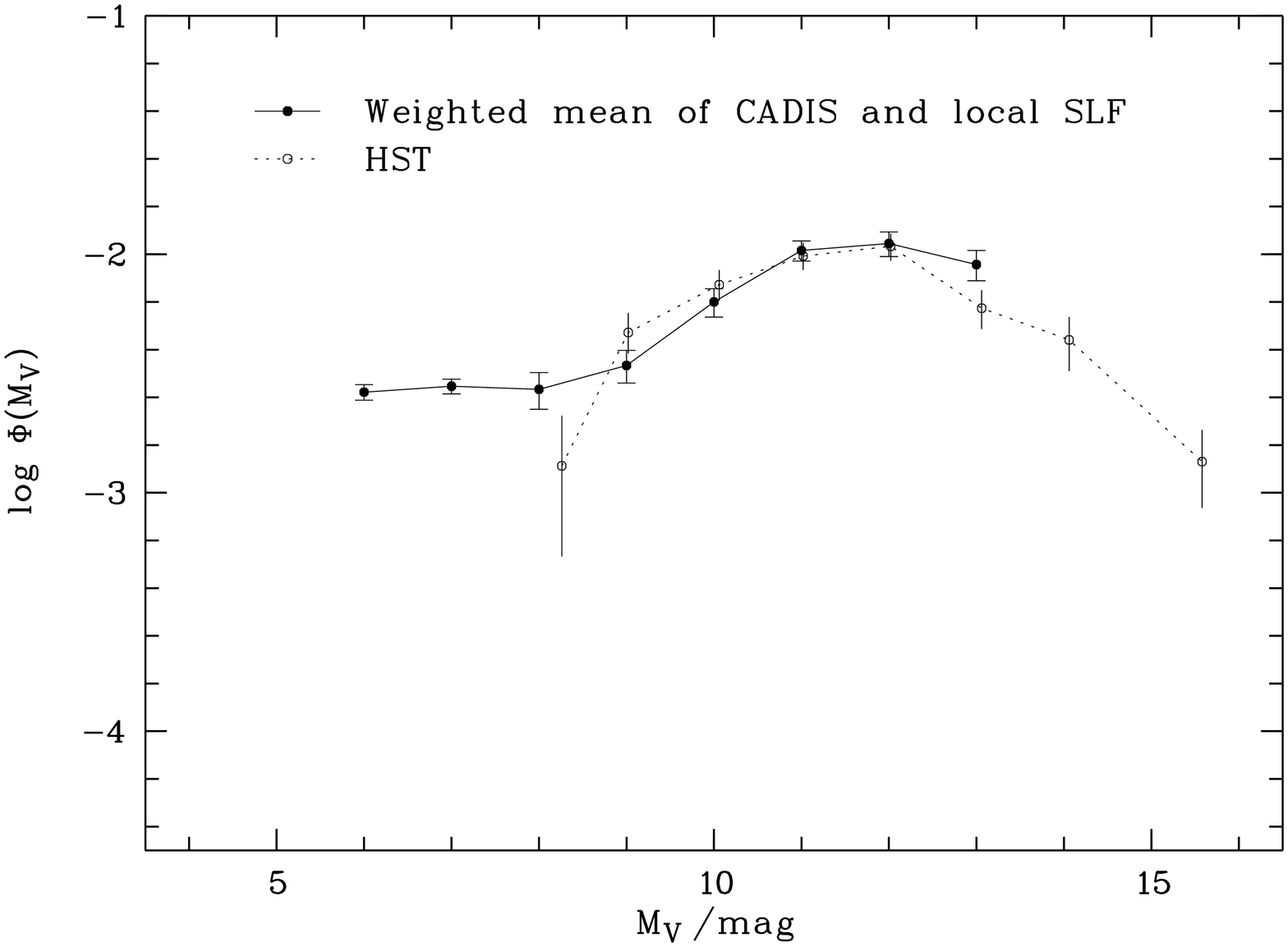,angle=270,clip=t,width=7.0cm}}
\caption[ ]{Upper panel: the mean SLF of the five CADIS fields (solid
line). For the calculation of the effective volumes we used the
fitted values of $h_1$ (see Table \ref{scaleheights}). The
dotted and dashed lines are the determination for $h_1=0.250$\,kpc, and
for $h_1=0.350$\,kpc, respectively. Middle panel: The solid line is
our current measurement, as in the upper panel. The dotted line is our first
measurement as published in Paper I, the dashed line is the local SLF
\citep{Venice}, which is based on HIPPARCOS parallaxes. Lower panel:
weighted mean of CADIS and local SLF, in comparison with a photometric SLF which is
based on HST observations \citep{Gould98}.\label{LKF}}
\end{figure}

The weighted mean of the SLFs of the five CADIS
fields is shown in Fig. \ref{LKF}. For each single field we also calculated
the effective volume by using the measured values of $h_1$. Also shown
in Fig. \ref{LKF} are the SLFs one would measure for $h_1=0.250$\,kpc
and $h_1=0.350$\,kpc (dotted and dashed lines in the upper panel), in
order to demonstrate how the SLF 
changes with the scaleheight assumed for the calculation of the
effective volume. The dash-dotted line in the middle shows our
old determination as published in Paper I (Fig. 10), the dashed line
is the local
SLF \citep{Venice} of stars inside a radius of 20\,pc around the sun,
which is based on HIPPARCOS parallaxes. In the range where the SLFs
overlap, they are consistent with each other. As in Paper I we calculated
the weighted mean of CADIS and local SLF, which is shown in the lower
panel of Fig. \ref{LKF}, in comparison with a photometric SLF which is
based on HST observations \citep{Gould98}.

We will use this luminosity function as input for the simulation,
which we describe in the following section.

\section{A Monte Carlo model of the Milky Way}\label{model}
The method described in the previous section has
the disadvantage that it involves a completeness correction, since
faint stars can only be traced at small distances. However, it is possible
to avoid completeness corrections, by modeling the {\it observed} colors and
apparent magnitudes of the stars in the fields, using Monte Carlo
simulations. The best-fitting model of the Galaxy can be determined by
$\chi^2$-fitting or a maximum likelihood analysis.

The two basic incredients are the stellar density distribution
function and the stellar luminosity function. We will keep the
luminosity function fixed, while varying the structure parameters of
the density distribution.

\subsection{The Density Distribution}

We use the same  model of the Milky Way as described
in Section (\ref{densitychapter}): a sum of two exponential disks with
scaleheights $h_1 < h_2$, and a power law halo. 

The local density is again given by the CNS4 stars (\citealp{Venice,FuchsHalo}). The
contribution of thick disk stars to the local normalisation, $n_2/(n_1+n_2)$, is left as
a free parameter, and is varied from 2\% to 18\% in steps of $1$\,\%. The scaleheights
$h_1$ and $h_2$ are varied from $150$\,pc to $400$\,pc in steps of $50$\,pc, and from
$200$\,pc to $1800$\,pc in steps of $100$\,pc, respectively. The
simulation was run for a radial scalelength $h_r=2000$\,pc, and for $h_r=3500$\,pc. The
power law index of the stellar halo was increased from $\alpha=2.0$ to
$\alpha=4.0$ in steps of $0.5$, for a spherical as well as for a flattened
halo with an axis ratio $(c/a)=0.6$.

For each field, the expected number of stars $N$ in the model under consideration is
calculated by integrating the distribution function along the line of sight:
\begin{eqnarray}
N=\int_0^\infty{\rho(r,l,b) ~\omega r^2} {\mathrm d}r~,
\end{eqnarray} 
where $\omega=7.6145\cdot 10^{-6}$ for one CADIS field and
$\rho(r,l,b)=\rho(r,l,b)_{\rm disk}+\rho(r,l,b)_{\rm halo}$.
For each of the $24\,480$ combinations of structure parameters we simulated
$1000$ times more stars than observed in each CADIS field, in order to avoid large poisson
noise in the model data.

\subsection{Magnitudes and colors}\label{simcolorsandmags}
In order to simulate ``observed'' apparent magnitudes and colors and
to keep the two different approaches consistent, we have essentially to
invert the steps described in Section \ref{starcounts}. First of all each
star which has been simulated according to the density distribution function is
assigned an absolute magnitude following the stellar luminosity
function determined from our sample (see Section
\ref{SLFsection}). The faint end of the luminosity function of {\it
halo} stars has only been 
determined with high uncertainties \citep{Gould03}. Although it has
been suggested that the luminosity function of the 
halo stars might be different from the one of the disk stars
\citep{Reyle01}, there is no accurate
measurement available: The measurement of the luminosity function of
halo subdwarfs is extremely sensible to the distance range in which
it is determined -- the SLFs measured from local subdwarfs (where the
number of stars is very small and limits the accuracy of the measurement) differ from
those measured in the outer halo. The reason for these discrepancies
might be the insufficient accuracy with which the stellar structure of the
halo is known \citep{Digby03}. Therefore, as long as the the  luminosity function
is not determined with higher accuracy and the density distribution of
the halo is known better, we  use our measured SLF  for
both disk and halo stars.  

In the range $6\leq M_V \leq 13$ the mean of the measured CADIS and the local SLF from
\citet{Venice} can be described quite precisely by a third-order 
polynomial in log-space: 
\begin{eqnarray}\label{slfform}
\log \Phi(M_V)=\sum_{i=0,3} c_i M^i~,
\end{eqnarray}
with $c_0=4.78676$, $c_1=-2.63075$, $c_2=0.297266$, $-0.0104039$.

Each star is assigned an absolute $V$ band
luminosity, randomly chosen from a distribution following equation
(\ref{slfform}).
Since there is no $V$ or comparable filter available in the CADIS
filter set, we have to convert the $M_V$ magnitudes into $M_{R_C}$
magnitudes (see Paper I), i.e. $M_{R_C}=0.945\cdot M_V$. CADIS
$(b-r)_C$ colors can be calculated from the absolute $R_C$ band
magnitudes using the main sequence relation described in Paper
I. The same relation was used in Section \ref{starcounts} to deduce
absolute magnitudes from the $(b-r)_C$ colors of the stars.

\subsubsection{Metallicities}
In Section \ref{starcounts} we have corrected for the influence of
different metallicities on the color-magnitude relation. Now we have
to simulate those effects in order to compare the model with the
data. Halo stars are again assumed to be fainter by $\Delta M_R=0.75$
as disk stars of the same color. In the direct determination we have 
not been able to separate thin from thick disk stars. In the simulation
however, we can account for the fact that thick disk stars have
subsolar metallicities, and hence have a different color-magnitude relation
than thin disk stars with solar metallicities. From a sample of 5
thick disk stars with $-0.3 < \left[{\rm Fe/H}\right] < -0.95$, where
HIPPARCOS parallaxes, spectroscopic metallicities and photometry in
blue and red passbands are available, we estimate the offset from the
solar main sequence valid for the thin disk stars to be $\Delta M_R\approx
0.35$ (C.~Flynn, priv.comm.). 

In order to take the
spread in metallicities into account, the simulated colors  are then smeared
out by adding a  value taken from a uniform distribution of numbers in
the range $-0.2\leq \delta (b-r) \leq +0.2$.

\subsubsection{Photometric errors}
Our observed data  have of course photometric errors, which are of the
order of $\sigma_R=0.01$ in case of the brighter ($R\leq 21$) and
$\sigma_R=0.15$ for the 
faintest ($R\approx 23$) stars. These have to be
simulated as well, in order to allow for  a reliable comparison with
the observations. Using the distance modulus $m-M=5\log r -5$, the
apparent magnitudes of the stars are calculated.  Each star is
assigned a photometric error which is chosen randomly out of a
gaussian  error distribution with a variance corresponding to its apparent
magnitude. The error is added to the apparent $R$ magnitude. The
procedure is repeated for the $B$ magnitudes, which have been
calculated from the simulated $(b-r)_C$ colors. The final colors are
then redetermined from the new, modified magnitudes.

\subsection{``Observing'' the Model}
Each simulated star has now an apparent $R_C$ magnitude and a
$(b-r)_C$ color, where metallicity effects and photometric errors are
incorporated in the simulation. These data can now be compared with the  observed
data.

The simulated stars are selected by their apparent magnitudes
($15.5\leq R_C \leq 23$). We bin the data, observed and simulated,
into a grid with steps of 
$\Delta R_C=0.5$ in magnitude and $\Delta (b-r)_C=0.2$ in
color.

Due to our survey selection criteria the bins at bright apparent
magnitudes contain only very few stars, which makes a $\chi^2$ fit of
the model to the data extremely difficult. 
Instead of using $\chi^2$ as an estimate of the goodness of fit, we
use the {\it C-statistic} for sparse sampling, which was developed by
\citet{Cash79}. It is based on the likelihood ratio, and is suited for
small number statistics. In the limit of large numbers it converges to
$\chi^2$.

If $n_i$ is the number of stars in the bin, $e_i$ the expectation
value predicted by the model and $N$ the number of bins, then
\begin{eqnarray}
C=\frac{2}{N}\sum_{i=1}^{N}{\left[n_i\ln
\left(\frac{n_i}{e_i}\right)-(n_i-e_i)\right]}
\end{eqnarray}

For each of the $24\,480$ models of each field we calculate $C$ by
summing over the bins in the magnitude-color grid.

\begin{figure*}[hp]
\centerline{\psfig{figure=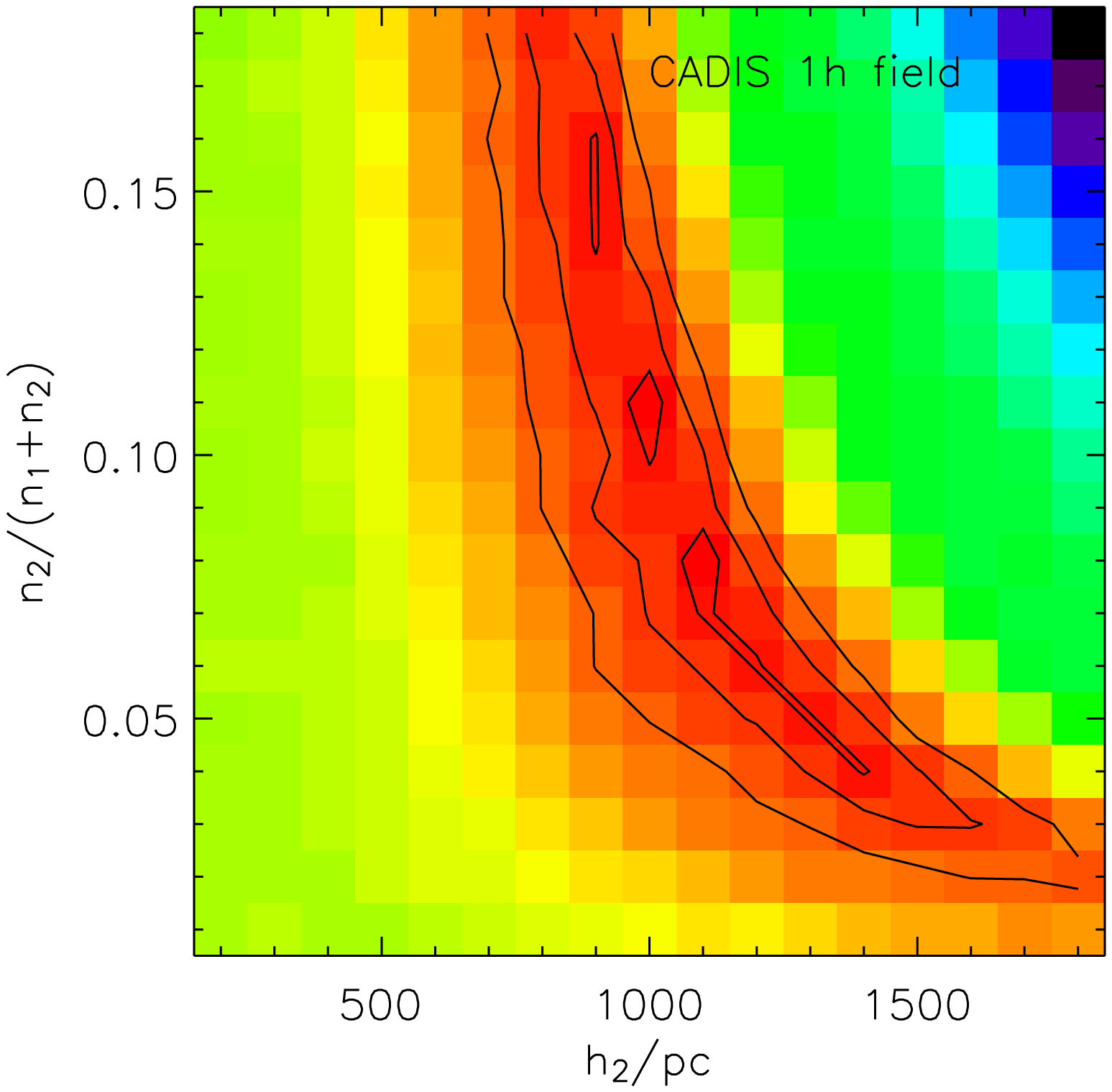,clip=t,width=8.0cm}
\psfig{figure=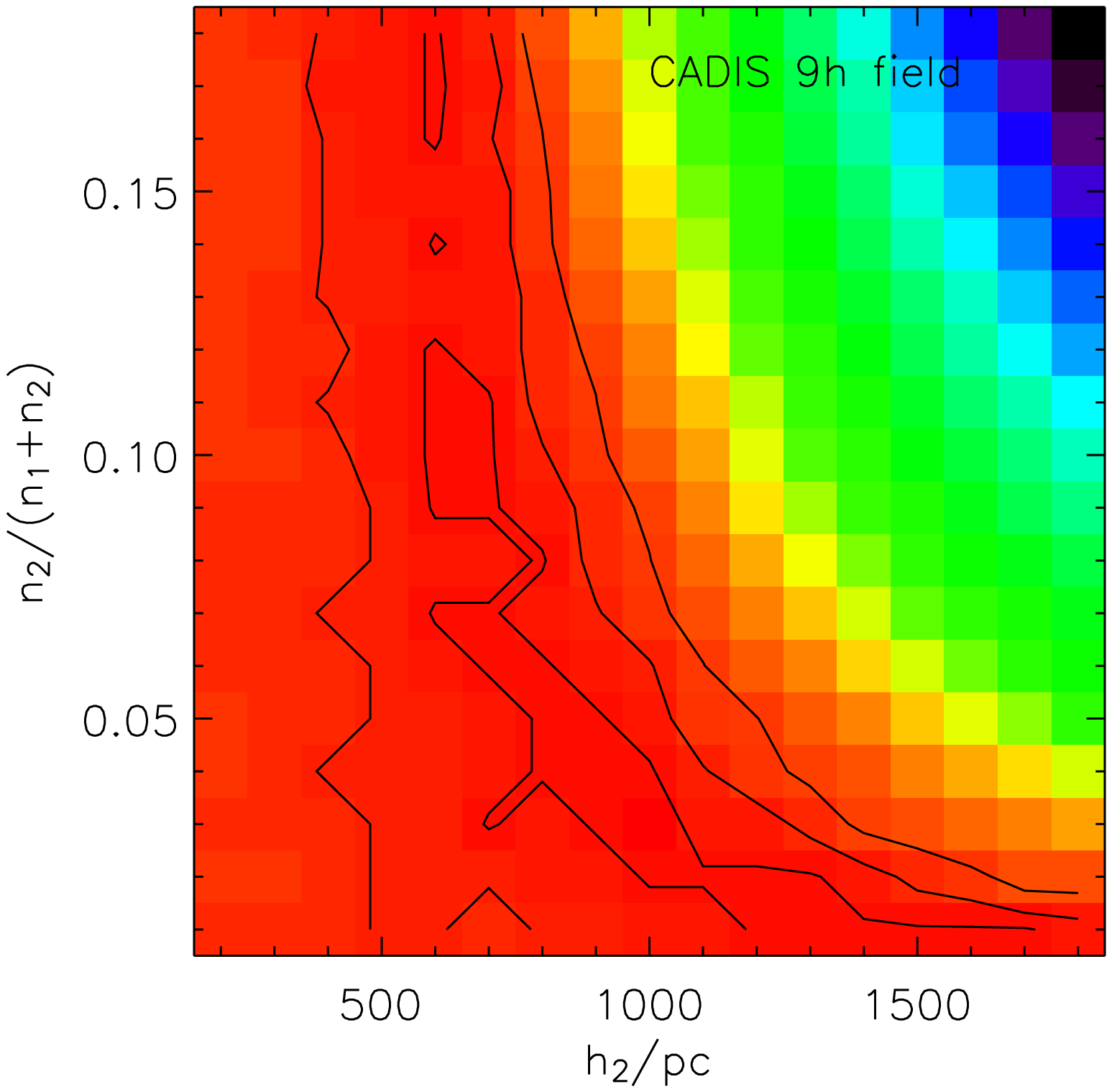,clip=t,width=8.0cm}}
\centerline{\psfig{figure=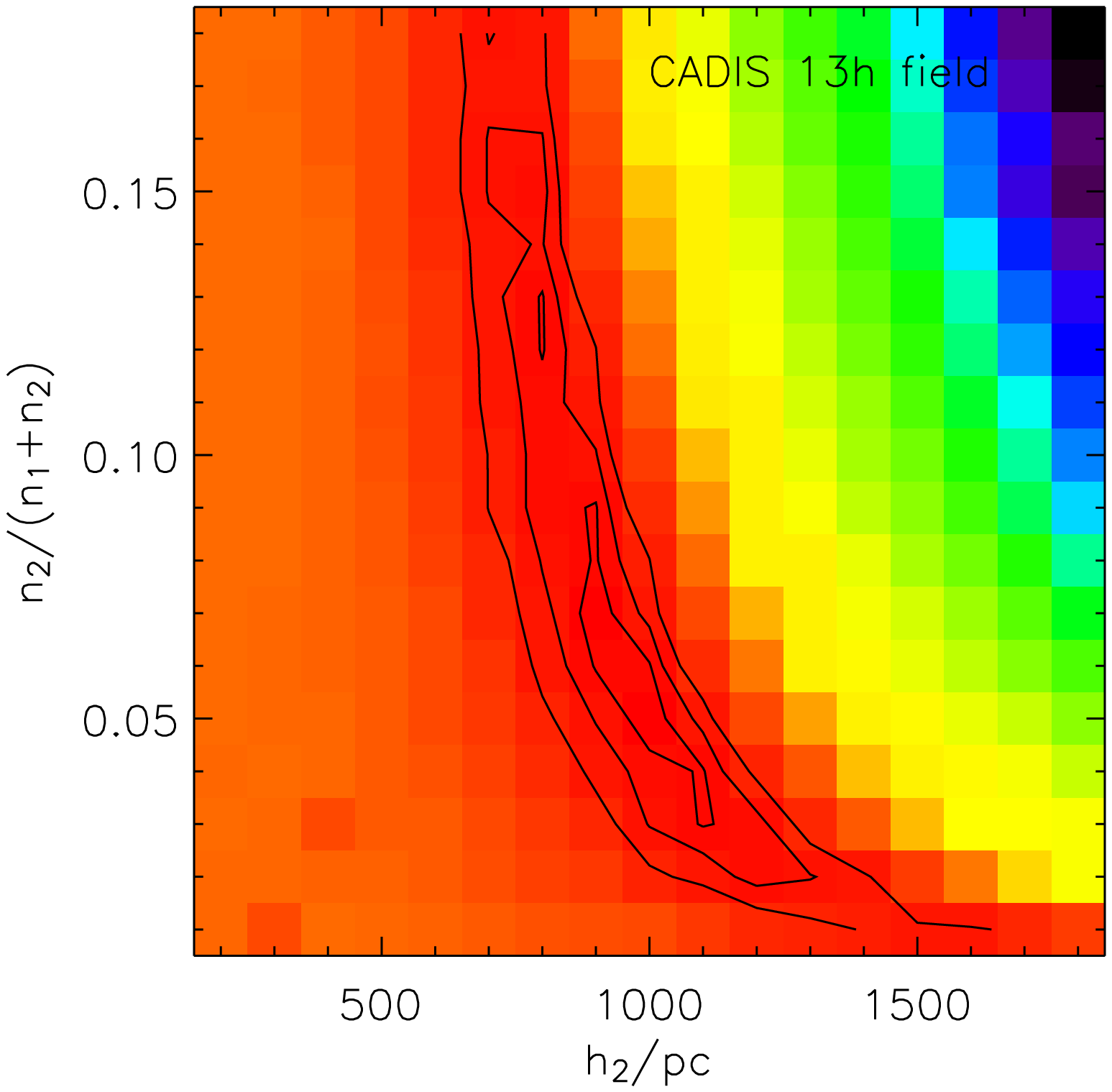,clip=t,width=8.0cm}
\psfig{figure=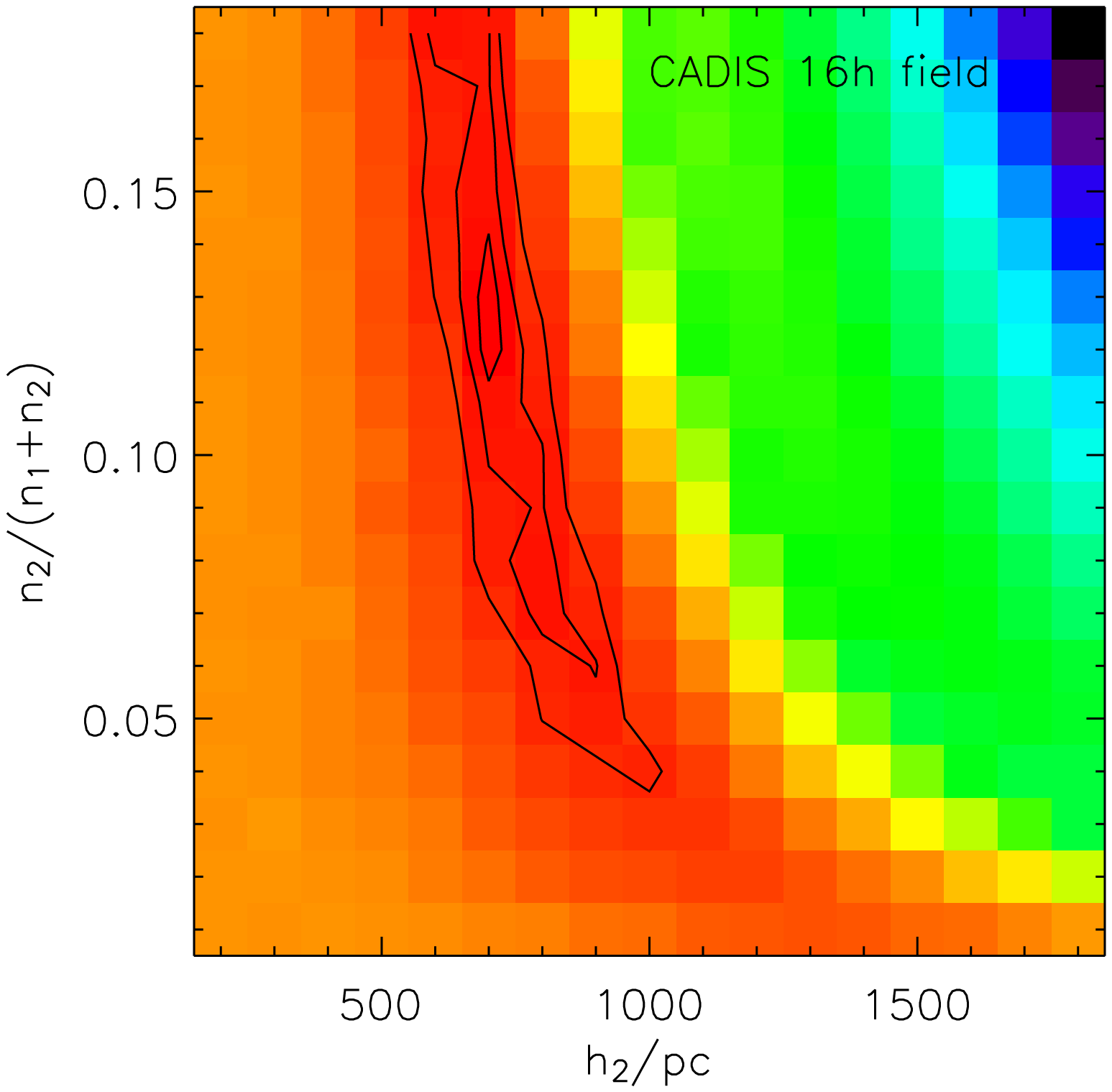,clip=t,width=8.0cm}}
\centerline{\psfig{figure=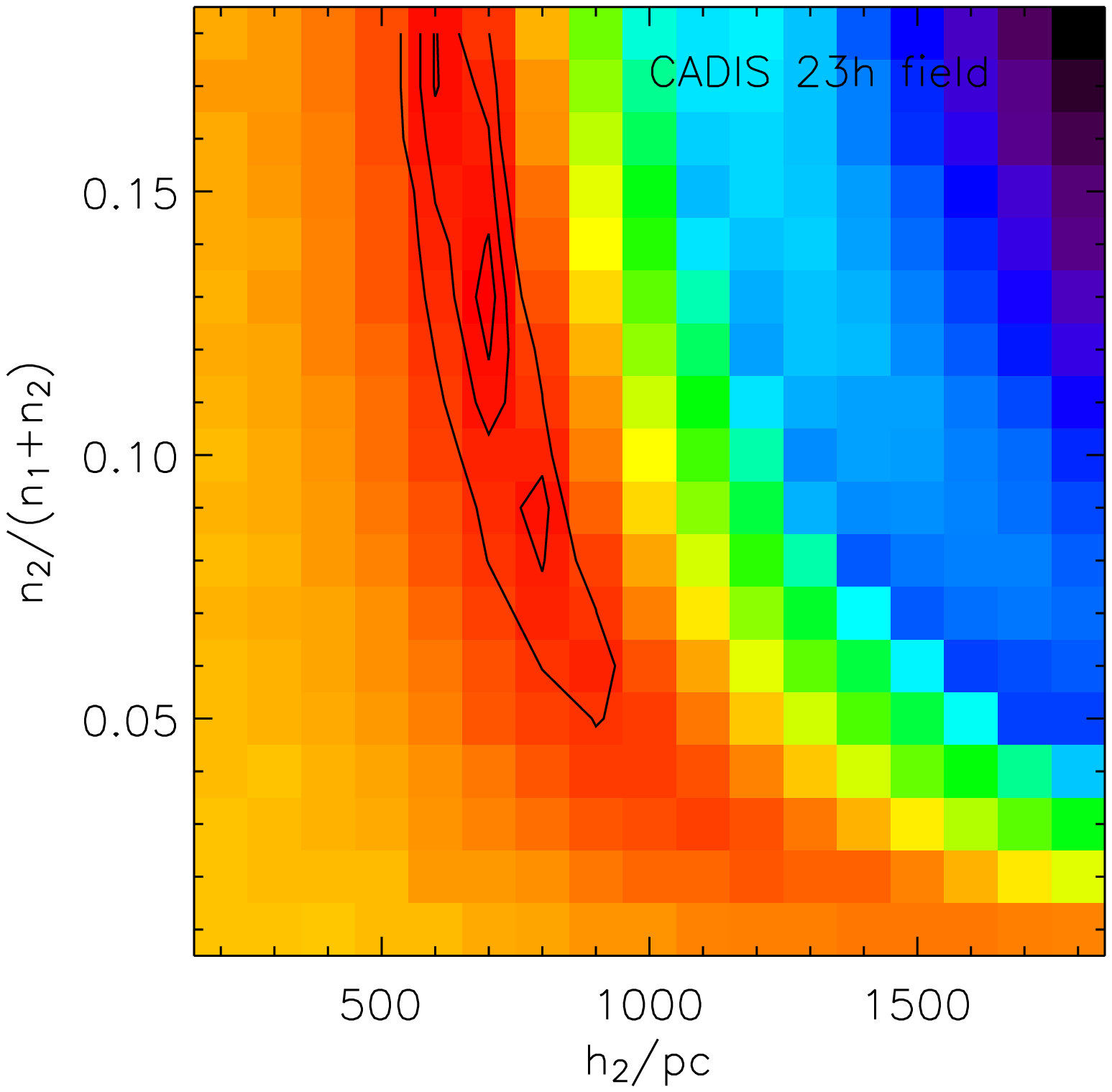,clip=t,width=8.0cm}
\psfig{figure=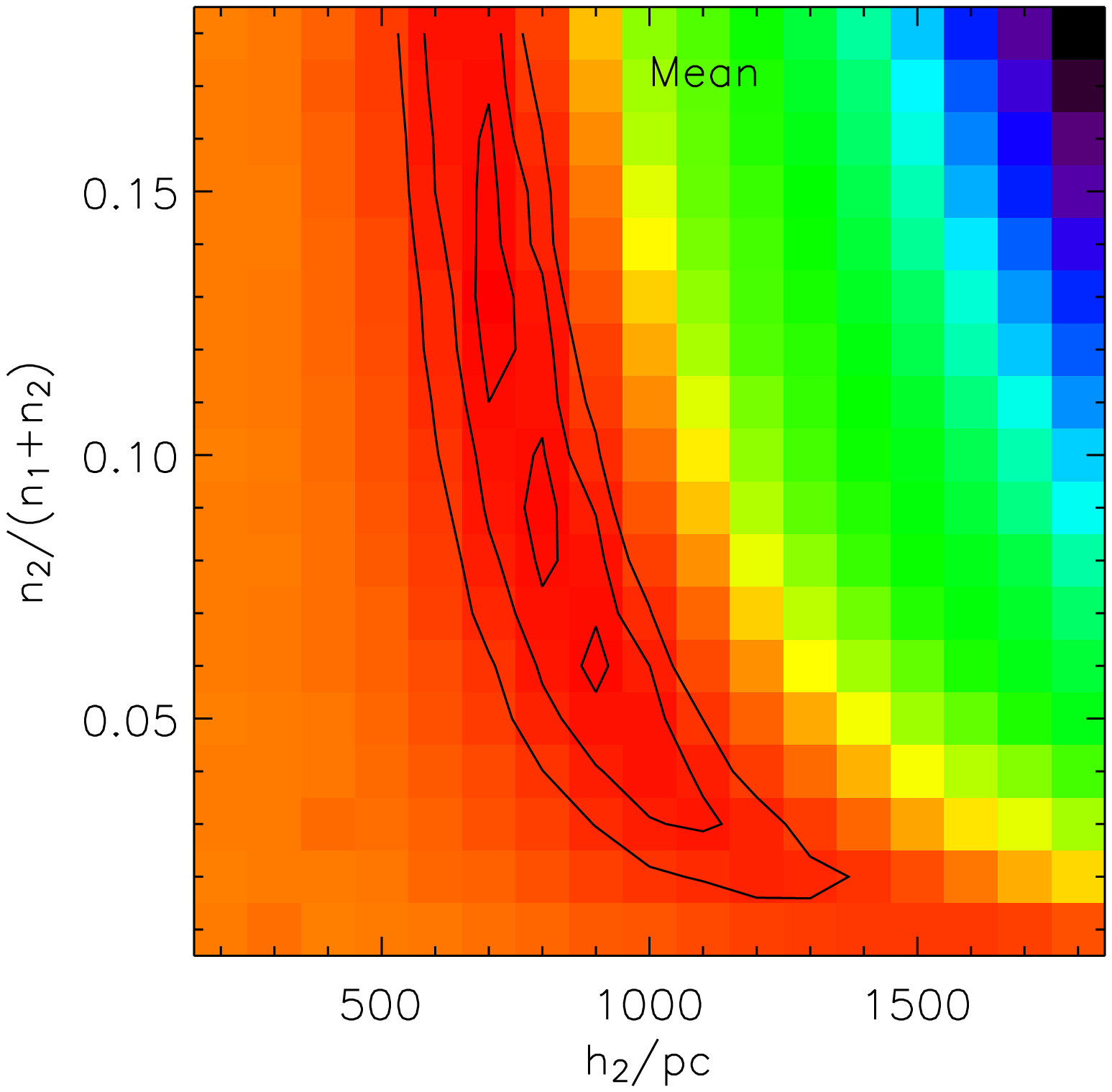,clip=t,width=8.0cm}}
\caption[ ]{The $C$ planes of the five fields for the
parameter $h_2$ and $n_2/(n_1+n_2)$ (with $h_r=3500$\,pc, $h_1=300$\,pc,
$(c/a)=0.6$,and $\alpha=2.5$). The 
plot in the lower right corner is the mean of the five indidual
fields. The contour lines show the $1\sigma$, $2\sigma$, and $3\sigma$
confidence levels.\label{chiq_reb}}
\end{figure*}

The overall $C$ is lowest for $\alpha=2.5$, $(c/a)=0.6$,
$h_r=3500$\,pc, and $h_1=300$\,pc, while $C$ is only marginally larger
if $(c/a)=1.0$ (and $\alpha=3.0$). There is no significant difference
between models with $h_r=2000$\,pc or $h_r=3500$\,pc. This is
consistent with the results from the previous section. Figure
(\ref{chiq_reb}) shows the
$C$ planes for $h_2$ and $n_2/(n_1+n_2)$ (with $h_r=3500$\,pc, $h_1=300$\,pc,
$(c/a)=0.6$, and $\alpha$ fixed at $2.5$), for each of the five
individual fields and the mean of all of them.

From the mean of the five fields we find the minimum of the $C$
statistic to be at $h_2=900$\,pc and $n_2/(n_1+n_2)\approx 0.09$. While the
value of $h_2$ is consistent with the value derived from the direct
measurement, the relative normalisation is  higher. However,
the fit is highly degenerated, and the minimum in the distribution of
$C$ is streched along the $n_2/(n_1+n_2)$-axis. Thus the relative
normalisation is only measurable with extremely large errors, and the
rather high value of $n_2/(n_1+n_2)=0.09$ should be regarded as an
upper limit.

\begin{figure*}[p]
\unitlength1cm
\begin{picture}(21,25)
\put(1,17){\epsfxsize=7.cm
\epsfbox{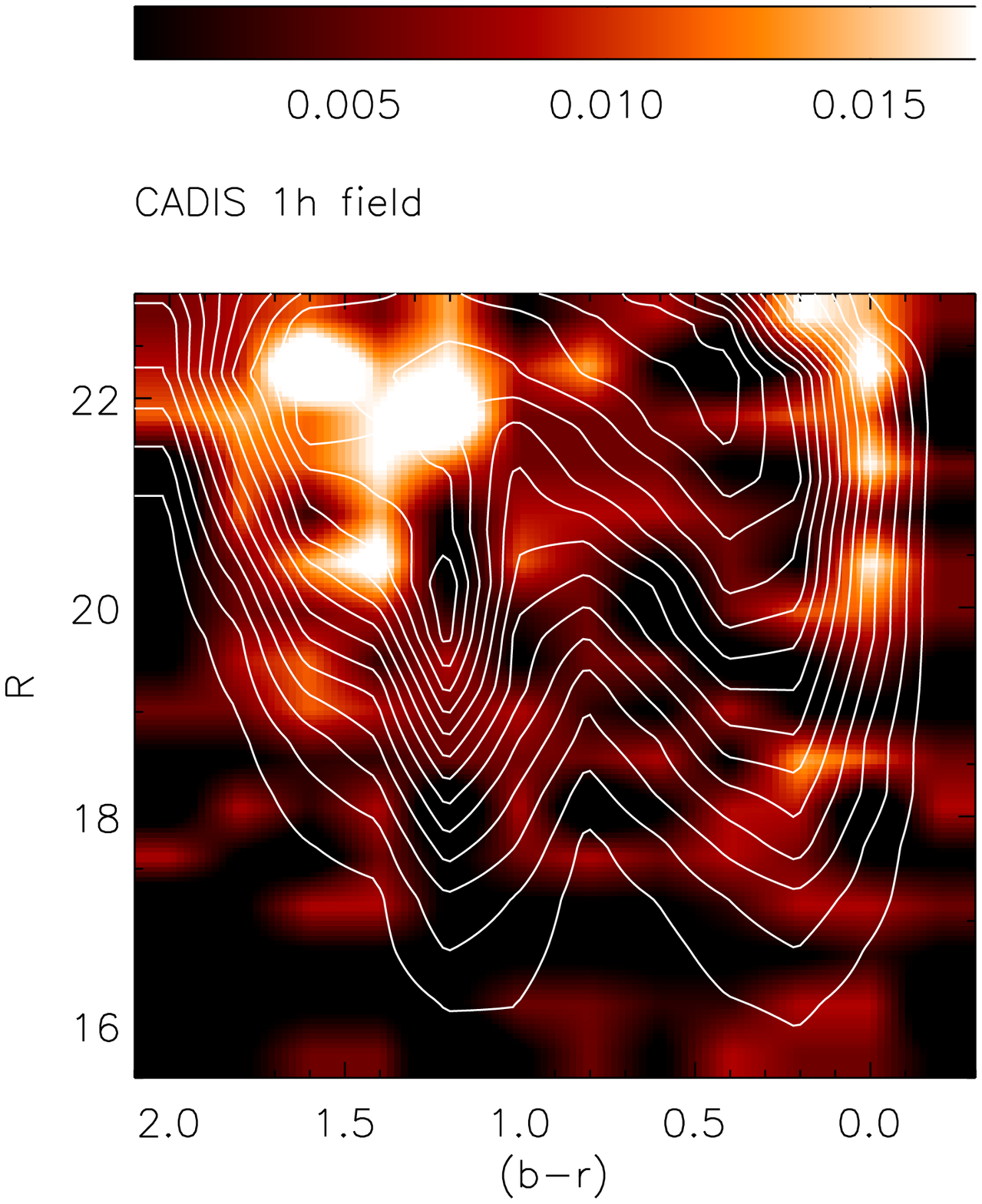}}
\put(1,9){\epsfxsize=7.cm
\epsfbox{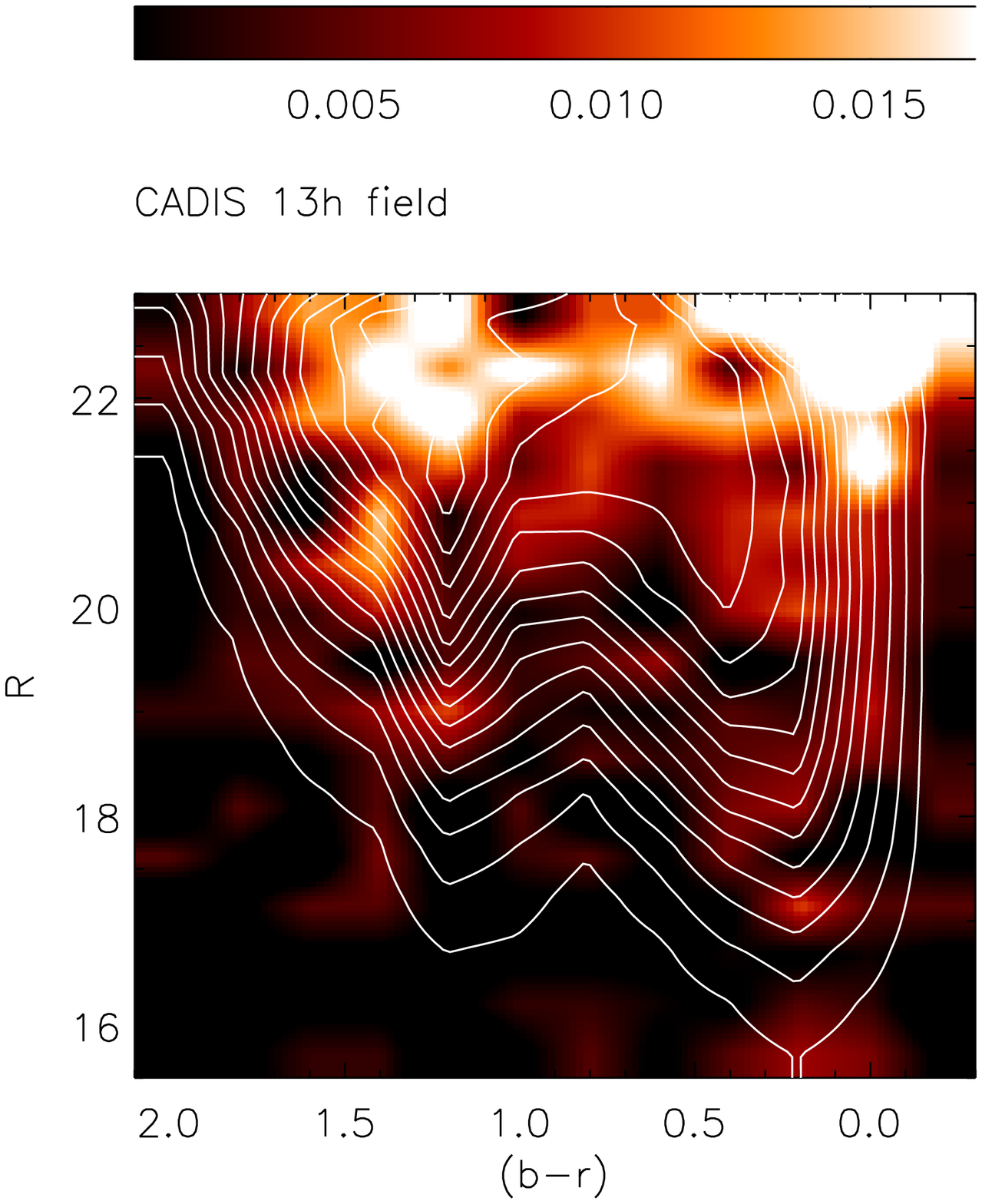}}
\put(9,17){\epsfxsize=7.cm
\epsfbox{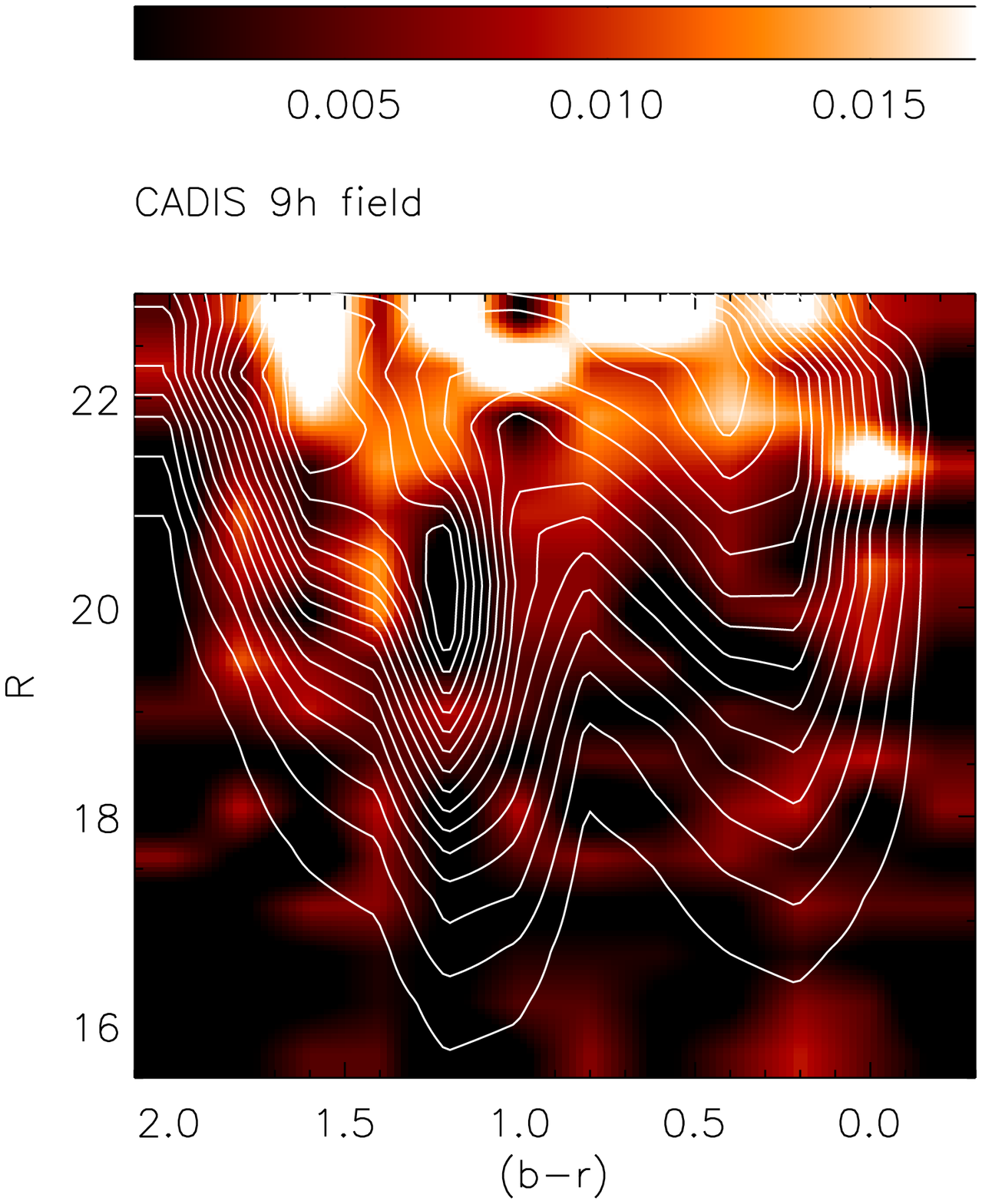}}
\put(9,9){\epsfxsize=7.cm
\epsfbox{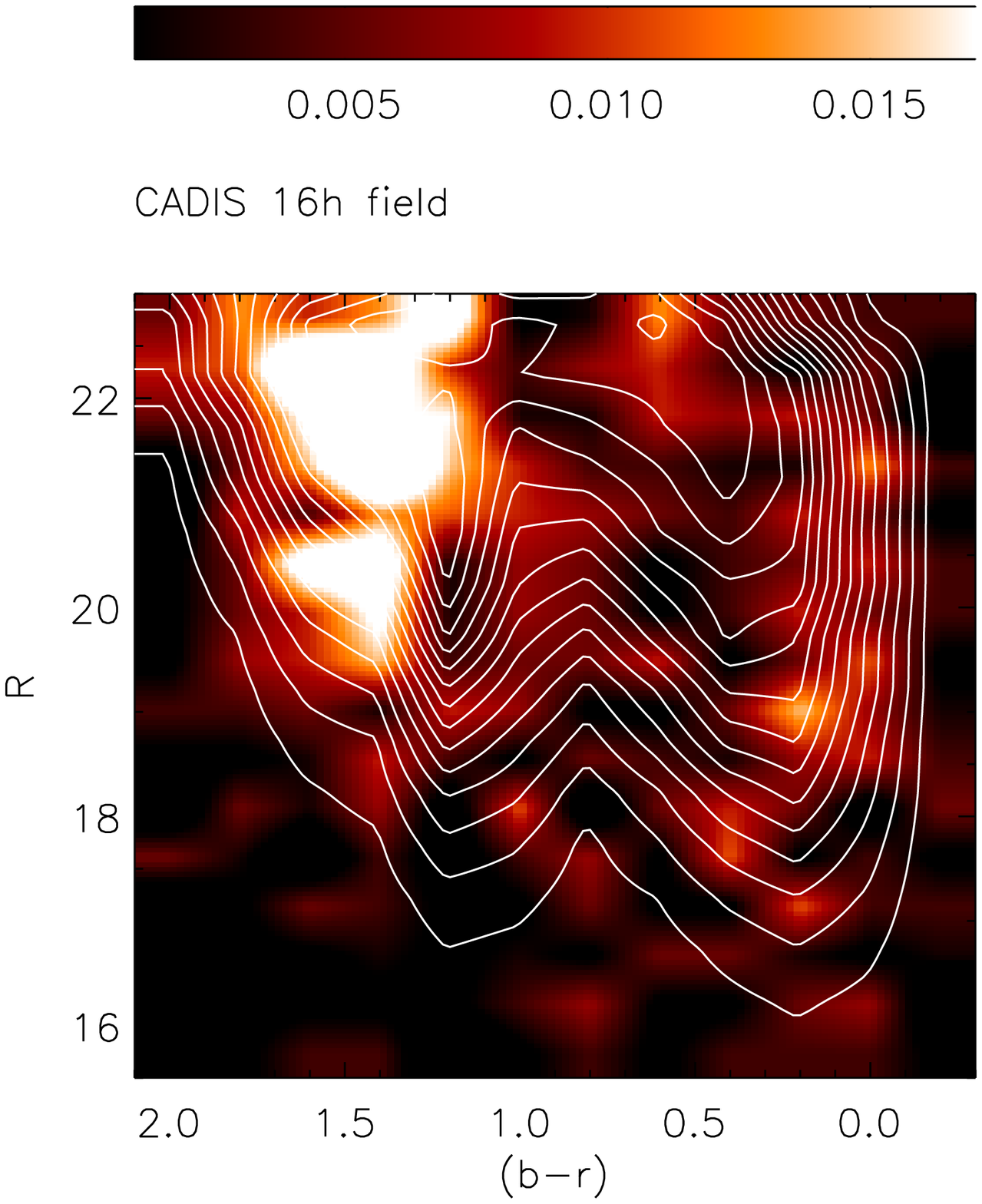}}
\put(1,1){\epsfxsize=7.cm
\epsfbox{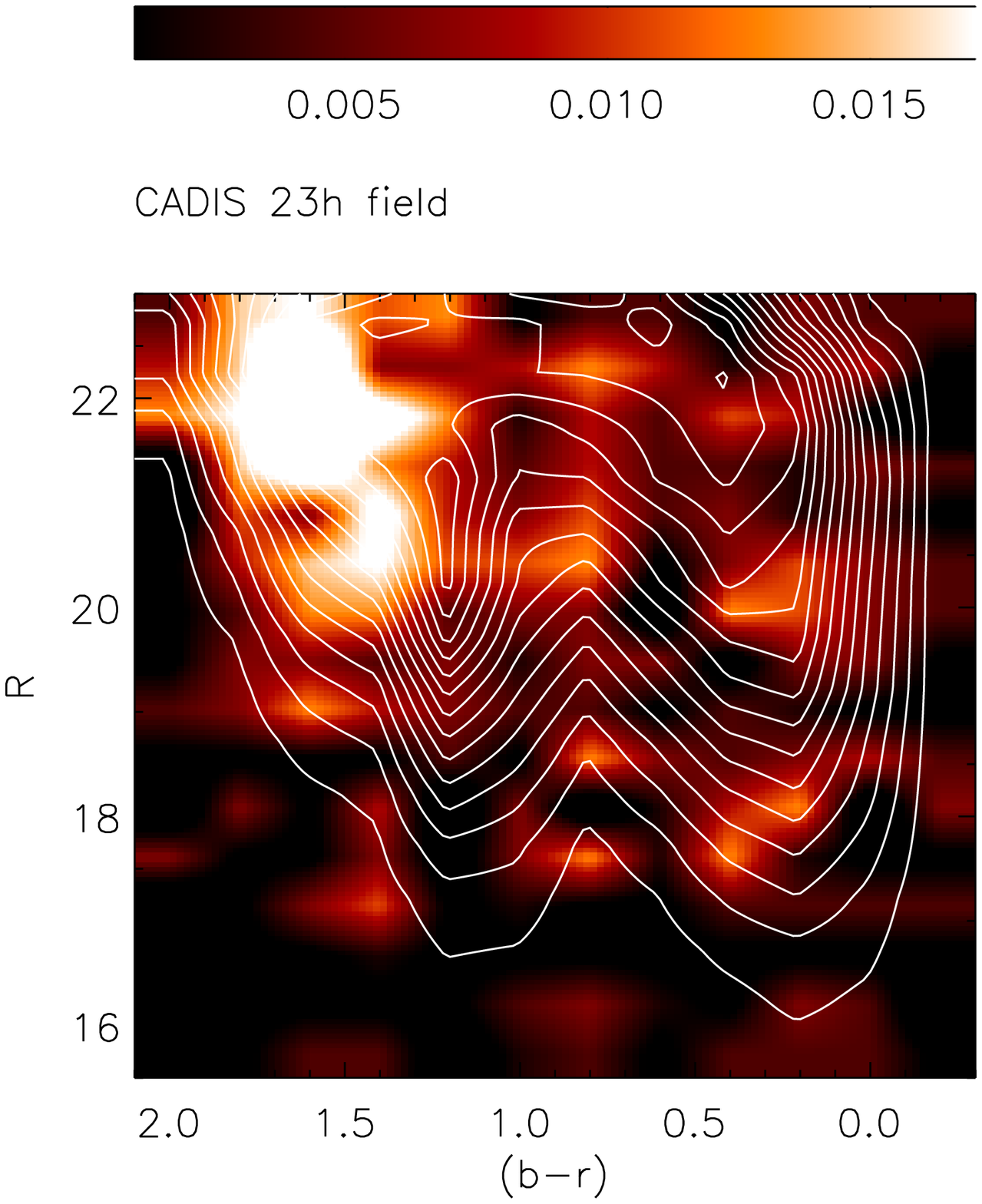}}
\end{picture}
\caption [ ] {The number density of stars in the $(b-r)_C$-$R_C$ plane, for the
five fields. The densities are normalised to the total number of stars
in each field. The images have been rebinned in steps of $0.1$ in
both color and magnitude. The contours overlaid are the simulated data for
 $h_r=3500$\,pc, $h_1=300$\,pc,
$h_2=900$\,pc, $n_2/(n_1+n_2)=0.09$,  $(c/a)=0.6$
and $\alpha=2.5$, also normalised to the number of simulated
stars. The contours are plotted for $0.001 \leq N \leq 0.015$ in steps
of $0.001$. 
\label{colormag}} 
\end{figure*}

Figure (\ref{colormag}) shows the color-coded representation of the
density of the stars in the magnitude-color diagram for the five
fields. The images have been rebinned in steps of $0.1$ in
both color and magnitude. The contour plots overlaid are the simulated
best-fit cases of the model ($h_r=3500$\,pc, $h_1=300$\,pc,
$h_2=900$\,pc, $n_2/(n_1+n_2)=0.09$,  $(c/a)=0.6$
and $\alpha=2.5$).

The prominent structure at the left of each image consists of disk
stars, whereas the halo stars are mainly distributed at the right. The
simulation matches the data very well except for the 13\,h field,
where the gap between halo and disk is not modelled. 
In all cases the  halo does not match the data very well, a possible
explanation might be that the real halo luminosity function, which we
do not know, is in fact significantly
different from the disk luminosity function which we use for the
simulation. 

\begin{figure}[h]
\centerline{\psfig{figure=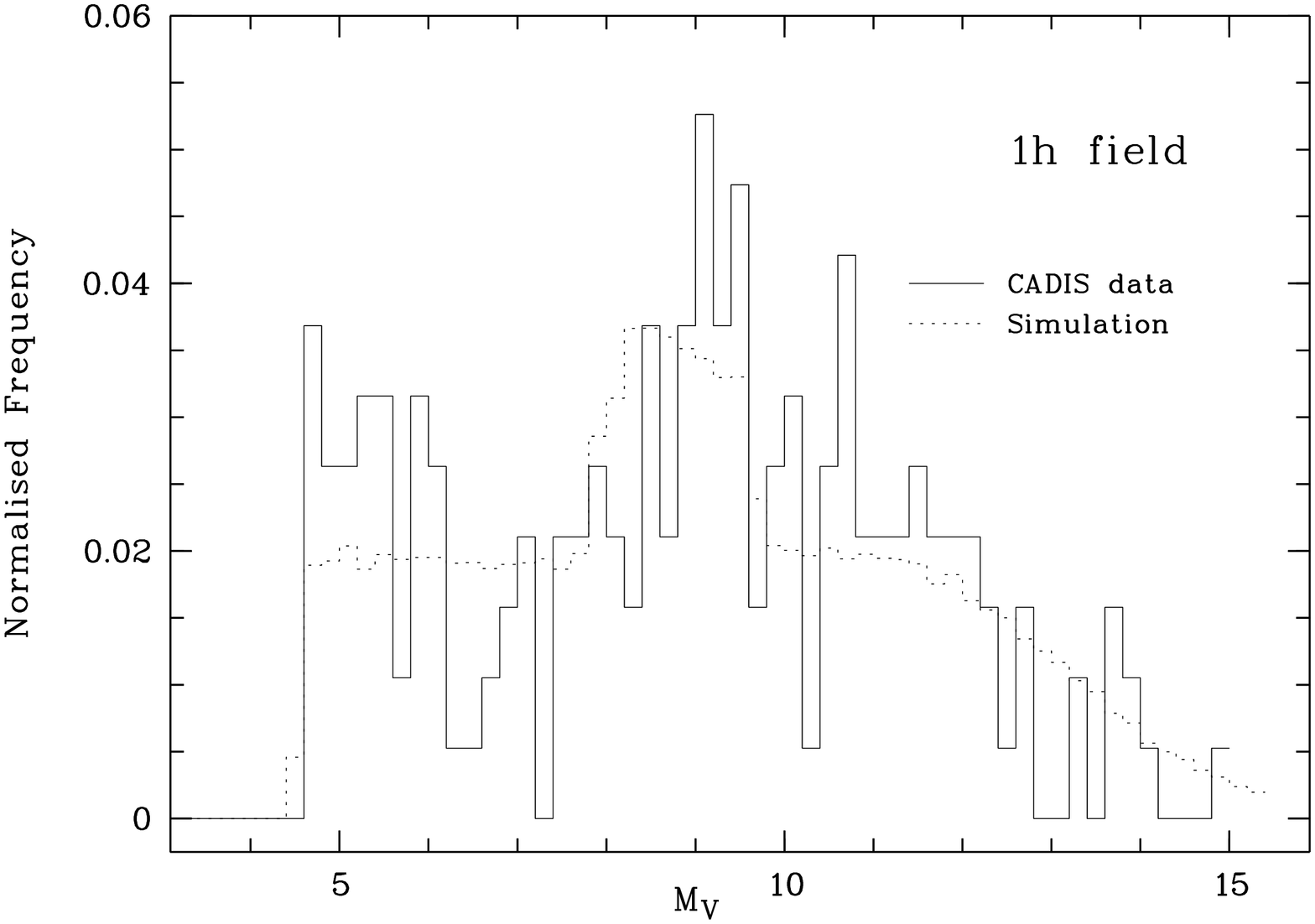,angle=270,clip=t,width=5.6cm}}
\centerline{\psfig{figure=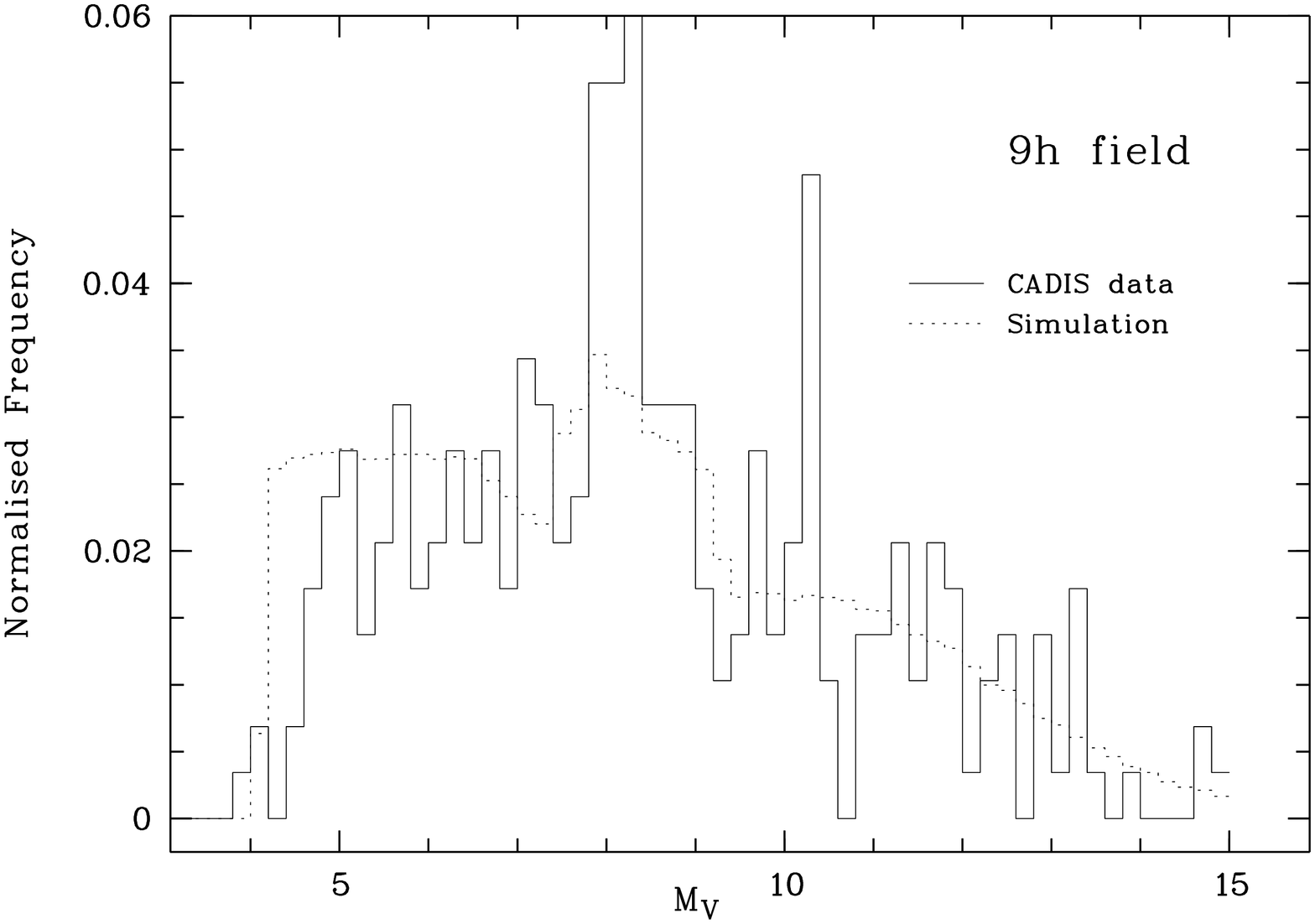,angle=270,clip=t,width=5.6cm}}
\centerline{\psfig{figure=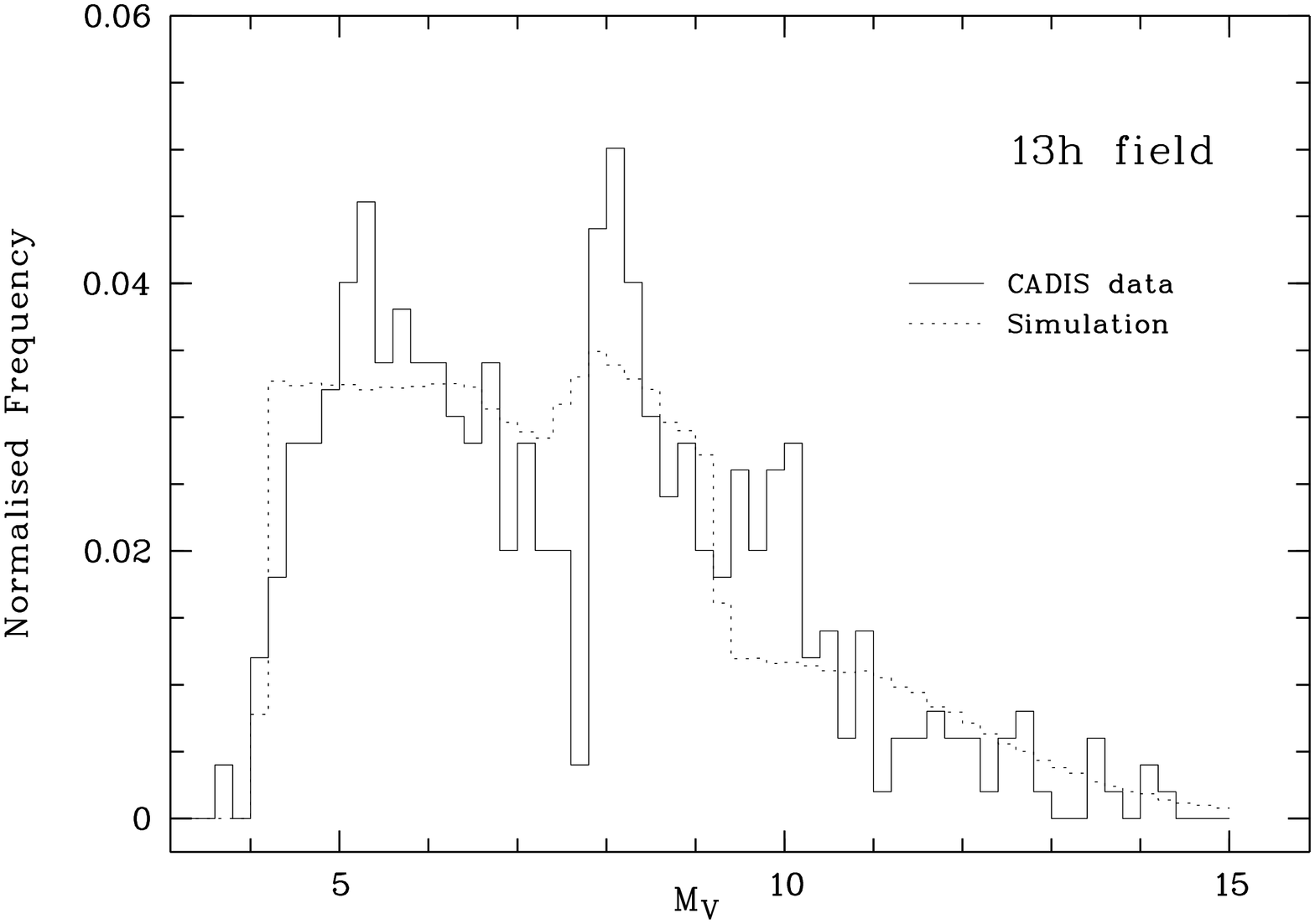,angle=270,clip=t,width=5.6cm}}
\centerline{\psfig{figure=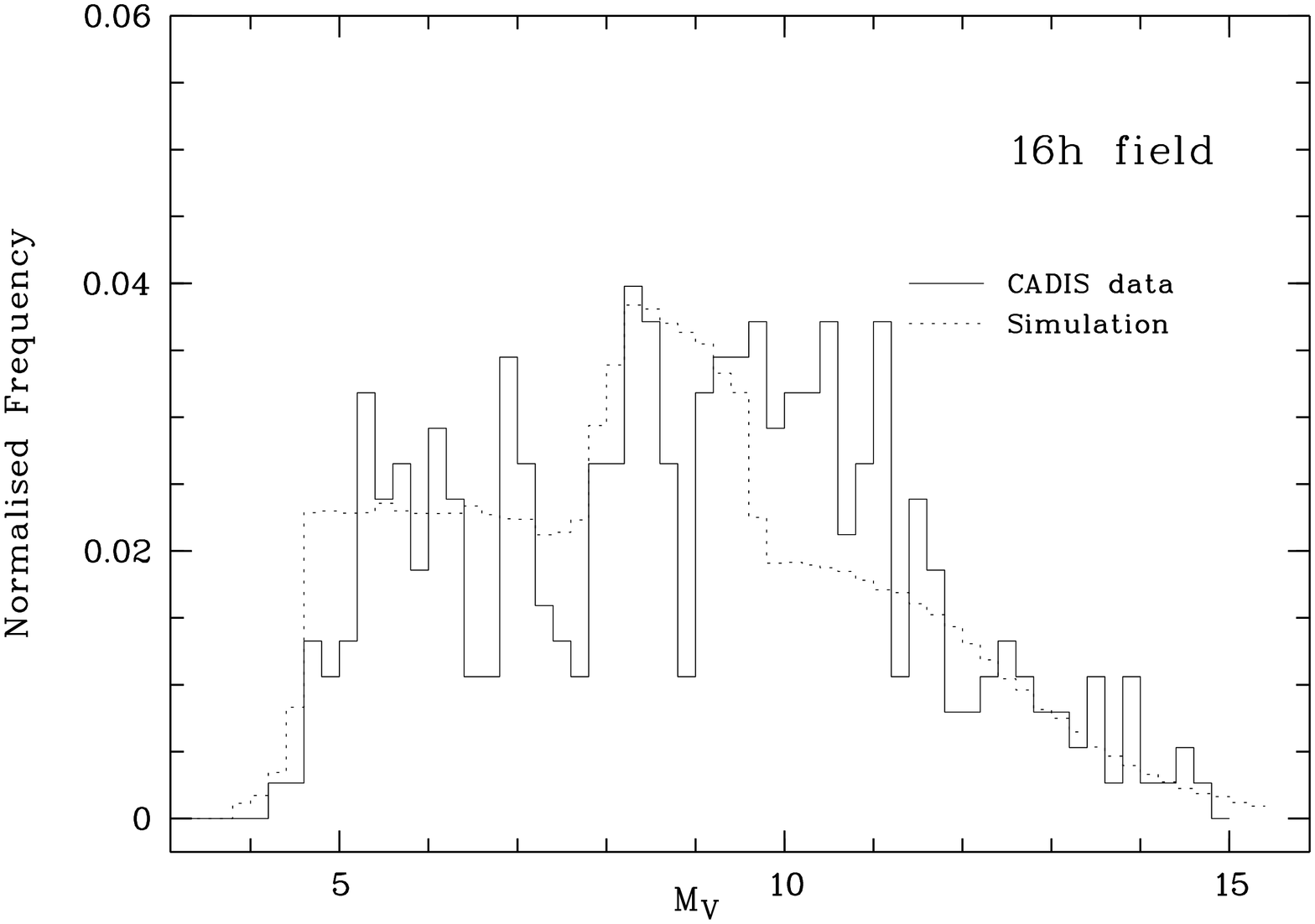,angle=270,clip=t,width=5.6cm}}
\centerline{\psfig{figure=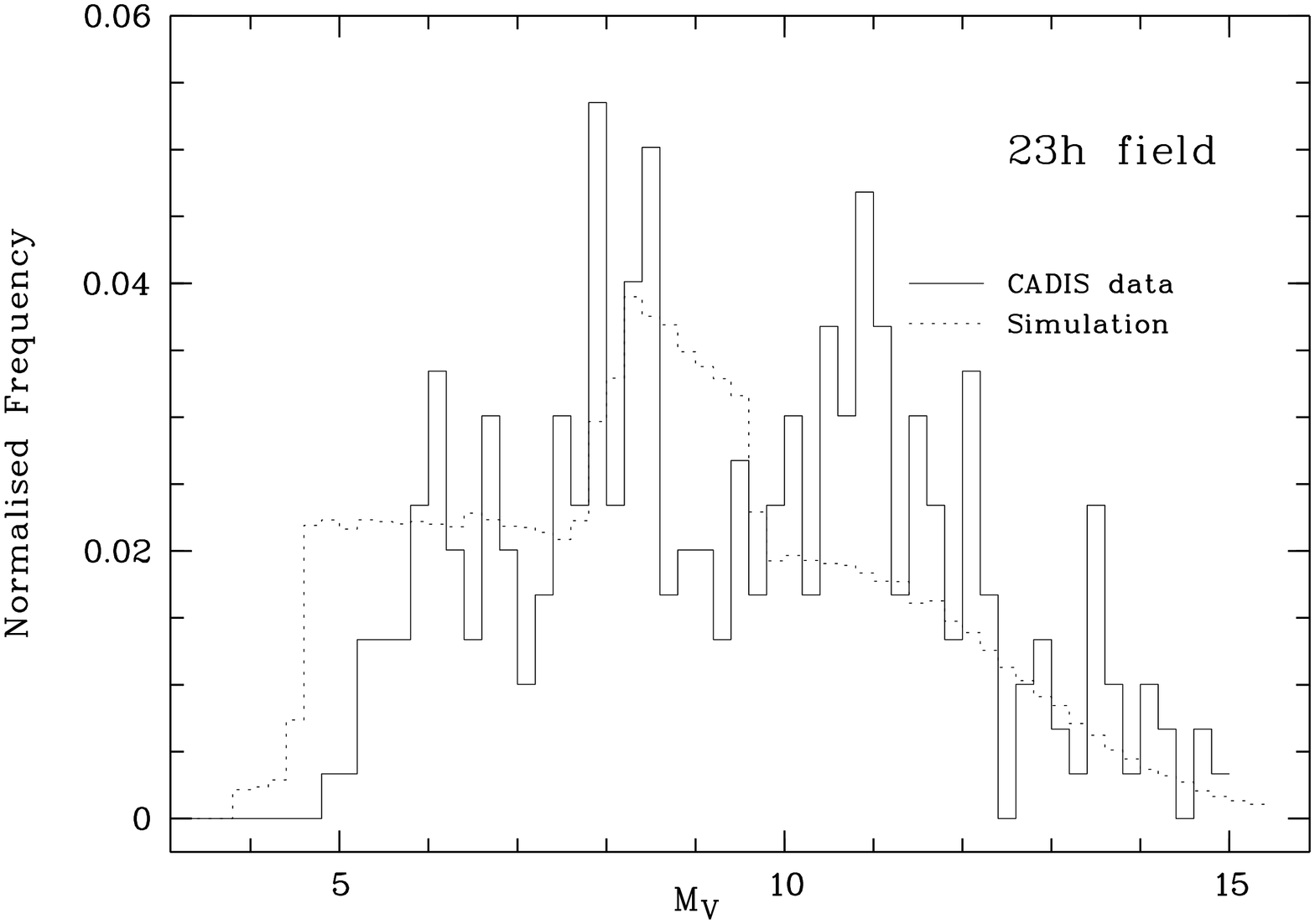,angle=270,clip=t,width=5.6cm}}
\caption[ ]{Comparison of absolute $V$ band magnitudes of the
simulation (dotted line) with the CADIS data (solid line), for the
best fit case.  \label{MVmag}}
\end{figure}

Figure (\ref{MVmag}) shows the distribution of absolute $V$ band
magnitudes of the CADIS data (calculated from the $(b-r)_C$ colors as described 
in
Section \ref{simcolorsandmags}) in comparison with the simulated data,
for the best fit cases. Both histograms have been normalised to the
total number of stars, in order that they can be compared directly. 

With exception of the CADIS  23\,h field the distribution of absolute
$V$ band magnitudes is matched very well by the simulation. In the
9\,h field the model can not account for the large peak at $M_V\sim 8$. It is not improbable
that the Galactic halo is clumpy \citep{Newberg02}, so the the feature 
may well be an overdensity  of stars with respect to the smooth
powerlaw halo.

In the 23\,h field the number of bright halo stars is
overpredicted. Since the 16\,h and the 23\,h  
fields are located at the same Galactic longitude and $|b_{16
h}|\approx |b_{23 h}|$, both fields should show the same 
distribution of color or absolute magnitudes, respectively. However,
as both distributions differ from each other at the bright end,
it seems more likely that the 23\,h field is 'missing' some stars,
rather than that the model (which matches the 16\,h field more closely) overpredicts
stars which are not supposed to be there.

\section{Summary and Discussion}\label{discussion}
Our model of the Galaxy consists of three components: a thin and a
thick disk (described by a sum of two exponentials), and a power law
halo. The local normalisation of disk and halo stars has been fixed by
stars selected from the CNS4 \citep{Venice} by their colors and apparent magnitudes.

We have estimated the structure parameters of the stellar density
distribution of the Milky Way using two different, complementary
methods: from a direct measurement of the density distribution of stars perpendicular to 
the Galactic plane ({\it Baconian ansatz}), and by modelling the
observed colors and apparent magnitudes of the stars and comparing
them to the data ({\it Kartesian ansatz}).

Both methods have their advantages and limitations: the direct
determination of the density involves completeness corrections because
faint stars are only traced to small distances. These rely on the
assumption that the stellar luminosity function (SLF) does not change
with distance and is the same for disk and halo stars. The same
assumption of course is used in the simulation. However, the
simulation avoids completeness corrections and thus the errors do not
depend on an iterative correction. A $\chi^2$ fitting of a simulated
color-magnitude distribution of stars works very well for a large
number of stars; with the CADIS data the method is at the edge of
feasibility. We therefore used the $C$ statistic developed by
\citet{Cash79} for sparse sampling to find the best fitting parameter
combination. 

From the $\chi^2$ fit to the direct measurement of the density
distribution of the stars, we find $h_1=281\pm13$\,pc if we assume
$h_r=2000$\,pc, and $h_1=283\pm13$\,pc if $h_r=3500$\,pc. Obviously,
due to the high Galactic latitude of the fields ($45\leq b\leq 60$),
we are not able to measure the scalelength (the halo becomes dominant
already at a projected radial distance of only about 7\,kpc).  The fit
of the scaleheight of the thick disk and the local percentage of thick
disk stars is highly degenerate. The best estimate is
$h_2\approx1000$\,kpc, and $n_2/(n1+n_2)\approx 0.04$. We also fitted
the density distribution in the halo and find $\alpha=2.51\pm0.07$ for
$(c/a)=1.0$, and $\alpha=2.12\pm0.06$ if we allow for a flattening of
the stellar halo with an axial ratio of $(c/a)=0.6$. The values of
$\chi^2$ formally favour the spherical halo, but this result is not
significant, since the power law slope and the axial ratio are highly
correlated. Furthermore the measured values of the exponent $\alpha$,
and the axial ratio $c/a$ depend on the local normalisation assumed
for the fit: slightly higher (lower) values of $\rho_0$ yield slightly
higher (lower) values of $\alpha$, and a flattened (spherical) halo is favoured. 

From the Cartesian method we derive  $h_2\approx900$\,kpc, $n_2/(n1+n_2)\approx 0.09$. 
We also find that the overall $\chi^2$ or $C$, respectively, is
smallest for $h_1=300$\,pc and an exponent $\alpha=3.0$. If we allow
for a flattening of the halo with an axial ratio of $(c/a)=0.6$, we
find that a slightly smaller value of $\alpha=2.5$ fits the data
better. The corresponding values of $\chi^2$ or $C$, respectively, are
only marginally smaller for a flattened halo. The values of $\alpha$
are generally slightly larger than in the direct measurement. A
possible explanation for this difference is that when we fit the
simulated distribution of stars in the color-magnitude diagram, we fit
all components of the galaxy simultaneously (whereas in the direct
measurement we fit disk and halo separately), and there might be a
degeneracy between halo and thick disk parameters.  

However, in general both methods yield essentially the same result, which we
regard as a corroboration that both work correctly within their
limitations.

\subsection{Comparison with other results}
A very detailed compilation of different results from different
authors can be found in \citet{Siegel02}. The values for the scaleheight
of the thick disk and the relative normalisation of thick disk stars
range from 900\,kpc to 2400\,kpc, and from 1\,\% to 6\,\%,
respectively. 

\citet{SloanStars01} carried out an investigation of Galactic
structure using $5.8\cdot 10^5$ stars brighter than $g'=21^{\rm mag}$
from the Sloan Digital Sky Survey (SDSS), covering a total area of
279\,\sqdegr. They presented models of the Milky Way for a large number
of free parameters, that is, they determine by means of a maximum
likelyhood analysis the scaleheights of the thin and thick disk,
respectively, the relative normalisation of both disk components in
the solar neighbourhood, the offset of
the sun above the Galactic plane, the exponent $\alpha$ of the power
law halo and the flattening of the halo. 

They find $h_1=330\pm3$\,pc, $580\la h_2\la 750$\,pc, $0.065\la
n_2/(n_1+n_2)\la 0.13$, and $\alpha=2.5\pm0.3$, and $c/a=0.55\pm
0.06$. \citet{Lemon04} also found a significant flattening of 
the halo ($c/a=0.56 \pm 0.01$).

While our results lie well within the range found by
most authors, our measurements for the scaleheights of the disk are
quite dissimilar from the Sloan results. 

A possible explanation might be that their determination, the results of which
differ significantly from all other determinations, suffers from the
degeneracy due to the large number of degrees of freedom in their
simulation: their fitting routine may have found a secondary minimum.

\citet{Robin00} found that there is a significant degeneracy between
the power law index $\alpha$ and the axis ratio of the halo. Their
best fit to the data yields $\alpha=2.44$ and $c/a=0.76$, but a
flatter spheroid with $c/a=0.6$ with $\alpha=2$ is not excluded either.
We find the same degeneracy in our data: If we assume $(c/a)=0.6$, we
find $\alpha=2.12\pm 0.06$ (or $\alpha=2.5$ in the simulation),  whereas for
$(c/a)=1.0$ we find that $\alpha=2.51\pm 0.07$ ( $\alpha=3.0$)  fits the data better. 

\subsection{Future prospects}
This investigation shows quite clearly that  the fit of a sum of two
exponentials is highly degenerate if the relative normalisation
$n_2/n_1$ or the scaleheights $h_1$ and $h_2$ is not known. Also the
measurement of $\alpha$, the power law index of the stellar halo, and
its axial ratio, $(c/a)$, are correlated and depend on the local normalisation.

Therefore it certainly depends on the assumptions and the number of
the degrees of freedom which values of $h_1$,  $h_2$,  $n_1$, $n_2$,
$\alpha$, and $(c/a)$  are determined in different analyses. The more
structure parameters are known and kept fixed, the narrower is the
$\chi^2$ (or $C$) distribution.  Thus it is important to combine results from different methods. 
 
The scaleheight of the thick disk for example will be determinable with high accuracy if the local
normalisation of thick disk stars is known better. The local density of
stars belonging to different components of the Galaxy has to be
determined by means of kinematics and metalicities of the stars in
order to break the degeneracy of the fit of a double exponential.

The degeneracy between the power law slope and the flattening of the
halo can be broken  by simultaneously  fitting density distributions estimated from deep star
counts in a large number of fields at different Galactic latitudes and
longitudes, respectively.  

When the ``size'' and exact shape of the stellar halo is known, it
will become much easier to determine the structure parameters of the
thick disk.

CADIS, although designed as an extragalactic survey, provided a
sufficient number of stars (and fields) to investigate the structure
of the Milky Way. However, not only  larger statistics, but also a
much better knowledge of the 
luminosity function of halo stars is required to increase the
accuracy of the results.

\begin{acknowledgements}

We thank all those involved in the Calar Alto Deep Imaging Survey,
especially H.-J. R{\"o}ser and C. Wolf, without whom carrying out the
whole project would have been impossible.\\
We also thank M. Alises and A. Aguirre for their help
and support during many nights at Calar Alto Observatory, and for carefully
carrying out observations in service mode.\\
We are greatly indepted to our referee, Annie. C. Robin, who pointed out several points
which had not received sufficient attention in the original
manuscript. This led to a substantial improvement of the paper.\\ 
S. Phleps acknowledges financial support
by the SISCO Network provided through the European
Community's Human Potential Programme under contract HPRN-CT-2002-00316.

\end{acknowledgements}

\bibliographystyle{aa}

\bibliography{lit}

\begin{thebibliography}{44}
\expandafter\ifx\csname natexlab\endcsname\relax\def\natexlab#1{#1}\fi

\bibitem[{{Bahcall} \& {Soneira}(1980{\natexlab{a}})}]{BahcallSoneiraII80}
{Bahcall}, J.~N. \& {Soneira}, R.~M. 1980{\natexlab{a}}, \apjl, 238, L17

\bibitem[{{Bahcall} \& {Soneira}(1980{\natexlab{b}})}]{BahcallSoneiraI80}
---. 1980{\natexlab{b}}, \apjs, 44, 73

\bibitem[{{Bahcall} \& {Soneira}(1981)}]{BahcallSoneira81}
---. 1981, \apjs, 47, 357

\bibitem[{{Bertin} \& {Arnouts}(1996)}]{Bertin96}
{Bertin}, E. \& {Arnouts}, S. 1996, \aaps, 117, 393

\bibitem[{{Bienaym{\' e}}(1999)}]{Bienayme99}
{Bienaym{\' e}}, O. 1999, \aap, 341, 86

\bibitem[{{Cash}(1979)}]{Cash79}
{Cash}, W. 1979, \apj, 228, 939

\bibitem[{{Chen} {et~al.}(1999){Chen}, {Figueras}, {Torra}, {Jordi}, {Luri}, \&
  {Galad{\'{\i}}-Enr{\'{\i}}quez}}]{Chen99}
{Chen}, B., {Figueras}, F., {Torra}, J., {et~al.} 1999, \aap, 352, 459

\bibitem[{{Chen} {et~al.}(2001){Chen}, {Stoughton}, {Smith}, {Uomoto}, {Pier},
  {Yanny}, {Ivezi{\' c}}, {York}, {Anderson}, {Annis}, {Brinkmann}, {Csabai},
  {Fukugita}, {Hindsley}, {Lupton}, {Munn}, \& {the SDSS
  Collaboration}}]{SloanStars01}
{Chen}, B., {Stoughton}, C., {Smith}, J.~A., {et~al.} 2001, \apj, 553, 184

\bibitem[{{Digby} {et~al.}(2003){Digby}, {Hambly}, {Cooke}, {Reid}, \&
  {Cannon}}]{Digby03}
{Digby}, A.~P., {Hambly}, N.~C., {Cooke}, J.~A., {Reid}, I.~N., \& {Cannon},
  R.~D. 2003, \mnras, 344, 583

\bibitem[{{Freeman}(1992)}]{Freeman92}
{Freeman}, K.~C. 1992, in IAU Symp. 149: The Stellar Populations of Galaxies,
  65

\bibitem[{{Fried} {et~al.}(2001){Fried}, {von Kuhlmann}, {Meisenheimer}, {Rix},
  {Wolf}, {Hippelein}, {K{\" u}mmel}, {Phleps}, {R{\" o}ser}, {Thierring}, \&
  {Maier}}]{Fried01}
{Fried}, J.~W., {von Kuhlmann}, B., {Meisenheimer}, K., {et~al.} 2001, \aap,
  367, 788

\bibitem[{{Fuchs} \& {Jahrei{\ss}}(1998)}]{FuchsHalo}
{Fuchs}, B. \& {Jahrei{\ss}}, H. 1998, \aap, 329, 81

\bibitem[{{Fuchs} \& {Wielen}(1993)}]{FuchsWielen93}
{Fuchs}, B. \& {Wielen}, R. 1993, in AIP Conf. Proc. 278: Back to the Galaxy,
  580

\bibitem[{{Gilmore}(1984)}]{Gilmore84}
{Gilmore}, G. 1984, \mnras, 207, 223

\bibitem[{{Gilmore} \& {Reid}(1983)}]{GilmoreReid83}
{Gilmore}, G. \& {Reid}, N. 1983, \mnras, 202, 1025

\bibitem[{{Gilmore} \& {Wyse}(1987)}]{GilmoreWyse87}
{Gilmore}, G. \& {Wyse}, R.~F.~G. 1987, in NATO ASIC Proc. 207: The Galaxy,
  247--279

\bibitem[{{Gould}(2003)}]{Gould03}
{Gould}, A. 2003, \apj, 583, 765

\bibitem[{{Gould} {et~al.}(1998){Gould}, {Flynn}, \& {Bahcall}}]{Gould98}
{Gould}, A., {Flynn}, C., \& {Bahcall}, J.~N. 1998, \apj, 503, 798

\bibitem[{{Jahrei{\ss}} \& {Wielen}(1997)}]{Venice}
{Jahrei{\ss}}, H. \& {Wielen}, R. 1997, Proceedings of the {ESA} {S}ymposium
  `{H}ipparcos - {V}enice '97', 13-16 May, {V}enice, {I}taly, {ESA} {SP}-402
  ({J}uly 1997), p. 675-680, 402, 675

\bibitem[{{Lang}(1992)}]{Lang92}
{Lang}, K.~R. 1992, in Astrophysical {D}ata {I}. {P}lanets and {S}tars, {X},
  937 pp. 33 figs.. {S}pringer-{V}erlag {B}erlin {H}eidelberg {N}ew {Y}ork

\bibitem[{{Lemon} {et~al.}(2004){Lemon}, {Wyse}, {Liske}, {Driver}, \&
  {Horne}}]{Lemon04}
{Lemon}, D.~J., {Wyse}, R.~F.~G., {Liske}, J., {Driver}, S.~P., \& {Horne}, K.
  2004, \mnras, 347, 1043

\bibitem[{{Meisenheimer} \& {R\"oser}(1986)}]{MeisenroeserEval}
{Meisenheimer}, K. \& {R\"oser}, H.-J. 1986, in Use of CCD Detectors in
  Astronomy, Baluteau J.-P. and D'Odorico S., (eds.), 227

\bibitem[{{Mendez} \& {van Altena}(1998)}]{MendezvanAltena98}
{Mendez}, R.~A. \& {van Altena}, W.~F. 1998, \aap, 330, 910

\bibitem[{{Newberg} {et~al.}(2002){Newberg}, {Yanny}, {Rockosi}, {Grebel},
  {Rix}, {Brinkmann}, {Csabai}, {Hennessy}, {Hindsley}, {Ibata}, {Ivezi{\' c}},
  {Lamb}, {Nash}, {Odenkirchen}, {Rave}, {Schneider}, {Smith}, {Stolte}, \&
  {York}}]{Newberg02}
{Newberg}, H.~J., {Yanny}, B., {Rockosi}, C., {et~al.} 2002, \apj, 569, 245

\bibitem[{{Norris}(1999)}]{Norris99}
{Norris}, J.~E. 1999, \apss, 265, 213

\bibitem[{{Ojha} {et~al.}(1996){Ojha}, {Bienayme}, {Robin}, {Creze}, \&
  {Mohan}}]{Ojha96}
{Ojha}, D.~K., {Bienayme}, O., {Robin}, A.~C., {Creze}, M., \& {Mohan}, V.
  1996, \aap, 311, 456

\bibitem[{{Oke}(1990)}]{Oke90}
{Oke}, J.~B. 1990, \aj, 99, 1621

\bibitem[{{Phleps} {et~al.}(2000){Phleps}, {Meisenheimer}, {Fuchs}, \&
  {Wolf}}]{Phleps00}
{Phleps}, S., {Meisenheimer}, K., {Fuchs}, B., \& {Wolf}, C. 2000, \aap, 356,
  108

\bibitem[{{Pickles}(1998)}]{Pickles98}
{Pickles}, A.~J. 1998, \pasp, 110, 863

\bibitem[{{Reid} {et~al.}(1997){Reid}, {Gizis}, {Cohen}, {Pahre}, {Hogg},
  {Cowie}, {Hu}, \& {Songaila}}]{Reid97}
{Reid}, I.~N., {Gizis}, J.~E., {Cohen}, J.~G., {et~al.} 1997, \pasp, 109, 559

\bibitem[{{Reid} {et~al.}(1996){Reid}, {Yan}, {Majewski}, {Thompson}, \&
  {Smail}}]{Reid96}
{Reid}, I.~N., {Yan}, L., {Majewski}, S., {Thompson}, I., \& {Smail}, I. 1996,
  \aj, 112, 1472

\bibitem[{{Reid} \& {Majewski}(1993)}]{ReidMajewski93}
{Reid}, N. \& {Majewski}, S.~R. 1993, \apj, 409, 635

\bibitem[{{Reyl{\' e}} \& {Robin}(2001)}]{Reyle01}
{Reyl{\' e}}, C. \& {Robin}, A.~C. 2001, \aap, 373, 886

\bibitem[{{Robin} \& {Creze}(1986)}]{Robin86}
{Robin}, A. \& {Creze}, M. 1986, \aap, 157, 71

\bibitem[{{Robin} {et~al.}(1992){Robin}, {Creze}, \& {Mohan}}]{Robin92}
{Robin}, A.~C., {Creze}, M., \& {Mohan}, V. 1992, \aap, 265, 32

\bibitem[{{Robin} {et~al.}(2000){Robin}, {Reyl{\' e}}, \& {Cr{\' e}z{\'
  e}}}]{Robin00}
{Robin}, A.~C., {Reyl{\' e}}, C., \& {Cr{\' e}z{\' e}}, M. 2000, \aap, 359, 103

\bibitem[{{R\"oser} \& {Meisenheimer}(1991)}]{Roeser91}
{R\"oser}, H.~. \& {Meisenheimer}, K. 1991, \aap, 252, 458

\bibitem[{{Ruphy} {et~al.}(1996){Ruphy}, {Robin}, {Epchtein}, {Copet},
  {Bertin}, {Fouque}, \& {Guglielmo}}]{Ruphy96}
{Ruphy}, S., {Robin}, A.~C., {Epchtein}, N., {et~al.} 1996, \aap, 313, L21

\bibitem[{{Siegel} {et~al.}(2002){Siegel}, {Majewski}, {Reid}, \&
  {Thompson}}]{Siegel02}
{Siegel}, M.~H., {Majewski}, S.~R., {Reid}, I.~N., \& {Thompson}, I.~B. 2002,
  \apj, 578, 151

\bibitem[{{Wainscoat} {et~al.}(1992){Wainscoat}, {Cohen}, {Volk}, {Walker}, \&
  {Schwartz}}]{Wainscoat92}
{Wainscoat}, R.~J., {Cohen}, M., {Volk}, K., {Walker}, H.~J., \& {Schwartz},
  D.~E. 1992, \apjs, 83, 111

\bibitem[{{Walsh}(1995)}]{eso}
{Walsh}, J. 1995, http://www.eso.org/observing/standards/spectra

\bibitem[{{Wolf} {et~al.}(2001{\natexlab{a}}){Wolf}, {Meisenheimer}, \&
  {R{\"o}ser}}]{Wolf01b}
{Wolf}, C., {Meisenheimer}, K., \& {R{\"o}ser}, H.~. 2001{\natexlab{a}}, \aap,
  365, 660

\bibitem[{{Wolf} {et~al.}(2001{\natexlab{b}}){Wolf}, {Meisenheimer},
  {R{\"o}ser}, {Beckwith}, {Chaffee}, {Fried}, {Hippelein}, {Huang},
  {K{\"u}mmel}, {von Kuhlmann}, {Maier}, {Phleps}, {Rix}, {Thommes}, \&
  {Thompson}}]{Wolf01a}
{Wolf}, C., {Meisenheimer}, K., {R{\"o}ser}, H.~., {et~al.} 2001{\natexlab{b}},
  \aap, 365, 681

\bibitem[{{Young}(1976)}]{Young76}
{Young}, P.~J. 1976, \aj, 81, 807

\end{thebibliography}
\end{document}